\documentclass[apj]{emulateapj}

\usepackage{graphicx,times}
\usepackage{subfigure}
\newcommand{\be}{\begin{equation}}
\usepackage{threeparttable}
\usepackage{booktabs}
\newcommand{\ee}{\end{equation}}
\newcommand{\bea}{\begin{eqnarray}}
\newcommand{\eea}{\end{eqnarray}}

\usepackage{amsmath}
\usepackage{cases}
\usepackage{longtable}
\usepackage{hyperref}
\usepackage{epstopdf}
\usepackage{amsmath,bm}
\usepackage{amssymb}
\usepackage{natbib}
\usepackage{morefloats}
\usepackage{multirow}
\usepackage{array}
\usepackage{verbatim}

\begin{document}

\title{Damping of MHD turbulence in partially ionized plasma: implications for cosmic ray propagation}
\author{Siyao Xu\altaffilmark{1}, Huirong Yan\altaffilmark{2} and A. Lazarian\altaffilmark{3}}

\altaffiltext{1}{Department of Astronomy, School of Physics, Peking University, Beijing 100871, China}
\altaffiltext{2}{Kavli Institute for Astronomy and Astrophysics, Peking University, Beijing 100871, China; hryan@pku.edu.cn}
\altaffiltext{3}{Department of Astronomy, University of Wisconsin, 475 North Charter Street, Madison, WI 53706, USA}

\begin{abstract}
We study the damping from neutral-ion collisions
of both incompressible and compressible magnetohydrodynamic (MHD) turbulence in partially ionized medium. 
We start from the linear analysis of MHD waves applying both single-fluid and two-fluid treatments. 
The damping rates derived from the linear analysis are then used in determining the damping scales of MHD turbulence. 
The physical connection between the damping scale of MHD turbulence and cutoff boundary of linear MHD waves is investigated. 
Our analytical results are shown to be applicable in a variety of partially ionized interstellar medium (ISM) phases and solar chromosphere. 
As a significant astrophysical utility, we introduce damping effects to propagation of cosmic rays in partially ionized ISM. 
The important role of turbulence damping in both transit-time damping and gyroresonance is identified. 
\end{abstract}

\keywords{magnetohydrodynamics (MHD)-turbulence-ISM: cosmic rays}

\section{Introduction}
Partially ionized plasmas are universal in both space and astrophysical environments. 
They fill the most volume throughout the Galaxy and build up molecular clouds from which stars form. 
The presence of neutrals affects the plasma dynamics indirectly, mediated by the friction between ions and neutrals
(see studies by e.g. \citealt{Pidd56, Kulsrud_Pearce}).

On the other hand, astrophysical plasmas are generally turbulent and magnetized
(see e.g. \citealt{Armstrong95, CheL10}). 
Both incompressible and compressible MHD turbulence in fully ionized medium have been ealier studied 
(\citealt{GS95}, hereafter GS95, \citealt{LG01, CL02_PRL, CL03}). 
The GS95 model of turbulence and its extensions in compressible fluids successfully reproduce the properties of MHD turbulence and 
are supported by numerical tests 
\citep{MG01, CL02_PRL, CL03, KowL10}.

The turbulence theory for a partially ionized medium should be constructed under the consideration of the damping by neutrals. 
There are extensive studies in existing literature regarding neutral-ion collisional damping, using either single-fluid 
(e.g. \citealt{Braginskii:1965, Bals96, Khod04, Fort07})
or two-fluid description 
(e.g. \citealt{Pudr90, Bals96, Kum03, Zaqa11, Mou11, Sol13, Soler13}).
The single-fluid approach has simpler formalism and hence physical transparency, but is restricted on the limited wave frequencies lower than 
the neutral-ion collisional frequency. While the two-fluid approach treats ion-electron and neutral fluids separately and is valid over a broad range of wave frequencies. 

But all these studies are restricted to linear MHD waves. 
The modern advances of MHD turbulence reveal it is a highly non-linear phenomenon, 
with the energy injected at a large scale cascading successively downwards at the turbulence eddy turnover rate
(see e.g. \citealt{CL03, CLV_lecnotes}). 
The effect of neutral-ion collisions on the cascade of compressible MHD turbulence was investigated by 
\citet{LG01}, 
but they mainly focused on a highly-ionized medium with gas pressure overwhelming magnetic pressure. 
The role of neutral viscous damping in a mostly neutral gas in turbulent motions was thoroughly studied in 
\citet{LVC04}. 
Combining the two damping effects, a recent work by 
\citet{XLY14} (hereafter Paper \uppercase\expandafter{\romannumeral1})
carried out a detailed study on damping of the Alfv\'{e}n component of MHD turbulence in a variety of turbulence regimes and demonstrated 
its significance in explaining the observed linewidth differences between neutrals and ions, and the potential value in measuring magnetic 
field in molecular clouds. 
In the current work, we extend the theory developed in Paper \uppercase\expandafter{\romannumeral1} to include compressible MHD turbulence.
The weak coupling of Alfv\'{e}n to compressible modes was suggested by 
GS95. 
The decomposition of MHD turbulence into Alfv\'{e}n, fast, and slow modes has been confirmed and realized through a statistical procedure 
in both Fourier and real spaces
\citep{CL02_PRL, CL03, KowL10}.
On this basis, we can deal with the three components of MHD turbulence separately. We take advantage of the well-established linear theory on 
wave propagation and dissipation in studying the damping processes of MHD turbulence. 

The output of this investigation has a wide range of astrophysical applications, for instance, cosmic ray (CR) propagation. 
The propagation of CRs in ISM is usually described in terms of diffusive transport of CRs in a turbulent magnetic field
(see e.g. \citealt{Schlickeiser02}).
The well established statistics on the scattering of CRs by magnetic irregularities can successfully explain the high-degree isotropy of the comic radiation. 
But the conventionally used, simple model of CR diffusion assuming a universal diffusion coefficient faces big challenges in interpreting
more and more observational findings
(e.g. \citealt{Hun97, Guil07, PAMELA11,Ackermann12}). 
With the development and acceptance of the modern concept of MHD turbulence
(see recent reviews by \citealt{BraL14, BL15}),
significant refinements and modifications have been progressively made in the picture of CR propagation, 
especially in the aspect of quantitatively relating the CR diffusion and the properties of the turbulent medium through which they pass. 

The consideration of turbulence damping is necessary for a proper description of the interaction between CR particles and MHD turbulence. 
The study in 
\citet{YL02,YL04, YL08} 
showed the important influence of turbulence properties and its damping on CR scattering and propagation in fully ionized plasma. 
The ionization fraction possessed by ISM varies over a vast expanse, ranging from fully ionized to almost entirely neutral media
\citep{Spit78, Drai11}.
In the presence of neutrals, the damping process in partially ionized medium is completely different from that in fully ionized medium.
Accordingly, the propagation of CRs in partially ionized medium is likely to present distinctive features and deserves a special attention. 
With this motivation, we proceed the earlier research in partially ionized medium and aim to achieve a realistic model of CR transport in ISM.

The organization of the paper is as follows. In Section 2, we introduce the concept of neutral-ion and ion-neutral decoupling. 
We then provide a linear analysis on MHD waves using both two-fluid and single-fluid approaches in Section 3. 
With the damping rates obtained from linear analysis, damping of MHD turbulence is presented in Section 4. The numerical tests of our analytical 
results in diverse partially ionized ISM conditions and solar chromosphere are given in Section 5. 
Section 6 contains an application on CR propagation in partially ionized medium. 
Section 7 presents further discussions. The summary is given in Section 8.

\section{Coupling between neutrals and ions} \label{sec: cpl}
The coupling of neutrals and ions depends on the wave frequency of interest in comparison with the neutral-ion collision frequency, 
\begin{equation}
\nu_{ni}=\gamma_d\rho_{i}.
\end{equation} 
Here $\rho_i$ is ion density and $\gamma_d$ is the drag coefficient defined in 
\citet{Shu92}.
$\nu_{ni}$ is used when we consider the drag force exerted on the neutrals by the ions. 
It is related with ion-neutral collision frequency $\nu_{in}$ by 
\begin{equation}
 \nu_{ni}\rho_{n}=\nu_{in}\rho_{i},
\end{equation}   
where $\rho_n$ is neutral density.
The condition $\omega \sim \nu_{ni}$ and $\omega \sim \nu_{in}$ correspond to neutral-ion decoupling scale $k_\text{dec,ni}^{-1}$ and 
ion-neutral decoupling scale $k_\text{dec,in}^{-1}$ respectively. 
In a predominantly neutral medium, $k_\text{dec,ni}^{-1}$ is much larger than $k_\text{dec,in}^{-1}$. 
At greater scales than $k_\text{dec,ni}^{-1}$, neutrals and ions are perfectly coupled and oscillate together. 
In this regime the single-fluid approximation is adequate to study the behavior and damping of MHD waves. 
Below $k_\text{dec,ni}^{-1}$, neutrals start to decouple from ions. Perturbations of magnetic field cannot be instantly transmitted to neutrals, thus 
Alfv\'{e}n waves are suppressed in neutral fluid. 

The expressions of $k_\text{dec}$ are different for different wave modes, depending on the form of their respective wave frequencies.
In the limit case of low-$\beta$, where $\beta=2c_s^2/V_A^2$ is defined as the ratio of gas and magnetic pressure,
both Alfv\'{e}n and fast modes propagate with Alfv\'{e}n speed. 
Their neutral-ion decoupling scales are determined by the conditions $V_A k\cos\theta=\nu_{ni}$ and $V_Ak=\nu_{ni}$ respectively,
where $\theta$ is the wave propagation angle relative to magnetic field. The $\cos\theta$ appears for Alfv\'{e}n modes because their waves travel 
along magnetic field, while fast waves propagate isotropically and their decoupling scales do not depend on propagation direction. 
The intrinsic nature of neutrals allows them to support their own sound waves after they decouple from ions. Once satisfying the condition
$c_{sn} k=\nu_{ni}$, neutrals are able to develop their sound waves and become independent from ions' modulation. Thus the $k_\text{dec,ni}^{-1}$ 
of slow modes depends on $\omega$ of sound waves. 
So we can summarize
in low-$\beta$ plasma, neutral-ion decoupling scale sets the largest scale for the absence of Alfv\'{e}n waves in neutral fluid 
for Alfv\'{e}n and fast modes, but the presence of sound waves in neutrals in the case of slow modes.

In the meantime, ions are still subject frequent collisions with surrounding neutrals above $k_\text{dec,in}^{-1}$. Therefore the remaining wave motions in ions can be strongly influenced by the collisional friction. 
On the other hand, the single-fluid approximation can still be used to describe the MHD waves since ions maintain the coupling state with neutrals. 
At shorter scales than $k_\text{dec,in}^{-1}$, ions also become ineffectively coupled with neutrals. 
The scale $k_\text{dec,in}^{-1}$ is determined by $\omega$ of MHD waves in ions alone. 
For the weakly coupled ions and neutrals, the assumption of strong coupling breaks down, so the two-fluid description with no restriction imposed on 
wave frequency ranges needs to be adopted.
In the following of the paper, we use "strongly" and "weakly" coupled regimes by meaning $k<k_\text{dec,ni}$ and $k>k_\text{dec,in}$ 
respectively. 

Based on above analysis, Table \ref{tab: decsc} lists the decoupling scales for different wave modes and regimes. 
\begin{table}[h]
\renewcommand\arraystretch{2}
\centering
\begin{threeparttable}
\caption[]{Decoupling scales of MHD waves.
}\label{tab: decsc} 
  \begin{tabular}{c|m{1.5cm}<{\centering}|m{1.5cm}<{\centering}}
      \toprule
               &    $k_\text{dec,ni}$     & $k_\text{dec,in}$  \\
    \hline
Alfv\'{e}n &       $\frac{\nu_{ni}}{V_A\cos\theta}$  &  $\frac{\nu_{in}}{V_{Ai}\cos\theta}$   \\
    \hline
 Fast        &       $\frac{\nu_{ni}}{V_A}$        &    $\frac{\nu_{in}}{V_{Ai}}$      \\
    \hline
 Slow      &       $\frac{\nu_{ni}}{c_{sn}}$  & $ \frac{\nu_{in}}{c_{si}\cos\theta}$   \\
\bottomrule
    \end{tabular}
 \end{threeparttable}
\end{table}
 
% we didn't consider decoupling corresponds to neutral viscosity 

\section{MHD waves in partially ionized plasma} \label{sec: linewave}
The study of damping process is based on the linear analysis of MHD waves in partially ionized plasma. 
We present the wave frequencies obtained from both two-fluid and single-fluid approaches. 
And we mainly focus on the solutions in a low-$\beta$ limit.

\subsection{Two-fluid approach}\label{sec: twoflu}

{\it Alfv\'{e}n waves}~~In a neutral-dominated medium, the dissipation of Alfv\'{e}n waves can be enhanced by both netral-ion collisional damping and 
the viscosity effects in neutrals. 
The relative importance of the two damping mechanisms can vary in different phases of ISM. 
When the neutral-ion collisional damping is much more efficient and viscous damping can be safely neglected, 
Alfv\'{e}n waves have a well-studied dispersion relation 
(see e.g., \citealt{Pidd56, Kulsrud_Pearce, Sol13}),
\begin{equation}\label{eq:dp}
    \omega^3+i(1+\chi)\nu_{ni}\omega^2-k^2\cos^2\theta V_{Ai}^2\omega-i\nu_{ni}k^2\cos^2\theta V_{Ai}^2=0, 
\end{equation}  
where $\chi=\rho_n/\rho_i$.
Under the assumption of weak damping $|\omega_I|\ll |\omega_R|$, the approximate solutions are 
 \begin{subequations} \label{eq: nicsol}
 \begin{align}
 &\omega_R^2=\frac{k^2\cos^2\theta V_{Ai}^2((1+\chi)\nu_{ni}^2+k^2\cos^2\theta V_{Ai}^2)}{(1+\chi)^2\nu_{ni}^2+k^2\cos^2\theta V_{Ai}^2},  \\
 &\omega_I=-\frac{\nu_{ni}\chi k^2\cos^2\theta V_{Ai}^2}{2((1+\chi)^2\nu_{ni}^2+k^2\cos^2\theta V_{Ai}^2)}. \label{eq: anasol} 
 \end{align}
 \end{subequations}
It is reduced to  
\begin{subequations} \label{eq: nisc}
 \begin{align}
 &\omega_R^2=V_{A}^2 k^2\cos^2\theta , \label{eq: anasolsca} \\
 &\omega_I=-\frac{\xi_n V_{A}^2 k^2\cos^2\theta }{2\nu_{ni}}  \label{eq: anasolsc} .
 \end{align}
\end{subequations}
in strongly coupled regime, i.e. $\omega \ll \nu_{in} $, and 
\begin{subequations} \label{eq: niwc}
 \begin{align}
 &\omega_R^2=V_{Ai}^2 k^2\cos^2\theta, \label{eq: comdrhwa} \\
 &\omega_I=-\frac{\nu_{in}}{2} \label{eq: comdrhw}
 \end{align}
\end{subequations}
in weakly coupled regime, i.e. $\omega \gg \nu_{in}$. 
%The factor $2$ appears because damping rate is half of the energy dissipation rate

The assumption of weak damping holds in both strongly and weakly coupled regimes, but fails at intermediate wave frequencies. 
The strongly damped region can be confined by the condition $|\omega_I|=|\omega_R|$. From Eq. \eqref{eq: nisc} and \eqref{eq: niwc},
it yields the boundary wavenumbers, 
\begin{subequations} \label{eq: alfcfsct}
\begin{align}
&  k_{c}^+=\frac{2\nu_{ni}}{ V_A \xi_n \cos\theta}, \label{eq: alfcfsc} \\
&  k_{c}^-=\frac{\nu_{in}}{ 2 V_{Ai} \cos\theta}.
\end{align}
\end{subequations} 
$|\omega_I|$ converges with $|\omega_R|$ at $k_c^\pm$. Within $[k_c^+, k_c^-]$, the Alfv\'{e}n waves become nonpropagating 
and have purely imaginary wave frequencies. 
Although the precise calculation of the cutoff interval of Alfv\'{e}n waves should involve the discriminant of Eq. \eqref{eq:dp} 
(e.g. \citealt{Sol13}),
numerical results confirm the boundaries $[k_{c}^+, k_{c}^-]$ obtained from the equality between $|\omega_I|$ and $|\omega_R|$ 
can very well confine the nonpropagating interval with $\omega_R=0$. 
Therefore we choose this simple derivation and use the resulting $k_{c}^\pm$ as the limit wavenumbers of the cutoff region. 

By comparing with Table \ref{tab: decsc}, we find the cutoff boundaries are connected with the decoupling scales by 
\begin{subequations} 
\begin{align}
& k_{c,\|}^+=\frac{2}{\xi_n} k_\text{dec,ni,$\|$}, \\
& k_{c,\|}^-=\frac{1}{2} k_\text{dec,in,$\|$}, 
\end{align}
\end{subequations} 
showing the cutoff region is slightly smaller than the $[k_\text{dec,ni}, k_\text{dec,in}]$ range.

When neutral viscosity dominates over the neutral-ion collisional damping, the approximate damping rate in strongly coupled regime is 
(\citealt{LVC04}; Paper \uppercase\expandafter{\romannumeral1})
\begin{equation}
   |\omega_I|=\tau_\upsilon^{-1}=k^2\nu_n, \label{eq: nvsappb}
\end{equation}
where $\nu_n=c_{sn}/(n_n\sigma_{nn})$ is the kinematic viscosity in neutrals. 
By equaling the above $|\omega_I|$ and $|\omega_R|$ from Eq. \eqref{eq: anasolsca}, the corresponding cutoff wavenumber is 
\begin{equation}\label{eq: alfnvkc}
  k_{c}^+= \frac{V_A\cos\theta}{\nu_n}.
\end{equation}
The wave frequencies and $k_{c}^-$ in weakly coupled regime are the same as those in the case of neutral-ion collisional damping.

{\it Fast and slow waves}~~We then turn to the magnetoacoustic waves. We consider neutral-ion collisional friction as the dominant damping mechanism. %why
We adopt the dispersion relation given by 
\citet{Soler13}
(see also \citealt{Zaqa11}). 
Taking advantage of the weak damping assumption, analytic solutions can be obtained in the limit case of strongly coupled fluids, 
\begin{subequations} 
\begin{align}
& \omega_R^2=\frac{1}{2}\left[(c_{s}^2+V_{A}^2)\pm\sqrt{(c_{s}^2+V_{A}^2)^2-4c_{s}^2V_{A}^2\cos^2\theta}\right]k^2, \\
& \omega_I=-\frac{k^2[\xi_nV_A^2(c_s^2k^2-\omega_R^2)+\xi_i c_s^2\omega_R^2]}{2\nu_{ni}[k^2(c_s^2+V_A^2)-2\omega_R^2]}. \label{eq: acdrgen}
\end{align}
\end{subequations}
Here $c_s=\sqrt{c_{si}^2\xi_i+c_{sn}^2\xi_n}$ is the sound speed in strongly coupled ions and neutrals.
The above solutions are consistent with those in earlier work, e.g. 
\citet{Ferr88}.
Different from Alfv\'{e}n waves, the compressibility introduces the important parameter $\beta$. Its definition was given in Section \ref{sec: cpl}.
In a low-$\beta$ plasma, the solutions in strongly coupled regime have simple expressions, 
\begin{subequations} \label{eq: scfadrt}
\begin{align}
& \omega_R^2=V_A^2 k^2, \label{eq: scfadra}\\
& \omega_I=-\frac{\xi_nV_A^2k^2}{2\nu_{ni}}, \label{eq: scfadr}
\end{align}
\end{subequations}
for fast modes, and 
\begin{subequations}  \label{eq: scsldrt}
\begin{align}
& \omega_R^2=c_s^2 k^2 \cos^2\theta, \label{eq: scsldra}\\
& \omega_I=-\frac{\xi_nc_s^2k_\perp^2 }{2\nu_{ni}}, \label{eq: scsldr}
\end{align}
\end{subequations}
for slow modes. 

In weakly coupled regime, under the low-$\beta$ condition, the propagating component of wave frequency becomes 
\begin{equation} \label{eq: falbwkr}
  \omega_R^2=V_{Ai}^2k^2
\end{equation}
for fast modes, and 
\begin{equation} \label{eq: slowlbwkr}
   \omega_R^2=c_{si}^2k^2\cos^2\theta
\end{equation}
for slow modes. 
And they have the same damping rate as Alfv\'{e}n waves at high wave frequencies (Eq. \eqref{eq: comdrhw}).

If we compare the expressions of $\omega_I$ of the three modes in strongly coupled regime, 
we find Alfv\'{e}n and fast modes both have the damping rate proportional to their quadratic wave frequencies (Eq. \eqref{eq: anasolsc}, \eqref{eq: scfadr}), as they can always induce oscillations of ions orthogonal to magnetic field. The additional magnetic restoring force 
exerted on ions results in relative drift between ions and neutrals and damping to the wave motions. 
It indicates faster propagating waves are more strongly damped. 
But for slow modes, we see from Eq. \eqref{eq: scsldr} that there is no damping to purely parallel propagation with $\theta=0$, 
since in low-$\beta$ plasma, slow modes propagating along magnetic field are analogous to sound waves in the absence of magnetic field, 
and hence undamped. 

We then compare the expressions of $\omega_I$ in both strongly and weakly coupled regimes. We find 
the neutral-ion collisions at low wave frequencies are responsible for coupling the two fluids together, and more 
frequent collisions bring weaker damping to MHD waves.  
Differently, the damping rate in the weak coupling regime (\eqref{eq: comdrhw}) purely depends on ion-neutral collisional frequency. 
The collisional friction plays the role of dissipating waves in ions.

Analogically, the cutoff boundaries can be determined from $|\omega_I|=|\omega_R|$. 
Fast modes have 
\begin{subequations} \label{eq: tffacfscto}
\begin{align}
  & k_{c}^+=\frac{2\nu_{ni}}{ V_A \xi_n}, \label{eq: tffacfsc} \\
  &  k_{c}^-=\frac{\nu_{in}}{ 2 V_{Ai} }, \label{eq: tffacfscb}
\end{align}
\end{subequations}
which are the same as Eq. \eqref{eq: alfcfsct} for Alfv\'{e}n waves except $k_{c, \|}^\pm$ are replaced by $k_{c}^\pm$. 
Slow modes have 
\begin{subequations}
\begin{align}
&  k_{c}^+=\frac{2\nu_{ni}\cos\theta}{ c_s \xi_n \sin^2\theta}, \label{eq: lwslcs} \\
&  k_c^-=\frac{\nu_{in}}{ 2 c_{si} \cos\theta}.
\end{align}
\end{subequations}
The relations between $k_c^\pm$ and the decoupling scales of fast waves (see Table \ref{tab: decsc}) are similar to those of Alfv\'{e}n waves 
in parallel direction to magnetic field, 
\begin{subequations} 
\begin{align}
& k_{c}^+=\frac{2}{\xi_n} k_\text{dec,ni}, \label{eq: cfdecfa} \\
& k_{c}^-=\frac{1}{2} k_\text{dec,in}. 
\end{align}
\end{subequations} 
Slow waves have the same $k_{c,\|}^-=\frac{1}{2} k_\text{dec,in,$\|$}$ as Alfv\'{e}n waves, but the relation between $k_c^+$ and $k_\text{dec,ni}$ is 
less straightforward. We will discuss this character of slow modes in Section \ref{sec: dmt}.

In fact, in the context of anisotropic MHD turbulence, the wave propagation angle $\theta$ appearing in $k_c^\pm$ and decoupling scales also 
depends on scales. In Section \ref{sec: dmt} we will present their exact expressions.

\subsection{Single-fluid approach}\label{sec: sgfapp}
Although the single-fluid approach has a restriction on the range of wave frequencies, i.e. $\omega<\nu_{in}$, 
it has the advantage of a simpler mathematical treatment. 
We take the dispersion relations from 
\citet{Bals96}
under the single-fluid approximation. 
We also follow the approximation of negligible ion density applied in their work for this subsection. 

The dispersion relation of Alfv\'{e}n waves is 
\begin{equation}\label{eq: balalfsg}
    \lambda^2+i\tilde{k}V_{An}\cos\theta \lambda- V_{An}^2\cos^2\theta=0.
\end{equation}
The same notations are used as defined in their paper, 
\begin{equation}\label{eq: balnor}
\lambda=\frac{\omega}{k}, \tilde{k}=k\tilde{L}, \tilde{L}=\frac{V_{An}\cos{\theta}}{\nu_{ni}}.
\end{equation}
Notice that Eq. \eqref{eq: balalfsg} can be directly obtained from Eq. \eqref{eq:dp} by omitting the first term $\omega^3$ in Eq. \eqref{eq:dp}, 
which is unimportant at low wave frequencies.  

The exact solutions to Eq. \eqref{eq: balalfsg} are straightforward, 
\begin{equation}
   \lambda=\frac{-i\tilde{k}\pm \sqrt{4-\tilde{k}^2}}{2} V_{An}\cos\theta.
\end{equation}
As is shown by 
\citet{Kulsrud_Pearce}, 
the solutions become purely imaginary when $\tilde{k}>2$, which is equivalent to $k_\|>2\nu_{ni}/V_{An}$. 
From Eq. \eqref{eq: alfcfsc}, we see $2\nu_{ni}/V_{An} \approx k_{c, \|}^+$ in mostly neutral plasma. 
It shows the single-fluid approach also can capture the lower cutoff boundary.

At $\tilde{k}<2$, we express $\lambda$ as $(\omega_R+i\omega_I)/k$. Together with Eq. \eqref{eq: balnor}, we get
\begin{subequations}\label{eq: sfalfwf}
 \begin{align}
 &\omega_R^2=\frac{4-\tilde{k}^2}{4} V_{An}^2 k^2 \cos^2\theta , \\
 &\omega_I=-\frac{1}{2}\tilde{k}k\cos\theta V_{An}= -\frac{V_{An}^2 k^2 \cos^2\theta }{2\nu_{ni}}. \label{eq: sfalfdr}
\end{align}
\end{subequations}
In comparison with the solutions to the two-fluid dispersion relation in strongly coupled regime (Eq. \eqref{eq: nisc}), 
with $\tilde{k}^2 \ll 4$ at low wave frequencies, Eq. \eqref{eq: sfalfwf} coincides with Eq. \eqref{eq: nisc} in weakly ionized plasma.

The dispersion relation of compressible modes is 
\citep{Bals96},
\begin{equation} \label{eq: sgfdsre}
\begin{aligned}
&   \lambda^4+i\tilde{k}V_{An} \sec \theta  \lambda^3 -(c_n^2+V_{An}^2) \lambda^2  \\
&   -i\tilde{k}V_{An}\sec \theta c_n^2 \lambda   +V_{An}^2 c_n^2\cos^2{\theta}=0.      
\end{aligned}
\end{equation}
Recall that we derived the analytical solutions to the two-fluid dispersion relation under the weak damping approximation in Section \ref{sec: twoflu}. 
To better capture the wave behavior beyond the weak damping limit, here we follow a different procedure called Newton's method 
to obtain the approximate solutions. To simplify the problem, we consider a paradigmatic case of a low-$\beta$ plasma. 
The approximate expressions of wave frequencies in some tractable limit cases are provided in Appendix \ref{app:b}.

We treat the left-hand side of Eq. \eqref{eq: sgfdsre} as a function $f$ and adopt the approximate roots of $f=0$ at the extreme $\beta \to 0$ as the starting point, 
\begin{subequations}
\begin{align}
  & \lambda_{0}^{1,2}=\frac{-i \tilde{k} \sec \theta \pm \sqrt{4-\tilde{k}^2\sec^2\theta}}{2}V_{An}, \label{eq: fafapp} \\ 
  & \lambda_{0}^{3,4}=\pm c_n \cos\theta. 
\end{align}   
\end{subequations}
$\lambda_{0}^{1,2}$ and $\lambda_{0}^{3,4}$ correspond to fast and slow waves respectively. 
Based on $\lambda_0$, a better approximation is given by 
\begin{equation}
    \lambda_1=\lambda_0- \frac{f(\lambda_0)}{f^\prime(\lambda_0)},
\end{equation}
where $f^\prime(\lambda_0)$ is the derivative of $f$ at $\lambda_0$. 
This one-step iteration can already converge to the actual solutions well. But the expressions of 
$\lambda_1$ can be too complicated to be illustrative. 
Therefore, instead of applying $\lambda_1$ (not shown for simplicity), we find $\lambda_1$ can be further reduced at low-$\beta$ condition and become
\begin{subequations}\label{eq: balcomsglb}
\begin{align}
  & \lambda_{1}^{1,2}\approx \frac{-i \tilde{k} \sec \theta \pm \sqrt{4-\tilde{k}^2\sec^2\theta}}{2}V_{An}, \label{eq: fafappbt} \\
  & \lambda_{1}^{3,4}\approx \pm c_n \cos\theta-\frac{c_n^2\tilde{k} \tan^2\theta \cos\theta}{2V_{An}}i.
\end{align}   
\end{subequations} 
We keep $\lambda_{1}^{1,2}$ the same as $\lambda_{0}^{1,2}$ since the modification is negligible. 
From Eq. \eqref{eq: fafappbt}, we see fast waves become nonpropagating when $\tilde{k}>2\cos\theta$. 
That is, $k>2\nu_{ni}/V_{An}$, in accordance with $k_c^+$ given by Eq. \eqref{eq: tffacfsc}.

When $\tilde{k}<2\cos\theta$, we rewrite Eq. \eqref{eq: balcomsglb} in terms of $\omega_R$ and $\omega_I$, and get 
\begin{subequations}
 \begin{align}
 & \omega_R^2=\frac{4-\tilde{k}^2\sec^2\theta}{4}V_{An}^2 k^2, \\
 & \omega_I=-\frac{\tilde{k}k\sec\theta}{2}V_{An}=-\frac{V_{An}^2k^2}{2\nu_{ni}} \label{eq: sgfadr}
 \end{align}
\end{subequations}
for fast waves, and 
\begin{subequations}
 \begin{align}
 & \omega_R^2=c_{sn}^2 k^2 \cos^2\theta, \\
 & \omega_I=-\frac{c_n^2\tilde{k} k \tan^2\theta \cos\theta }{2V_{An}}=-\frac{c_n^2 k_\perp^2 }{2\nu_{ni}} \label{eq: sgsldr}
 \end{align}
\end{subequations}
for slow waves. In weakly ionized medium, they are in good agreement with Eq. \eqref{eq: scfadrt} and \eqref{eq: scsldrt} derived using two-fluid approach. 

\citet{Zaq12} argued that the appearance of the cutoff wavenumber in Alfv\'{e}n waves is not a physical phenomenon under the single-fluid description. 
Nevertheless, above analysis demonstrates in strongly coupled regime, the single-fluid approach provides the same analytical wave frequencies 
as those obtained using two-fluid approach in neutral dominated medium. 
Furthermore, the different analytical procedure we applied for single-fluid MHD can inform us with the lower cutoff boundary directly from the 
expression of $\omega_R$ in the cases of Alfv\'{e}n and fast waves. 
It indicates the single-fluid approach is adequate to deal with turbulence damping when damping occurs in strongly coupled regime. 

In addition, we also need to point out, 
similar to the simplified expression of the two-fluid $\omega_I$ at low-$\beta$ condition (Eq. \eqref{eq: scsldr}), 
we also see from the single-fluid $\omega_I$ (Eq. \eqref{eq: sgsldr}) that 
damping is absent in the case of purely parallel propagation of slow modes. 
This conclusion was earlier reached through normal mode analysis in single-fluid approach by 
\citet{Fort07}, 
whereas the two-fluid approach by 
\citet{Zaqa11}
and the energy equations given by 
\citet{Braginskii:1965}
both show non-zero damping rate for purely parallel propagation of slow modes. 
Indeed, according to the general expression of $\omega_I$ in Eq. \eqref{eq: acdrgen}, $\omega_I$ is not zero at $\theta=0$. But it does not 
mean the simplified expression of $\omega_I$ at low-$\beta$ medium (Eq. \eqref{eq: scsldr} and \eqref{eq: sgsldr}) is invalid.
Because first, from Eq. \eqref{eq: acdrgen} we see the term with $k_\|$ is subdominant and can be safely neglected at low-$\beta$ condition; 
and second, $k_\|$ becomes more and more insignificant compared with $k_\perp$ towards smaller scales due to the turbulence anisotropy (see 
next section). 
The above analytical damping rates will be numerically tested in Section \ref{sec: num}.

\section{Damping of MHD turbulence}\label{sec: dmt}
MHD turbulence is a highly non-linear phenomenon, as a cascade of turbulent energy towards the undamped smallest scale. 
The remarkable feature of the GS95 model of turbulence is the hydrodynamic-like mixing motions of magnetic field lines in perpendicular 
direction have the same time scale as that of the wave-like motions along field lines. That is, 
Alfv\'{e}nic turbulence cascades over one wave period, $k_\| V_A \sim k_\perp v_\perp$, which is called the critical balance condition. 
Combining the Kolmogorov scaling for the hydrodynamic-like motions $v_\perp \propto k_\perp ^{-1/3}$, the critical balance condition naturally leads 
to the scale-dependent anisotropy, $k_\| \sim k_\perp^{2/3}$.
It cannot be overemphasized that the GS95 laws are true only in local reference system in respect with the local configuration of magnetic field lines. 
The importance of the {\it local} notion was not pointed out by GS95, but demonstrated in 
\citet{LV99} 
and later numerically testified in 
\citet{CV00, MG01, CLV_incomp}.
This specification on reference system makes the original GS95 picture more self-consistent 
and entails the dependence on scales of turbulence anisotropy. 
For the sake of tradition and uniformity, we use the notations $k_\|$ and $k_\perp$ when applying the scaling relations, 
but keep in mind that instead of ordinary Fourier components of wave vectors, they should be treated as inverse of 
scales measured parallel and perpendicular to the local mean magnetic field.

The scaling relations were first introduced for trans-Alfv\'{e}nic turbulence in GS95, and later generalized for different regimes of MHD turbulence 
including sub-Alfv\'{e}nic turbulence 
\citep{LV99}.
With the scale-dependent anisotropy of MHD turbulence taken into account, 
the above linear analysis of wave damping can be used to set the damping scales of MHD turbulence. 
We next briefly describe the cascading of MHD turbulence and list the corresponding damping scales derived in Paper \uppercase\expandafter{\romannumeral1}.

\subsection{Alfv\'{e}n modes}
Alfv\'{e}n modes cascade independently from compressible modes and have cascading rate 
(GS95, \citealt{Lazarian06})
\begin{subnumcases}
 {\tau_{cas}^{-1}=\label{eq: supcara}}
k^{2/3}L^{-1/3}V_L,~~~~~~~l_A<1/k<L, \label{eq: supcaraa}\\
k_{\perp}^{2/3}L^{-1/3}V_L,~~~~~~~1/k<l_A, \label{eq: supcarab}
\end{subnumcases}
for super-Alfv\'{e}nic turbulence, i.e., $M_A=V_L/V_A>1$. Here $l_A$ is the scale where turbulence anisotropy starts to develop and increases towards smaller scales. At scales smaller than $l_A$, 
the parallel and perpendicular components of the wave vector with respect to the local direction of magnetic field are related by 
\begin{equation} \label{eq: supscal}
  k_\|\sim l_A^{-1}(k_\perp l_A)^{2/3}.
\end{equation}

Sub-Alfv\'{e}nic turbulence has $M_A<1$, and a cascading rate 
\citep{LV99}
\begin{subnumcases}
{\tau_{cas}^{-1}=\label{eq: subcara}} 
\frac{(kv_l)^2}{\omega_A}=\frac{k^2V_L^2}{k_\perp V_A}, ~~~~~~~~l_{tr}<1/k<L, \label{eq: subcaraa}\\
v_l/l_\perp=k_\perp^{\frac{2}{3}}L^{-\frac{1}{3}}V_LM_A^{\frac{1}{3}}, 
~~~~~1/k<l_{tr}, \label{eq: subcarab}
\end{subnumcases}
where $l_{tr}$ is the transition scale from weak to strong turbulence. 
Weak turbulence only evolves in perpendicular direction with $k_\perp$ increasing during cascade, while strong turbulence 
has the scaling relation
\begin{equation}
 k_\|\sim L^{-1}(k_\perp L)^{2/3}M_A^{4/3}.
\label{eq: subscal}
\end{equation}

The damping scale of turbulence can be approximately determined by setting $|\omega_I| \sim \tau_{cas}^{-1}$. 
With both damping effects included, the general expression of damping scale is 
\begin{subequations}
\label{eq: dssup2dam}
\begin{align}
 & k_{\text{dam},\|}=\frac{-(\nu_n+\frac{V_{Ai}^2}{\nu_{in}})+\sqrt{(\nu_n+\frac{V_{Ai}^2}{\nu_{in}})^2+\frac{8V_A\nu_n l_A}{\xi_n}}}{2\nu_n l_A},  \label{eq: supkpar}\\
 & k_\text{dam}=k_{\text{dam},\|} \sqrt{1+l_A k_{\text{dam},\|}}, ~~~~~~~~~1/k_\text{dam}<l_A.
\end{align}
\end{subequations}
for super-Alfv\'{e}nic turbulence, and 
\begin{equation}
\begin{split}
\label{eq: subgsbc}
& k_{\text{dam},\|}=\frac{-(\nu_n+\frac{V_{Ai}^2}{\nu_{in}})+\sqrt{(\nu_n+\frac{V_{Ai}^2}{\nu_{in}})^2+\frac{8V_A\nu_n LM_A^{-4}}{\xi_n}}}{2\nu_n L M_A^{-4}}, \\
& k_\text{dam}=k_{\text{dam},\|} \sqrt{1+LM_A^{-4} k_{\text{dam},\|}}, ~~~~~~~~~1/k_\text{dam}<l_{tr}.
\end{split}
\end{equation}
for sub-Alfv\'{e}nic turbulence. In most situations, only the dominant damping effect needs to be considered. 
In the case when damping is dominated by neutral-ion collisions, from Eq. \eqref{eq: anasolsc}, \eqref{eq: supcarab}, \eqref{eq: supscal}, \eqref{eq: subcarab} and \eqref{eq: subscal}, we obtain 
\begin{equation} \label{eq: mtnisupds}
     k_\text{dam}=\bigg(\frac{2\nu_{ni}}{\xi_n}\bigg)^{\frac{3}{2}}L^{\frac{1}{2}}V_L^{-\frac{3}{2}}
\end{equation}
for super-Alfv\'{e}nic turbulence when $k_\text{dam}^{-1}<l_A$, and 
\begin{equation}\label{eq: mtnisubds}
   k_\text{dam}=\bigg(\frac{2\nu_{ni}}{\xi_n}\bigg)^{\frac{3}{2}}L^{\frac{1}{2}}V_L^{-\frac{3}{2}}M_A^{-\frac{1}{2}}
\end{equation}
for sub-Alfv\'{e}nic turbulence when $k_\text{dam}^{-1}<l_{tr}$. 

By applying the scaling relations Eq. \eqref{eq: supscal} and \eqref{eq: subscal} to the decoupling scales of Alfv\'{e}n modes in Table \ref{tab: decsc}, we find 
$k_\text{dam}$ in above equations are directly related with $k_\text{dec,ni}$ by 
\begin{equation}\label{eq: damdec}
    k_\text{dam}=(\frac{2}{\xi_n})^\frac{3}{2}k_\text{dec,ni}
\end{equation}
for both super- and sub-Alfv\'{e}nic turbulence. And $k_\text{dec,in}$ takes the form 
\begin{equation}\label{eq: decsup}
    k_\text{dec,in}=(\frac{\nu_{in}}{V_{Ai}})^\frac{3}{2}L^\frac{1}{2}M_A^{-\frac{3}{2}}
\end{equation}
in super-Alfv\'{e}nic turbulence, and 
\begin{equation}\label{eq: decsub}
    k_\text{dec,in}=(\frac{\nu_{in}}{V_{Ai}})^\frac{3}{2}L^\frac{1}{2}M_A^{-2}
\end{equation}
in sub-Alfv\'{e}nic turbulence.
Here we use the perpendicular component $k_\perp$ to represent total $k$, since we 
assume the decoupling and damping scales are sufficiently small where turbulence anisotropy is prominent, which is a common occurrence in astrophysical environments.

When neutral viscous damping plays a more important role, 
a combination of Eq. \eqref{eq: nvsappb}, \eqref{eq: supcarab}, \eqref{eq: supscal}, \eqref{eq: subcarab} and \eqref{eq: subscal} yields 
\begin{equation}\label{eq: mtnvsupds}
   k_\text{dam} = \nu_n^{-\frac{3}{4}}L^{-\frac{1}{4}}V_L^\frac{3}{4}, ~~~1/k_\text{dam}<l_A, 
\end{equation}
for super-Alfv\'{e}nic turbulence, and 
\begin{equation}\label{eq: mtnvsubds}
k_\text{dam} = \nu_{n}^{-3/4}L^{-1/4}V_L^{3/4}M_A^{1/4}, ~~~1/k_\text{dam}<l_{tr}, 
\end{equation}
for sub-Alfv\'{e}nic turbulence. The assumption made here is the mean free path of neutral-neutral collisions $l_n$ is relatively small compared with 
$l_A$ in super-Alfv\'{e}nic turbulence and $L$ in sub-Alfv\'{e}nic turbulence, which is reasonable in most ISM conditions.

It is instructive to compare the cutoff wave numbers derived from linear MHD waves and the damping scales of MHD turbulence. 
Due to the critical balance,  $|\omega_R|=\tau_{cas}^{-1}$ holds for Alfv\'{e}nic turbulence. 
Consequently, the cutoff condition $|\omega_R|=|\omega_I|$ of MHD waves is equivalent to the damping condition $\tau_{cas}^{-1}=|\omega_I|$ of MHD turbulence. 
Notice that the wave propagation direction, $\cos\theta=k_\|/k$, follows the scaling relations given by Eq. \eqref{eq: supscal} and \eqref{eq: subscal}, and is scale-dependent. By taking this into account, the $k_c^+$ corresponding to Eq. \eqref{eq: alfcfsc} and \eqref{eq: alfnvkc} 
are fully consistent with $k_\text{dam}$ expressed by Eq. \eqref{eq: mtnisupds}, \eqref{eq: mtnisubds}, \eqref{eq: mtnvsupds}, and \eqref{eq: mtnvsubds}. 
In Table \ref{Tab: ctof}, we summarize the expressions for the cutoff boundaries $k_c^\pm$ of the three turbulence modes by taking the 
scaling relations of MHD turbulence into account. "NI" and "NV" represent neutral-ion collisional and neutral viscous damping. 
The asterisks indicate $k_c^+$ coincides with $k_\text{dam}$ in corresponding situations. 

Regarding the relation between the cutoff boundaries and decoupling scales, from Eq. \eqref{eq: damdec}, \eqref{eq: decsup} and \eqref{eq: decsub}, 
we deduce 
\begin{subequations}
 \begin{align}
 & k_c^+=(\frac{2}{\xi_n})^\frac{3}{2}k_\text{dec,ni}, \label{eq: kpdecnia}\\
 & k_c^-=2^{-\frac{3}{2}} k_\text{dec,in}. \label{eq: kmdin}
\end{align}
\end{subequations}
in the case of neutral-ion collisional damping. 
The difference between them is independent of turbulence properties. 

\begin{table}[h]
\renewcommand\arraystretch{2}
\centering
\begin{threeparttable}
\caption[]{Cutoff boundaries of MHD modes
}\label{Tab: ctof} 
\begin{tabular}{c|c|c|c|c}
 \toprule
                                        & \multicolumn{3}{c|}{ $k_c^+$     }                                                                   &  $k_c^-$       \\
             \hline
 \multirow{4}*{Alfv\'{e}n}  &  \multirow{2}*{Super} & NI &  Eq. \eqref{eq: mtnisupds}$^*$  & \multirow{2}*{$\big(\frac{\nu_{in}}{2V_{Ai}}\big)^\frac{3}{2}L^\frac{1}{2}M_A^{-\frac{3}{2}}$} \\
    \cline{3-4}
                                       &                                 & NV & Eq. \eqref{eq: mtnvsupds}$^*$ &   \\
    \cline{2-5}
                                       &  \multirow{2}*{Sub}    & NI &  Eq. \eqref{eq: mtnisubds}$^*$  & \multirow{2}*{$\big(\frac{\nu_{in}}{2V_{Ai}}\big)^\frac{3}{2}L^\frac{1}{2}M_A^{-2}$} \\
     \cline{3-4}           
                                       &                                 & NV & Eq. \eqref{eq: mtnvsubds}$^*$   &  \\                        
                                       
         \hline
              Fast                   &  \multicolumn{3}{c|}{Eq. \eqref{eq: tffacfsc} }    &       Eq. \eqref{eq: tffacfscb}      \\
         \hline             
   \multirow{2}*{Slow}      & Super    & \multicolumn{2}{c|}{$\big(\frac{2\nu_{ni}}{c_s \xi_n}\big)^\frac{3}{4}L^{-\frac{1}{4}}M_A^\frac{3}{4}$$^*$}  & $\big(\frac{\nu_{in}}{2c_{si}}\big)^\frac{3}{2}L^\frac{1}{2}M_A^{-\frac{3}{2}}$ \\
    \cline{2-5}
                                       & Sub       & \multicolumn{2}{c|}{$\big(\frac{2\nu_{ni}}{c_s \xi_n}\big)^\frac{3}{4}L^{-\frac{1}{4}}M_A$$^*$}  & $\big(\frac{\nu_{in}}{2c_{si}}\big)^\frac{3}{2}L^\frac{1}{2}M_A^{-2}$ \\
\bottomrule
    \end{tabular}
 \end{threeparttable}
\end{table}

\subsection{Compressible modes}
\citet{CL03} found fast modes form in both high- and low-$\beta$ media and follow a cascade similar to the one in acoustic turbulence.
The cascading rate of fast modes is
\citep{CL03, YL04}
\begin{subnumcases}
 {\tau_{cas}^{-1}= \label{eq: carfm}}
(\frac{k}{l_A})^{\frac{1}{2}}\frac{V_A^2}{V_f},~~~~~~~M_A>1, \label{eq: carfma} \\
(\frac{k}{L})^{\frac{1}{2}}\frac{V_L^2}{V_f},~~~~~~~~M_A<1,
\end{subnumcases}
where $V_f$ is the phase speed of fast modes. It takes the form 
\begin{equation}\label{eq: faphsc}
  V_f=\sqrt{\frac{1}{2}(c_{s}^2+V_{A}^2)+\frac{1}{2}\sqrt{(c_{s}^2+V_{A}^2)^2-4c_{s}^2V_{A}^2\cos^2\theta}}
\end{equation}
in strongly coupled regime, and 
\begin{equation}
 V_f=\sqrt{\frac{1}{2}(c_{si}^2+V_{Ai}^2)+\frac{1}{2}\sqrt{(c_{si}^2+V_{Ai}^2)^2-4c_{si}^2V_{Ai}^2\cos^2\theta}}
\end{equation}
in weakly coupled regime. The transition of $V_f$ takes place at $k_\text{dec,in}$ if the turbulence survives damping.
Fast modes cascade radially, so the wave propagation direction can be considered fixed during the cascade. 
The intersection scale between $\tau_{cas}^{-1}$ with $V_f$ from Eq. \eqref{eq: faphsc} and $|\omega_I|$ (Eq. \eqref{eq: acdrgen})
corresponds to the damping scale 
\begin{subnumcases} 
{k_\text{dam}=\label{eq: fdsang}}
l_A^{-\frac{1}{3}}\left(\frac{2\nu_{ni}V_A^2(c_s^2+V_A^2-2V_f^2)}{V_f\left[\xi_n V_A^2(c_s^2-V_f^2)+\xi_i c_s^2V_f^2\right]}\right)^{\frac{2}{3}},\\
L^{-\frac{1}{3}}\left(\frac{2\nu_{ni}V_L^2(c_s^2+V_A^2-2V_f^2)}{V_f\left[\xi_n V_A^2(c_s^2-V_f^2)+\xi_i c_s^2V_f^2\right]}\right)^{\frac{2}{3}}, \label{eq: fdsangsub}
\end{subnumcases} 
for super- and sub-Alfv\'{e}nic turbulence respectively. 
In low-$\beta$ plasma, the above expressions can be written as 
\begin{subnumcases} 
  {k_\text{dam} = \label{eq: fdssimlb}}
  \bigg(\frac{2\nu_{ni}}{\xi_n}\bigg)^{2/3} l_A^{-1/3} V_A^{-2/3}, ~~~~~~~~M_A>1, \\
  \bigg(\frac{2\nu_{ni}}{\xi_n}\bigg)^{2/3} V_L^{4/3} L^{-1/3} V_A^{-2}, ~~M_A<1.
\end{subnumcases} 
By comparing the damping scale with the neutral-ion decoupling scale (Table \ref{tab: decsc}), we find 
\begin{subnumcases} 
  {\frac{k_\text{dam}}{k_\text{dec,ni}} = }
  \bigg(\frac{2}{\xi_n}\bigg)^\frac{2}{3} (k_\text{dec,ni} l_A)^{-\frac{1}{3}}, ~~~~~~~~M_A>1, \\
  \bigg(\frac{2}{\xi_n}\bigg)^\frac{2}{3} (k_\text{dec,ni} L)^{-\frac{1}{3}}M_A^\frac{4}{3}, ~~~M_A<1. 
\end{subnumcases} 
Since the terms $(k_\text{dec,ni} l_A)^{-1}$ and $(k_\text{dec,ni} L)^{-1}$ are usually much smaller than $1$, $k_\text{dam}<k_\text{dec,ni}$
is a common occurrence. It means the damping of fast modes takes place before neutral-ion decoupling. 
It is worthwhile to mention that since the cutoff scale is always smaller than the neutral-ion decoupling scale for fast modes (Eq. \eqref{eq: cfdecfa}), 
the cutoff does not play a role in the damping of fast modes.

Slow modes cascade passively and have the same $\tau_{cas}^{-1}$ as Alfv\'{e}n modes (GS95). 
According to the critical balance, the damping condition $\tau_{cas}^{-1}=|\omega_I|$ is equivalent to $k_\| V_A=|\omega_I|$, 
while the cutoff condition $|\omega_R|=|\omega_I|$ gives $k_\| c_s=|\omega_I|$.
In a low-$\beta$ medium, $\tau_{cas}^{-1}$ is always larger than $|\omega_R|$ on a certain scale. 
It means the cutoff of slow waves happens before the equilibrium between $\tau_{cas}^{-1}$ and $|\omega_I|$ is reached. 
Therefore the damping scale of slow modes depends on the cutoff scale $1/k_c^+$ (see Table \ref{Tab: ctof} for the expressions). 
On the other hand, slow modes dissipate subsequently after Alfv\'{e}n modes are damped out, so their damping scale should not be smaller than 
that of Alfv\'{e}n modes. 
Accordingly, in combination with Eq. \eqref{eq: mtnisupds} and \eqref{eq: mtnisubds}, we have $k_\text{dam}$ of slow modes as 
\begin{equation} \label{eq: slofinds}
   k_\text{dam}=\text{min}\Bigg[\bigg(\frac{2\nu_{ni}}{c_s \xi_n}\bigg)^\frac{3}{4}L^{-\frac{1}{4}}M_A^\frac{3}{4}, ~\bigg(\frac{2\nu_{ni}}{\xi_n}\bigg)^{\frac{3}{2}}L^{\frac{1}{2}}V_L^{-\frac{3}{2}}\Bigg].
\end{equation}
when $1/k_\text{dam}<l_A$ in super-Alfv\'{e}nic turbulence,  and 
\begin{equation} \label{eq: slofindsub}
   k_\text{dam}=\text{min}\Bigg[\bigg(\frac{2\nu_{ni}}{c_s \xi_n}\bigg)^\frac{3}{4}L^{-\frac{1}{4}}M_A, ~\bigg(\frac{2\nu_{ni}}{\xi_n}\bigg)^{\frac{3}{2}}L^{\frac{1}{2}}V_L^{-\frac{3}{2}}M_A^{-\frac{1}{2}}\Bigg].
\end{equation}
when $1/k_\text{dam}<l_{tr}$ in sub-Alfv\'{e}nic turbulence. 
In fact, the ratio between the two terms, $k_c^+$ of slow modes and $k_\text{dam}$ of Alfv\'{e}n modes in Eq. \eqref{eq: slofinds} and \eqref{eq: slofindsub} reveals 
\begin{subnumcases} 
  {\frac{k_{c,s}^+}{k_\text{dam,A}} = }
   \bigg[\bigg(\frac{\xi_n}{2}\bigg) \bigg(\frac{V_A}{c_s}\bigg) (k_{\text{dec,ni},\|} l_A)^{-1}\bigg]^\frac{3}{4}, ~M_A>1, \nonumber  \\
   ~~\\
   \bigg[\bigg(\frac{\xi_n}{2}\bigg) \bigg(\frac{V_A}{c_s}\bigg) (k_{\text{dec,ni},\|} L)^{-1} M_A^4\bigg]^\frac{3}{4}, ~M_A<1, \nonumber \\
   ~~
\end{subnumcases} 
where $k_{\text{dec,ni},\|}=\nu_{ni}/V_A$. The terms $(k_{\text{dec,ni},\|} l_A)^{-1}$ and $(k_{\text{dec,ni},\|} L)^{-1}$ are much smaller than $1$. 
If the $\beta$ value of the medium is not extremely small, $k_{c,s}^+ < k_\text{dam,A}$ usually stands. Thus we can safely take $k_\text{dam}=k_c^+$
for slow modes in most cases. 

Slow modes have the same relation between $k_c^-$ and $k_\text{dec,in}$ as Alfv\'{e}n modes (see Eq. \eqref{eq: kmdin}).
We then examine the relation between $k_c^+$ and $k_\text{dec,ni}$. 
The neutral-ion decoupling condition $c_{sn}k=\nu_{ni}$ can be written as 
\begin{equation}\label{eq: comsdc}
    \frac{\xi_nc_{sn}^2k^2}{2\nu_{ni}}=\frac{\xi_nc_{sn}k}{2}, 
\end{equation}
while the cutoff condition is (see Eq. \eqref{eq: scsldrt})
\begin{equation}\label{eq: comscf}
   \frac{\xi_nc_s^2k_\perp^2 }{2\nu_{ni}}=c_s k_\|. 
\end{equation} 
Under the condition of strong turbulence anisotropy, i.e. $k \sim k_\perp$ and $k_\perp \gg k_\|$, we see the cutoff condition can be reached 
on a larger scale than the neutral-ion decoupling scale. 
Thus slow modes fall into a different situation from fast and Alfv\'{e}n modes (see Eq. \eqref{eq: cfdecfa} and \eqref{eq: kpdecnia}) and can 
have $k_c^+<k_\text{dec,ni}$.
% how to explain 
But in terms of damping, due to $k_\text{dam}=k_c^+$, slow modes are similar to fast modes and also damped out when neutrals and ions 
are strongly coupled together.

Comparing this section with Section \ref{sec: linewave}, we can see 
the main difference between the damping processes of linear MHD waves and MHD turbulence is the consideration of scale-dependent anisotropy, 
which depends on turbulence properties, e.g. super- or sub-Alfv\'{e}nic turbulence, and turbulence regimes, e.g. weak or strong turbulence. 
For our study on turbulence damping of Alfv\'{e}n and slow modes, a fixed wave propagation direction, e.g. purely parallel or perpendicular propagation, 
has no physical meaning. 
Furthermore, as regards scaling estimations, e.g. evaluation of the damping scales, turbulence anisotropy is the essential physical ingredient. 
Ignorance of it can lead to severely wrong conclusions in related astrophysical applications.

\section{Numerical tests in different phases of partially ionized ISM and solar chromosphere} \label{sec: num}    

The generality of the two-fluid approximation allows us to apply our results to a wide variety of astrophysical situations. 
We next apply the analysis on turbulence damping in different phases of partially ionized ISM. 
To avoid confusion, we stress that by numerical tests, we mean numerically solving the full dispersion relations with given physical parameters, 
but not performing any MHD simulations.

\begin{table}[h]
\renewcommand\arraystretch{1.3}
\centering
\begin{threeparttable}
\caption[]{Parameters used for different phases of partially ionized ISM and solar chromosphere.
}\label{Tab: ism} 
  \begin{tabular}{ccccccc}
      \toprule
 & WNM & CNM & MC & DC & SC\\
      \midrule
$n_\text{H}$[$cm^{-3}$]  & $0.4$ & $30$ & $300$ & $10^4$ & $4.2\times10^{12}$ \\
$n_e/n_\text{H}$  & $0.1$ & $10^{-3}$ & $10^{-4}$ & $10^{-6}$ & $1.78\times10^{-2}$ \\
$T$[K]  & $6000$ & $100$ & $20$ & $10$ & $6220$ \\
$B$[$\mu$ G]  & $8.66$ & $8.66$ & $8.66$ & $86.6$ & $6.96\times10^7$ \\
$\beta$  & $0.22$ & $0.23$ & $0.20$ & $0.03$ & $0.03$ \\
$M_A$ & $0.4$ & $2.9$ & $9.2$ & $5.3$ & $0.4$\\
 \bottomrule
    \end{tabular}
 \end{threeparttable}
\end{table}

Table \ref{Tab: ism} lists the typical physical parameters for warm neutral medium (WNM), cold neutral medium (CNM), molecular cloud (MC) 
and dense core in a molecular cloud (DC). The parameters $n_\text{H}$, $n_e/n_\text{H}$, and $T$ are taken from 
\citet{LVC04}. 
We assume temperatures are equal for all species. 
Magnetic field strengths are $\sqrt{3}\times$ the half maximum values of the Zeeman measurements from figure 1 in 
\citet{Crut10}
as the total strength of the three-dimensional magnetic field.
In addition, we adopt the drag coefficient as $\gamma_d=3.5\times10^{13}$cm$^3$g$^{-1}$s$^{-1}$ from 
\citet{Drai83}\footnote{This constant $\gamma_d$ applies in conditions like molecular clouds where the relative velocities between 
neutrals and ions are relatively low and the effective cross-section of neutral-ion collisions is much larger than geometric cross section
(see \citealt{Shu92}).}, 
and the driving condition of turbulence 
\begin{equation}
       L=30\, \text{pc},   ~V_L=10 \,\text{km s}^{-1}.
\end{equation}
The masses of ions and neutrals are $m_i=m_n=m_\text{H}$ for WNM and CNM, and $m_i=29 m_\text{H}$, $m_n=2.3 m_\text{H}$ for the other environments
\citep{Shu92}.

Not only ISM, observations confirm MHD waves are also ubiquitous in the solar atmosphere
(e.g. \citealt{DeP07, DeP12}). 
The damping process of MHD waves in the partially ionized solar atmosphere, i.e. photosphere and chromosphere, has been widely 
suggested as a heating mechanism of the external solar atmosphere
(e.g. \citealt{Ost61, Hol86, NU96, Goo00, Goo01}).
Thus we also extend our analysis to the solar chromosphere (SC) environment. 
In reality, the physical parameters vary with altitude from the photosphere level. For the sake of simplicity, we neglect the vertical variation, but consider a toy model in a homogeneous medium 
with all parameters taken at a median height. This simplified analysis serves as a paradigmatic case of damping in chromosphere-like environment. 
A more sophisticated model may be required when making comparisons with observational data. 

We adopt model C for quiet sun in 
\citet{Vern81}
at a height $1280$ km and only consider hydrogen. Magnetic field strength is obtained by assuming 
$B=B_\text{ph} (\rho/\rho_\text{ph})^{0.3}$
\citep{LeA06}.
With $B_\text{ph}=1.5\times 10^3$ G and $\rho_\text{ph}=2.73\times 10^{-7}$ g cm$^{-3}$ at the photospheric level used, we get $B=69.6$ G. 
For the neutral-ion collisional cross-section, we use the value ($\approx 1 \times 10^{-14}$ cm$^2$) proposed by 
\citet{VrKr13}. 
Their approach with energy dependence and quantum effects contained yields a collisional cross-section two orders of magnitude larger than that obtained from the hard sphere model
\citep{Braginskii:1965}.
Then drag coefficient can be calculated with the expression
\citep{Braginskii:1965, Soler13}, 
\begin{equation}
\gamma_d=\frac{1}{2m_H}\sqrt{\frac{16k_BT}{\pi m_H}} \sigma_{ni}.
\end{equation}
Besides, we use the height of chromosphere as the injection scale $L=2500$ km, and assume turbulent velocity at $L$ as $V_L=30$ km s$^{-1}$ based on the velocity amplitude measurement of 
the chromospheric waves by 
\citet{Oka11}. 
The parameters are also listed in Table \ref{Tab: ism}. 

According to the values of $\beta$ listed in Table \ref{Tab: ism}, in the following comparisons with numerical results, we will use the simplified 
analytical results introduced in previous sections for dealing with low-$\beta$ media.

\subsection{Alfv\'{e}n modes}
\begin{figure*}[htbp]
\centering
\subfigure[WNM]{
   \includegraphics[width=8cm]{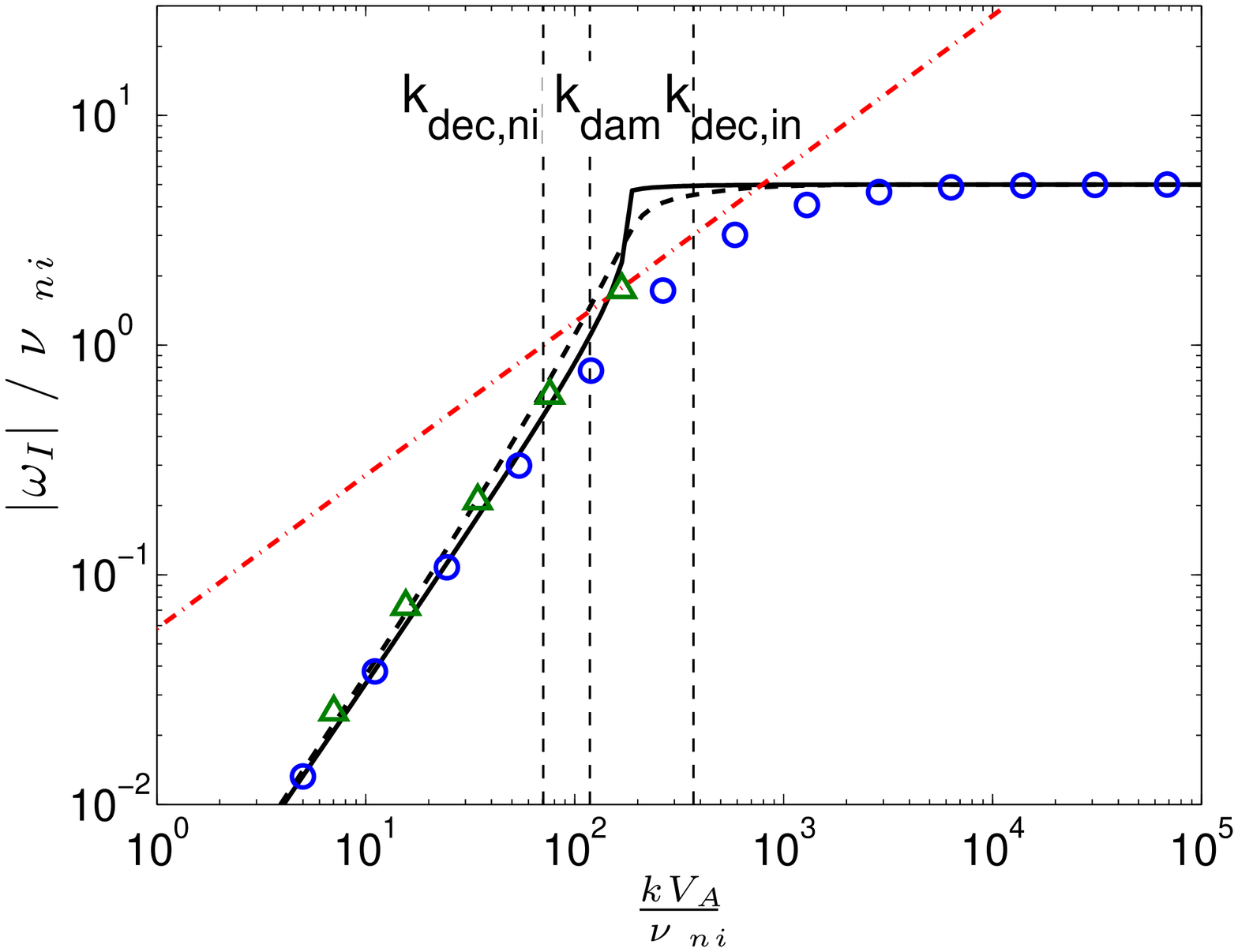}\label{figalfwnm}}
\subfigure[CNM]{
   \includegraphics[width=8cm]{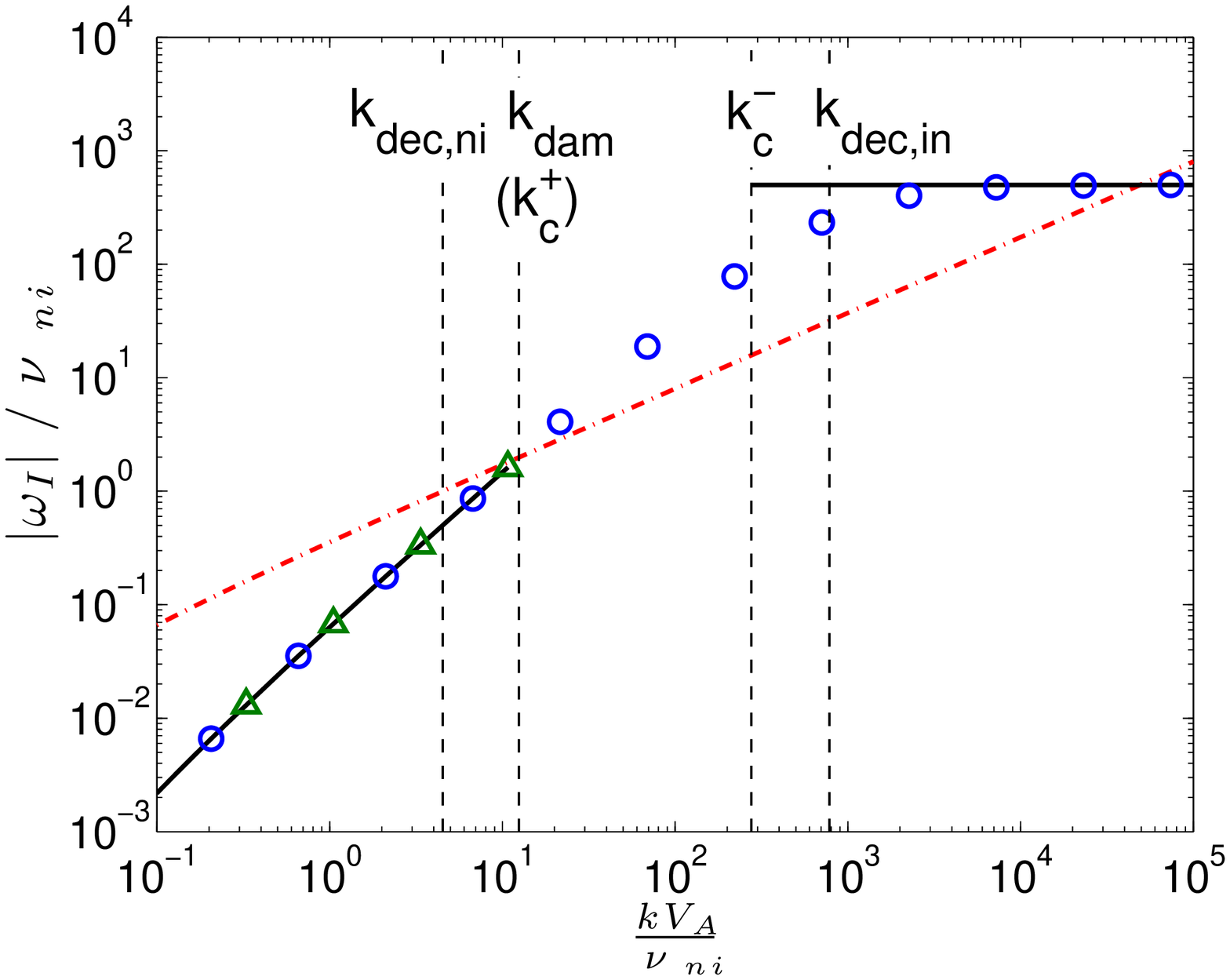}\label{figalfcnm}}
%\subfigure[MC]{
 %  \includegraphics[width=8cm]{mcalf.eps}\label{figalfmc}}
\subfigure[DC]{
   \includegraphics[width=8cm]{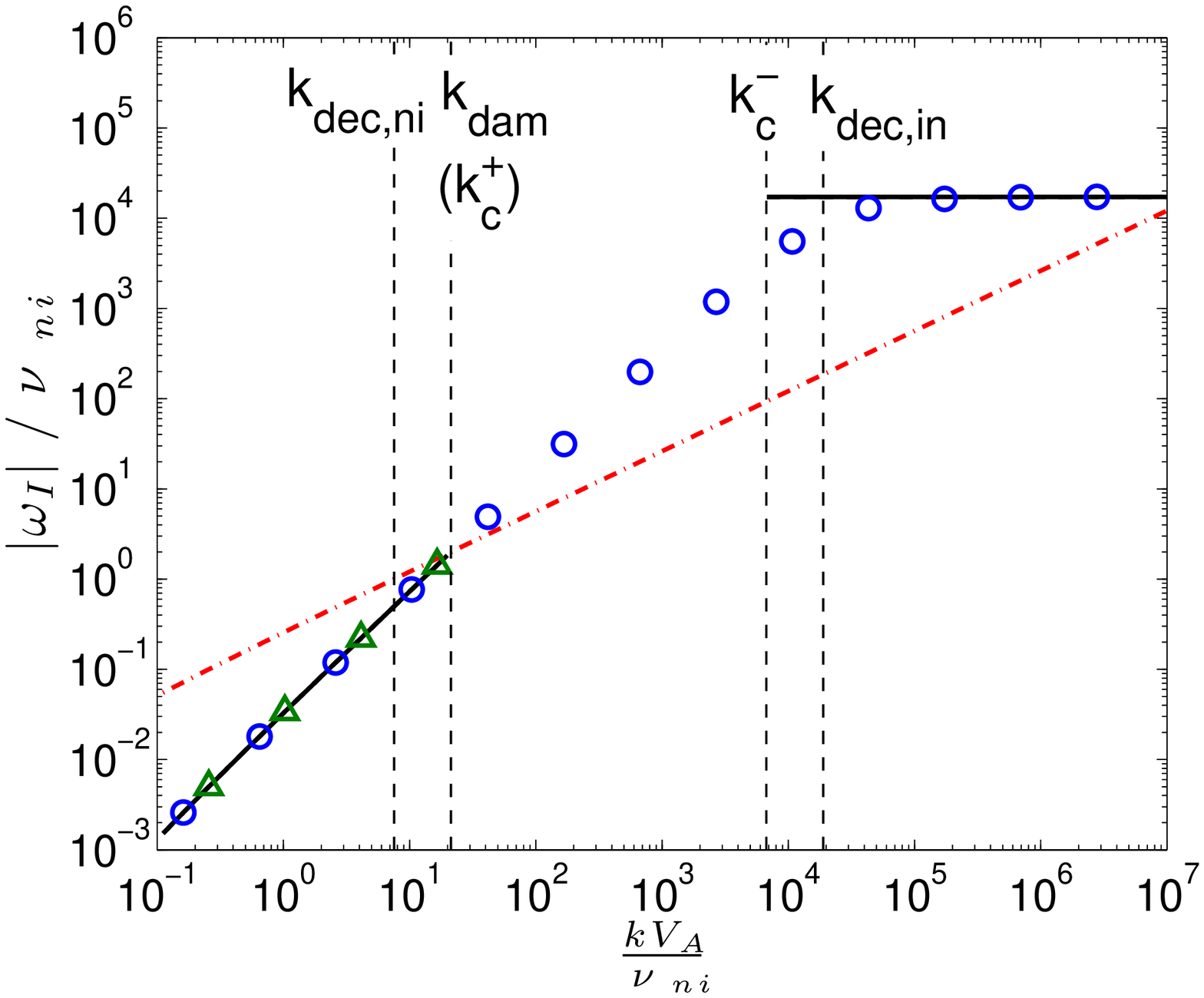}\label{figalfdc}}
\subfigure[SC]{
   \includegraphics[width=8cm]{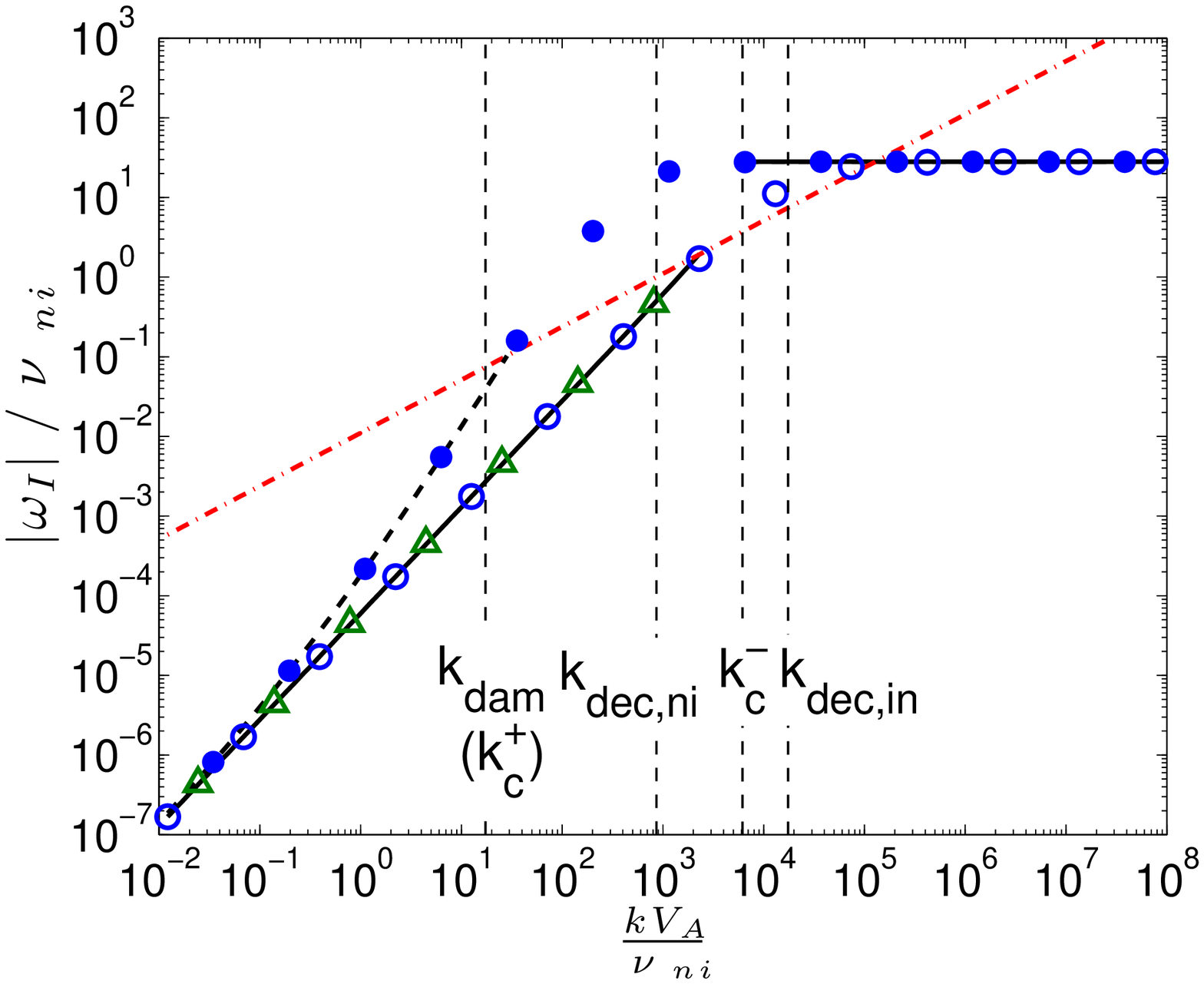}\label{figalfsc}}   
\caption{$|\omega_I|/\nu_{ni}$ vs. $kV_A/\nu_{ni}$ of Alfv\'{e}n modes in various environments. 
Solid lines are numerical damping rates of neutral-ion collisional damping. 
Dashed lines are numerical damping rates including both neutral-ion collisional and viscous damping effects. 
Open circles are analytical damping rates corresponding to neutral-ion collisional damping. 
Triangles show the same as open circles, but for single-fluid approach. 
The filled circles in (d) show the analytical damping rate considering both damping effects.
Dash-dotted lines are the cascading rates of Alfv\'{e}n modes. They intersect with the damping rates at $k_\text{dam}$. 
The analytical wavenumbers $k_\text{dec,ni}$, $k_\text{dec,in}$, $k_\text{dam}$, and $k_c^\pm$ are denoted by vertical dashed lines. }
\label{fig: damalf}
\end{figure*}

We assume the energy is equally distributed into Alfv\'{e}n, fast, and slow modes at the scale of turbulence driving. 
Fig.~\ref{figalfwnm} illustrates the damping of the sub-Alfv\'{e}nic turbulence in WNM. 
The cascading rate is given by Eq. \eqref{eq: subcarab}. Notice the scales shown are within the strong 
turbulence regime. 
We show the damping rates obtained by numerically solving the two-fluid dispersion relations with only neutral-ion collisional damping 
(Eq. \eqref{eq:dp}) and 
both neutral-ion collisional and viscous damping (see Paper \uppercase\expandafter{\romannumeral1}). 
In this case, neutral-ion collisions and neutral viscosity have comparable damping efficiencies in the strongly coupled regime, 
so the general expression of $k_\text{dam}$ 
(Eq. \eqref{eq: subgsbc}) applies. 
The analytical damping rate is from Eq. \eqref{eq: anasol}, which coincides with the numerical result at both small and large
wavenumbers, but underestimates the numerical result at intermediate wavenumbers when $|\omega_I|$ and $|\omega_R|$ are of the same order. 
The single-fluid damping rate is given by Eq. \eqref{eq: sfalfdr}, which traces the numerical one well at small wavenumbers.
The numerical solutions do not exhibit a nonpropagating interval with $\omega_R=0$.
%but the boundary of the highly damped region is indicated by $k_\text{dam}$ (the same as $k_c^+$). 

Fig.~\ref{figalfcnm} and \ref{figalfdc} display the damping of super-Alfv\'{e}nic turbulence in CNM and DC.  
The solutions with both damping effects are in full coincidence with those considering only neutral-ion collisional damping
due to negligible viscous damping. 
Accordingly, $k_\text{dam}$ for neutral-ion collisional damping (Eq. \eqref{eq: mtnisupds}) applies, which coincides with the
lower cutoff boundary $k_c^+$. 
%From Eq. \eqref{eq: mtnisupds} we see,  given the driving condition of turbulence, $k_\text{dam}$ only depends on the density and 
%ionization fraction of the medium. 
By comparing with numerical results, we are convinced that the single-fluid approach can provide a good approximation of the 
actual damping rate down to the damping scale (or cutoff boundary) of Alfv\'{e}n modes. 
The pure imaginary solutions are omitted from the numerical results, corresponding to the cutoff regions. 

Fig. \ref{figalfsc} shows the damping of Alfv\'{e}n modes in sub-Alfv\'{e}nic SC. 
The filled circles represent the analytical damping rate including both damping effects (see Paper \uppercase\expandafter{\romannumeral1}), 
\begin{equation}
   |\omega_I|=\frac{\left[\tau_\upsilon^{-1}(\tau_\upsilon^{-1}+(1+\chi)\nu_{ni})+k^2\cos^2\theta V_{Ai}^2\right]  \chi\nu_{ni} }{2[k^2\cos^2\theta V_{Ai}^2+\chi \tau_\upsilon^{-1}\nu_{ni}+(\tau_\upsilon^{-1}+(1+\chi)\nu_{ni})^2]}.
\end{equation}
Obviously neutral viscosity dominates over neutral-ions collisions in damping Alfv\'{e}n modes. 
It meets the prediction in Paper \uppercase\expandafter{\romannumeral1}
that neutral viscosity can play a more important role of damping in sub-Alfv\'{e}nic turbulence than in super-Alfv\'{e}nic turbulence.
The exact criteria for determining the relative importance between neutral viscosity and neutral-ion collisions in damping super- and 
sub-Alfv\'{e}nic turbulence can be found in Paper \uppercase\expandafter{\romannumeral1}.
As in the case of WNM, the general expression by Eq. \eqref{eq: subgsbc} including both damping effects 
can give a good approximation of $k_\text{dam}$. 
But here we present $k_\text{dam}$ with a simpler form (Eq. \eqref{eq: mtnvsubds}) derived from viscous damping alone 
as an approximate $k_\text{dam}$. 
We see in SC condition, owing to the existence of neutral viscosity, the damping scale is considerably larger than that predicted by 
neutral-ion collisional damping, 
and is also larger than the neutral-ion decoupling scale. 
It shows not only neutral-ion collisions, neutral viscosity should also be considered as an important mechanism in damping Alfv\'{e}n modes.

\begin{figure}[htbp]
\centering
\includegraphics[width=9cm]{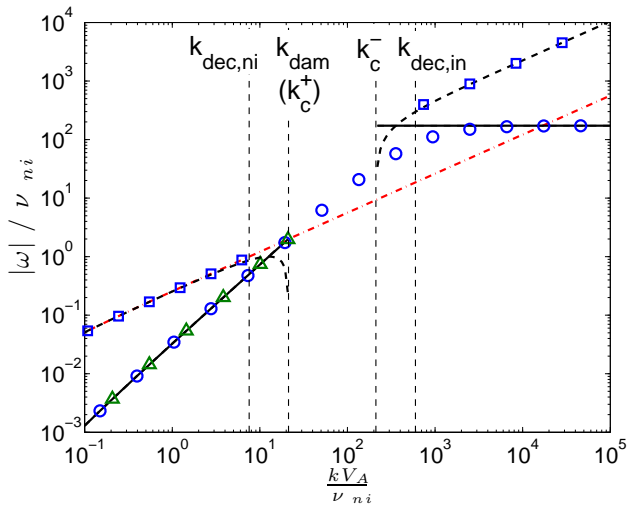}
\caption{$|\omega_R|/\nu_{ni}$ and $|\omega_I|/\nu_{ni}$ vs. $kV_A/\nu_{ni}$ of Alfv\'{e}n modes in MC. 
The same symbols as in Fig. \ref{fig: damalf} are used, except that 
the dashed line and open squares are the numerical and analytical $|\omega_R|$.}
\label{figalfmc}
\end{figure}

Besides damping rate, the propagating component of wave frequency
$|\omega_R|$ is also present in Fig. \ref{figalfmc} as an example in the case of Alfv\'{e}n modes in MC. 
The same symbols are used as in Fig. \ref{fig: damalf},
except that the dashed line and open squares are the numerical and analytical (Eq. \eqref{eq: anasolsca} and \eqref{eq: comdrhwa}) $|\omega_R|$ respectively.
Notice that as pointed out earlier in Section \ref{sec: dmt}, due to the critical balance, $|\omega_R|$ in strongly coupled regime coincides with 
the cascading rate.
We find $|\omega_R|$ starts to decay at $k_\text{dec,ni}$, and is cutoff at $k_c^+$. 
Then $|\omega_R|$ arises again at $k_c^-$, but is only fully resumed after reaching $k_\text{dec,in}$.
Thus the region $[k_\text{dec,ni}, k_\text{dec,in}]$ contains the cutoff zone $[k_c^+, k_c^-]$. 
Different from wave damping, which has damping rate insensitive to decoupling scales, 
the propagating behavior of waves depends on properties of 
fluid coupling, and can only be studied outside $[k_\text{dec,ni}, k_\text{dec,in}]$ region.

\subsection{Fast modes}
\begin{figure*}[htbp]
\centering
\subfigure[WNM]{
   \includegraphics[width=8cm]{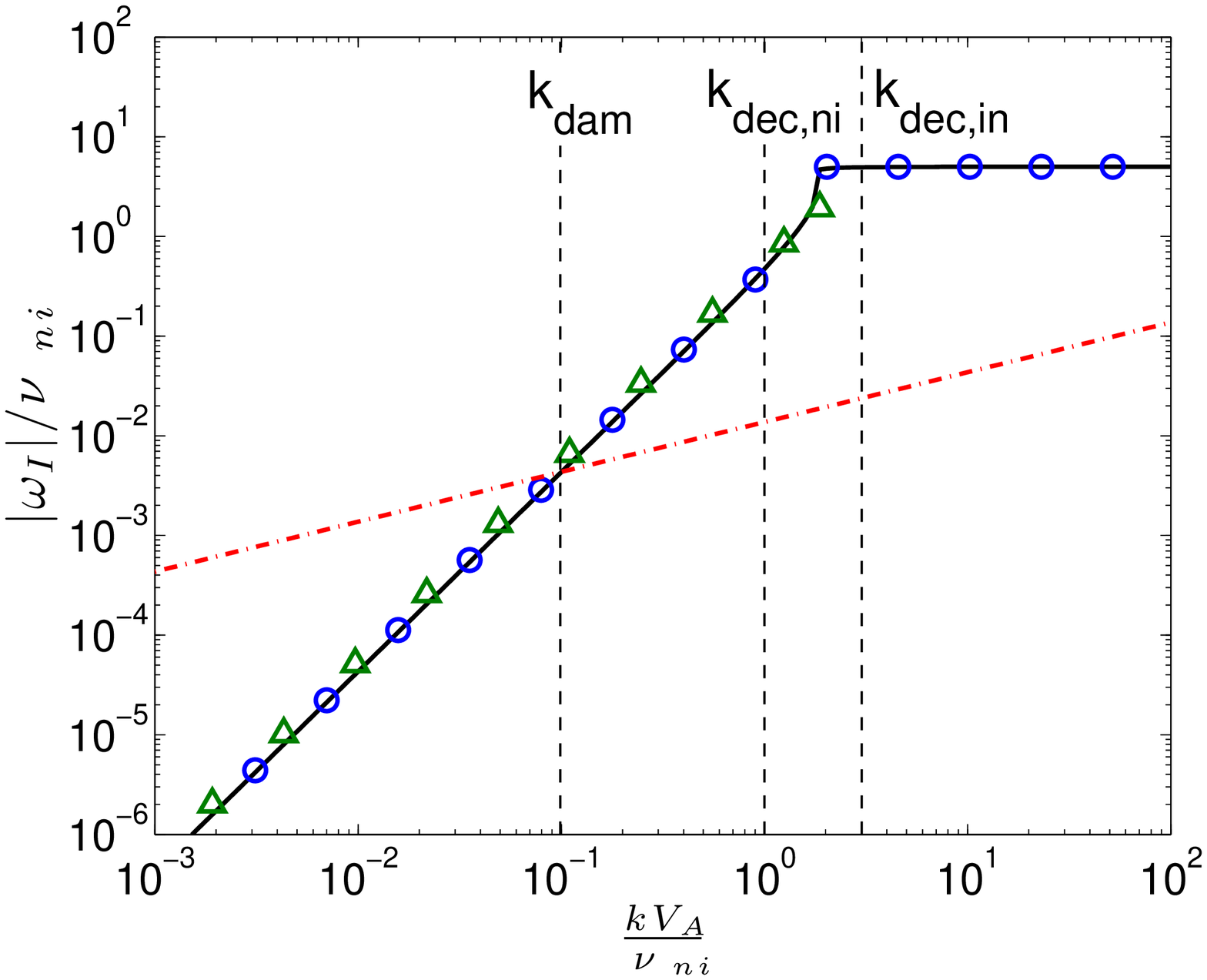}\label{figfastwnm}}
\subfigure[CNM]{
   \includegraphics[width=8cm]{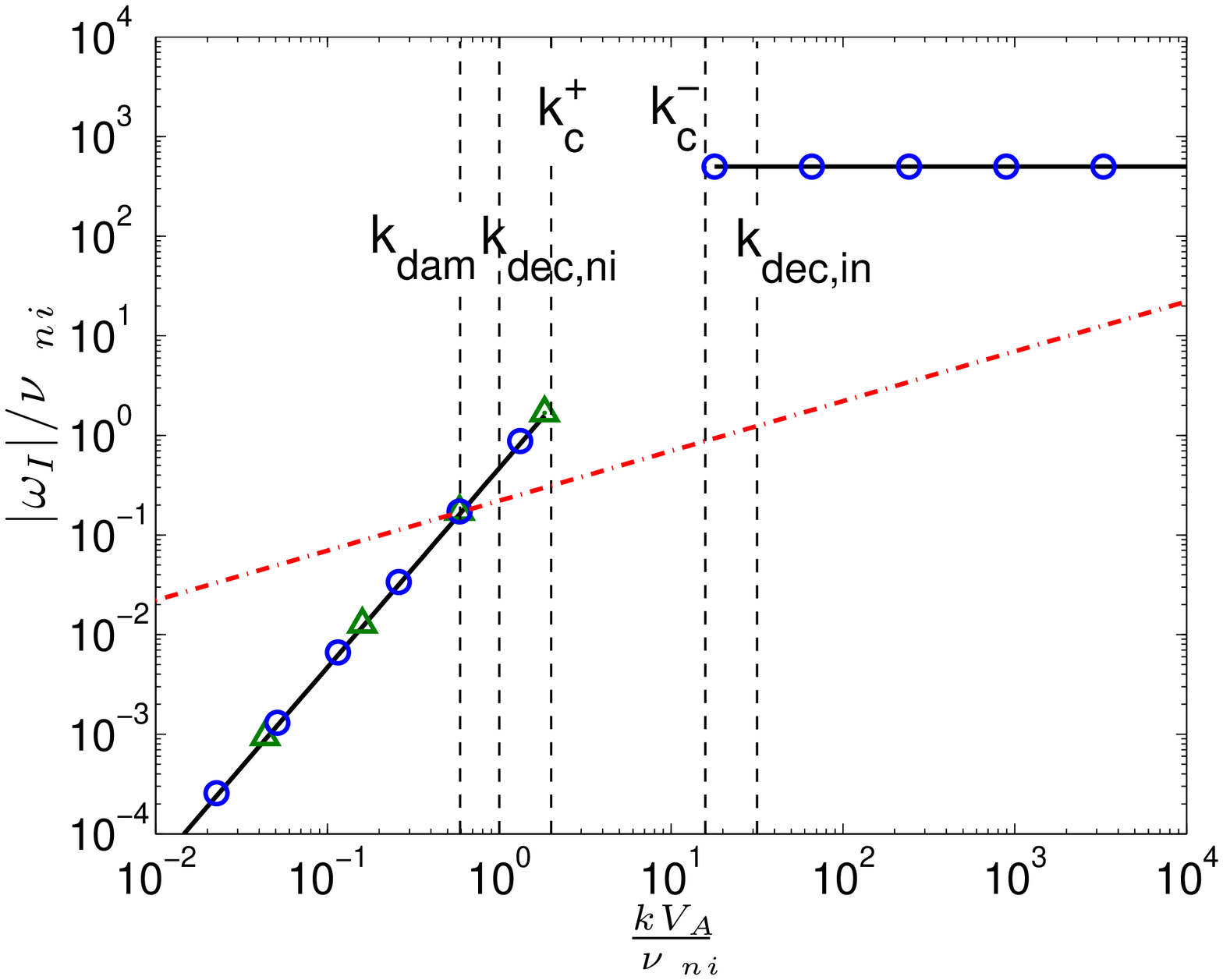}\label{figfastcnm}}
%\subfigure[MC]{
%   \includegraphics[width=8cm]{mcfast.eps}\label{figfastmc}}
\subfigure[DC]{
   \includegraphics[width=8cm]{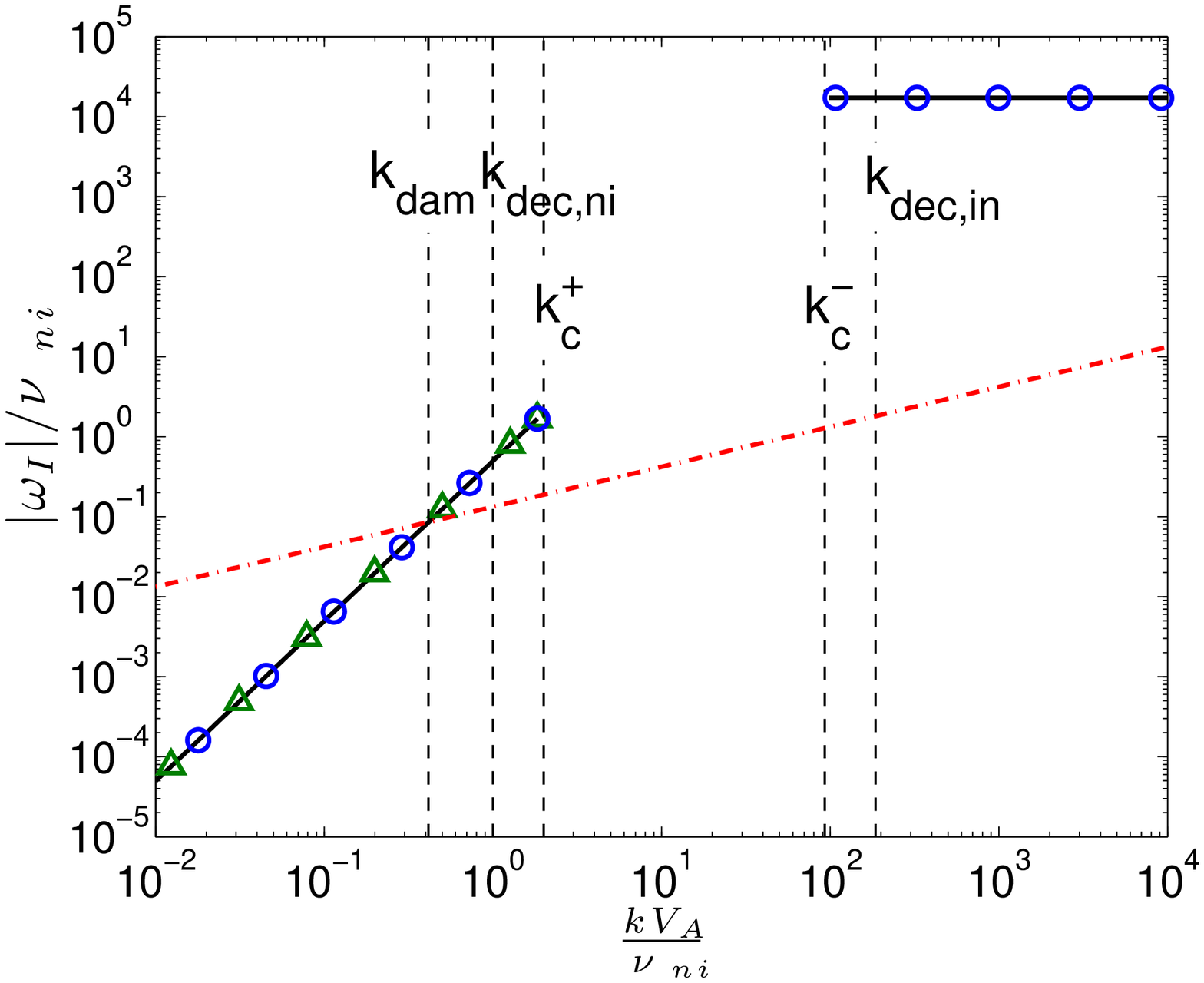}\label{figfastdc}}
\subfigure[SC]{
   \includegraphics[width=8cm]{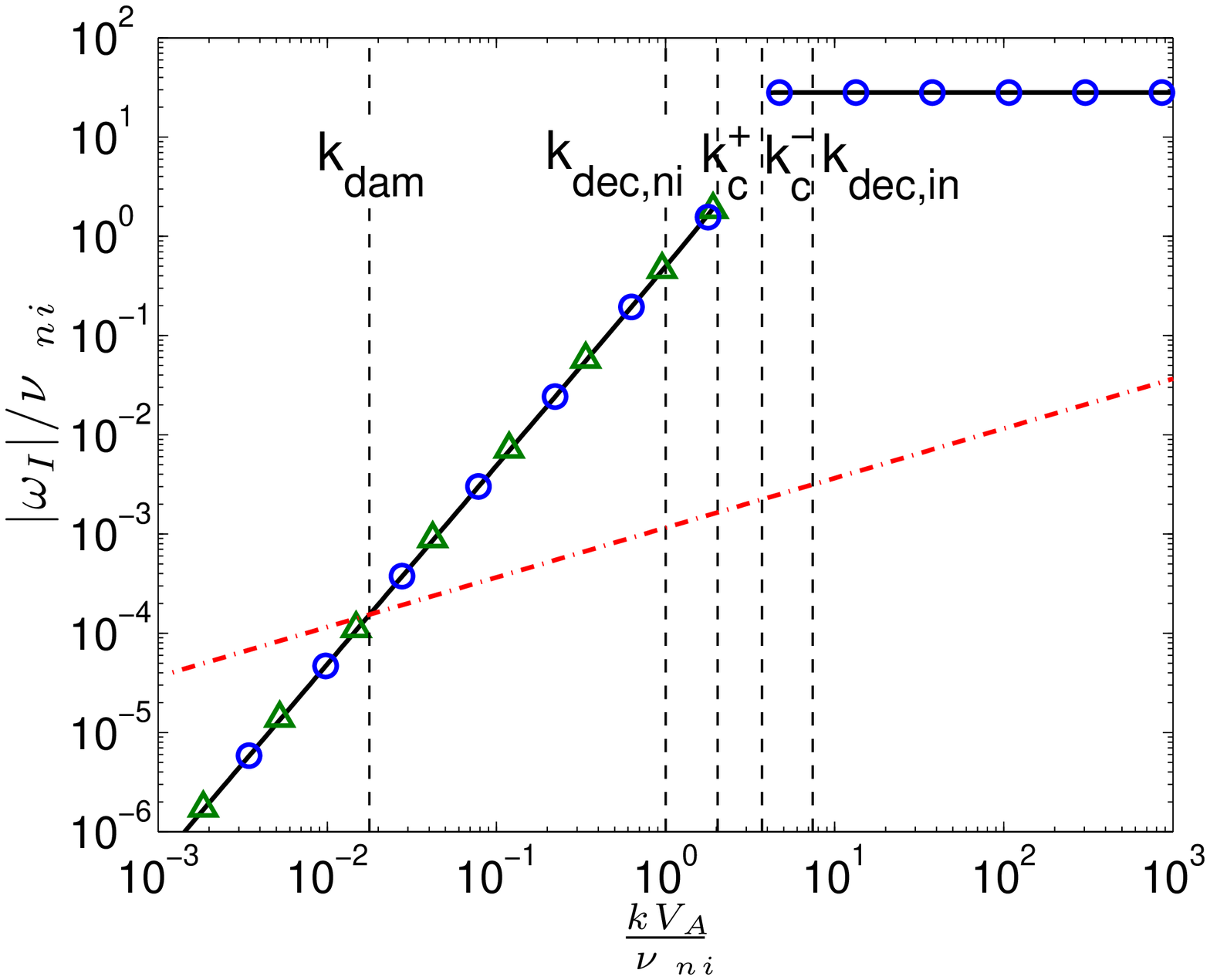}\label{figfastsc}}   
\caption{Same as Fig. \ref{fig: damalf} but for fast modes. Here we set $\theta=45^{\circ}$ as an example.}
\label{fig: damfa}
\end{figure*}

Fig. \ref{figfastwnm}-\ref{figfastsc} display the comparison between the analytical and numerical damping rates of fast modes in different 
ISM conditions and SC. 
The cascading rate is from Eq. \eqref{eq: carfm}, with the wave propagation angle fixed at $45^\circ$. 
The analytical two-fluid (Eq. \eqref{eq: scfadr} and \eqref{eq: comdrhw}) and single-fluid (Eq. \eqref{eq: sgfadr}) damping rates
are in excellent agreement with the numerical result on scales larger than the cutoff boundary $1/k_c^+$. 

\begin{figure}[htbp]
\centering
\includegraphics[width=9cm]{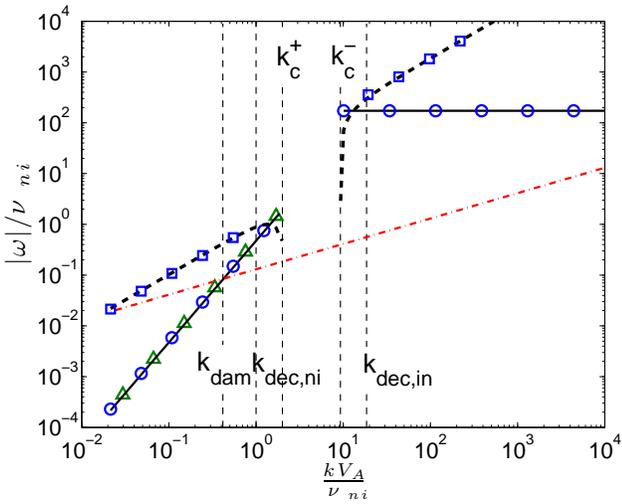}
\caption{Same as Fig. \ref{figalfmc} but for fast modes.}
\label{figfastmc}
\end{figure}

Both $|\omega_R|$ and $|\omega_I|$ are shown in Fig. \ref{figfastmc} for fast modes in MC. The numerical $|\omega_R|$ is obtained 
by numerically solving the dispersion relation (equation (51) in \citealt{Soler13}).
Analytical $|\omega_R|$ is from Eq. \eqref{eq: scfadra} and \eqref{eq: falbwkr}. 
Similar to the case of Alfv\'{e}n modes, the cutoff interval is contained within $[k_\text{dec,ni}, k_\text{dec,in}]$, 
and only outside $[k_\text{dec,ni}, k_\text{dec,in}]$, fast waves have phase speed equal to the Alfv\'{e}n speed.

\begin{figure}[htbp]
\centering
\includegraphics[width=9cm]{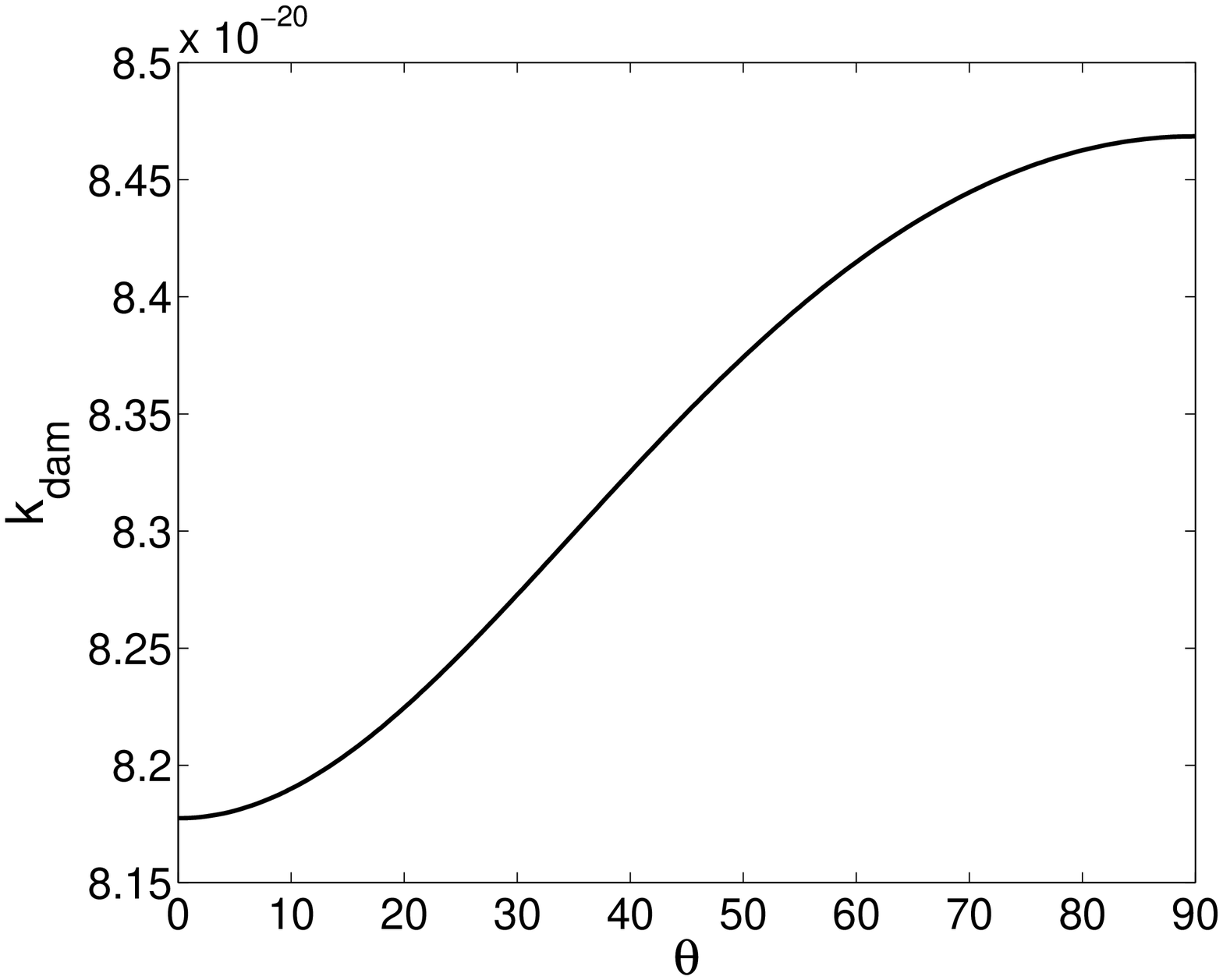}
\caption{$k_\text{dam}$ of fast modes as a function of $\theta$ in WNM. }
\label{figfdswnm}
\end{figure}

In all the conditions present, fast modes are damped out when neutrals and ions are strongly coupled, i.e. $k_\text{dam}<k_\text{dec,ni}$. 
In particular, fast modes in WNM are the most severely damped among the ISM phases. 
WNM has much lower total density than other ISM conditions, which leads to a faster wave phase speed and lower 
neutral-ion collisional frequency. Accordingly, fast modes in WNM have a lower cascading rate (Eq. \eqref{eq: carfm}) but a higher 
damping rate (see discussions in Section \ref{sec: twoflu}). It means the intersection between the two rates can take place at a large scale. 
In SC, high wave phase speed comes from strong magnetic field strength and leads to a low cascading rate. 
Although large ion density yields a high neutral-ion collisional frequency and low damping rate (Eq. \eqref{eq: scfadr}), it can still exceed the 
substantially slow cascade on a large scale. 
Therefore fast modes in SC are also severely damped. 
The vertical dashed lines indicate the damping scale given by Eq. \eqref{eq: fdssimlb}. 
It shows the approximate $k_\text{dam}$ at low-$\beta$ limit is consistent with numerical results. 
In fact, according to Eq. \eqref{eq: fdsang}, $k_\text{dam}$ has a dependence on the wave propagation direction. 
Fig.~\ref{figfdswnm} presents the damping scale given by Eq. \eqref{eq: fdsangsub} as a function of $\theta$ in WNM as an example. 
The damping scale in other cases show similar results. 
We find despite the existence of anisotropy, the dependence of $k_\text{dam}$ on $\theta$ is so weak that a constant damping scale can be applied. 
It leads to a conclusion that the energy distribution of fast modes keeps isotropic over all existent scales.

% fig a, cascading rate of neutrals should be modified 

\subsection{Slow modes}
\label{sec: slowsr}
\begin{figure*}[htbp]
\centering
\subfigure[WNM]{
   \includegraphics[width=8cm]{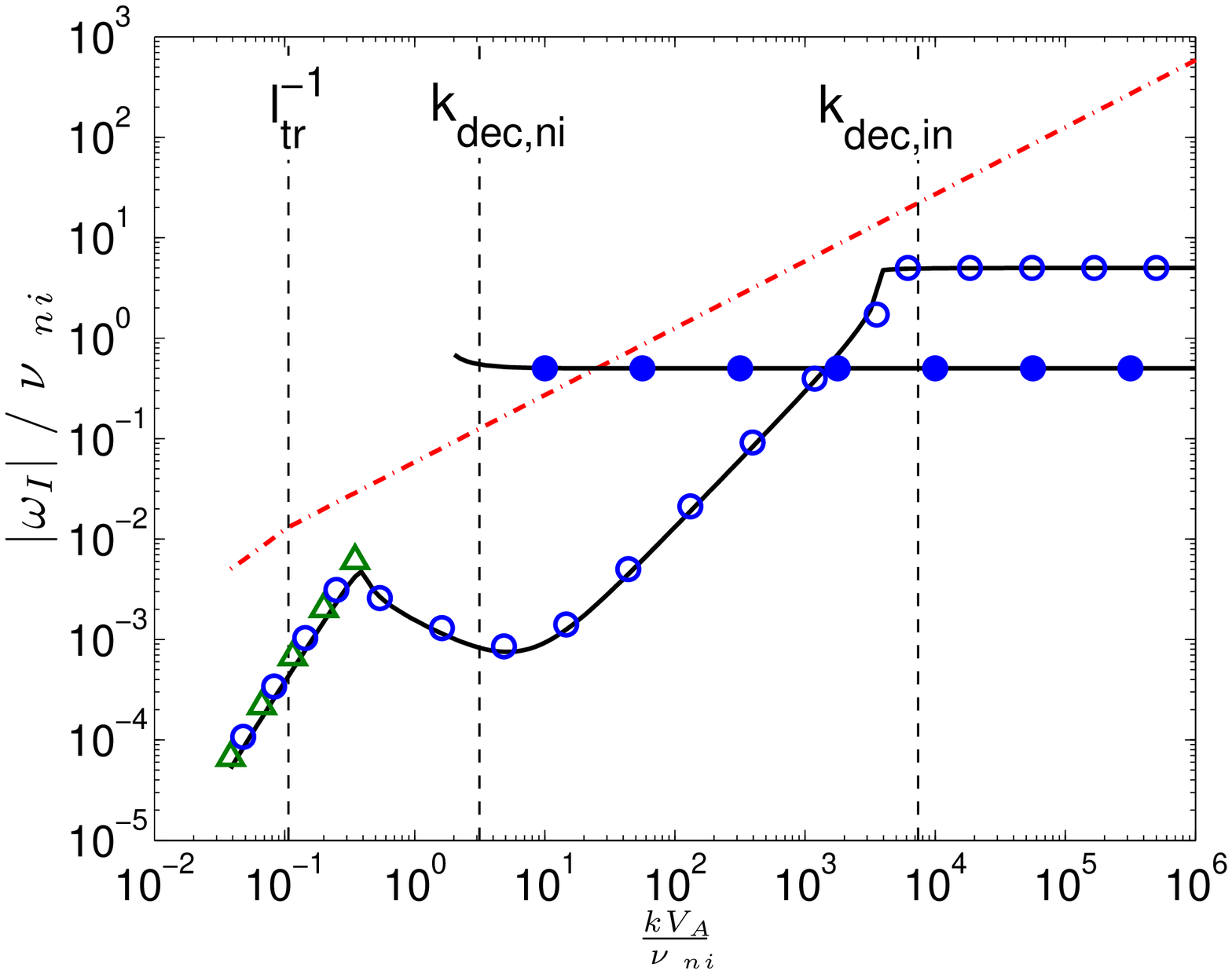}\label{figslowwnm}}
\subfigure[CNM]{
   \includegraphics[width=8cm]{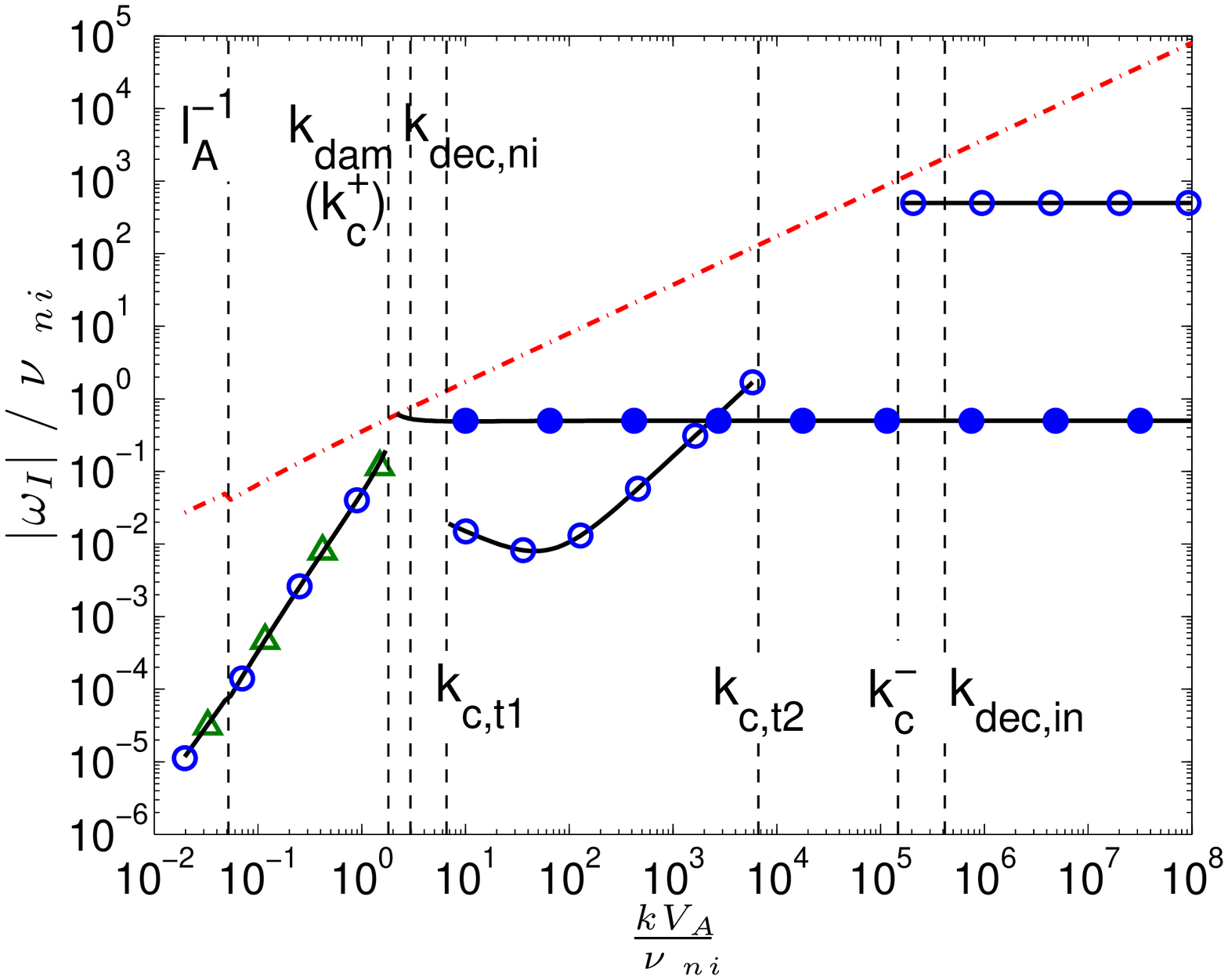}\label{figslowcnm}}
%\subfigure[MC]{
%   \includegraphics[width=8cm]{mcslow.eps}\label{figslowmc}}
\subfigure[DC]{
   \includegraphics[width=8cm]{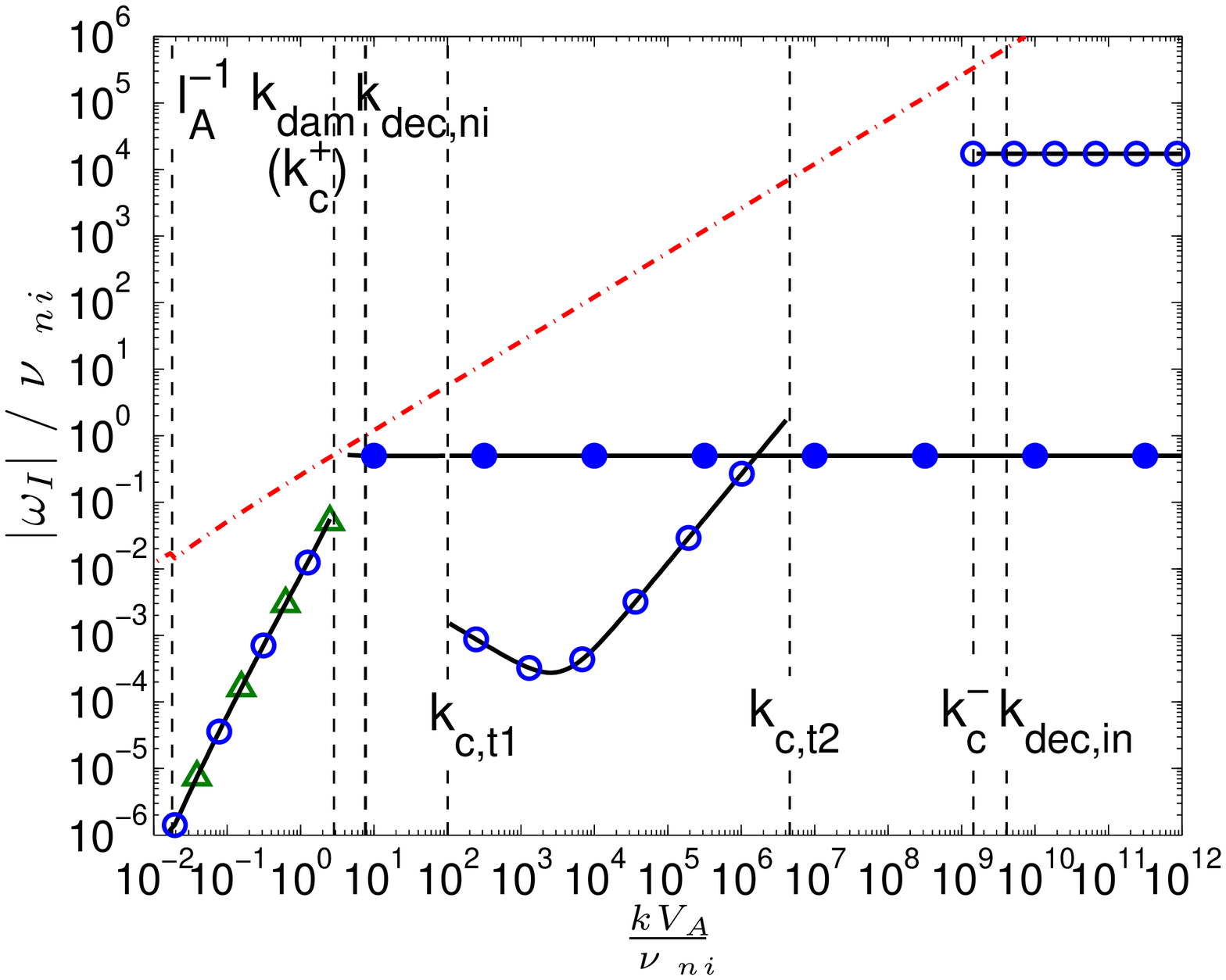}\label{figslowdc}}
\subfigure[SC]{
   \includegraphics[width=8cm]{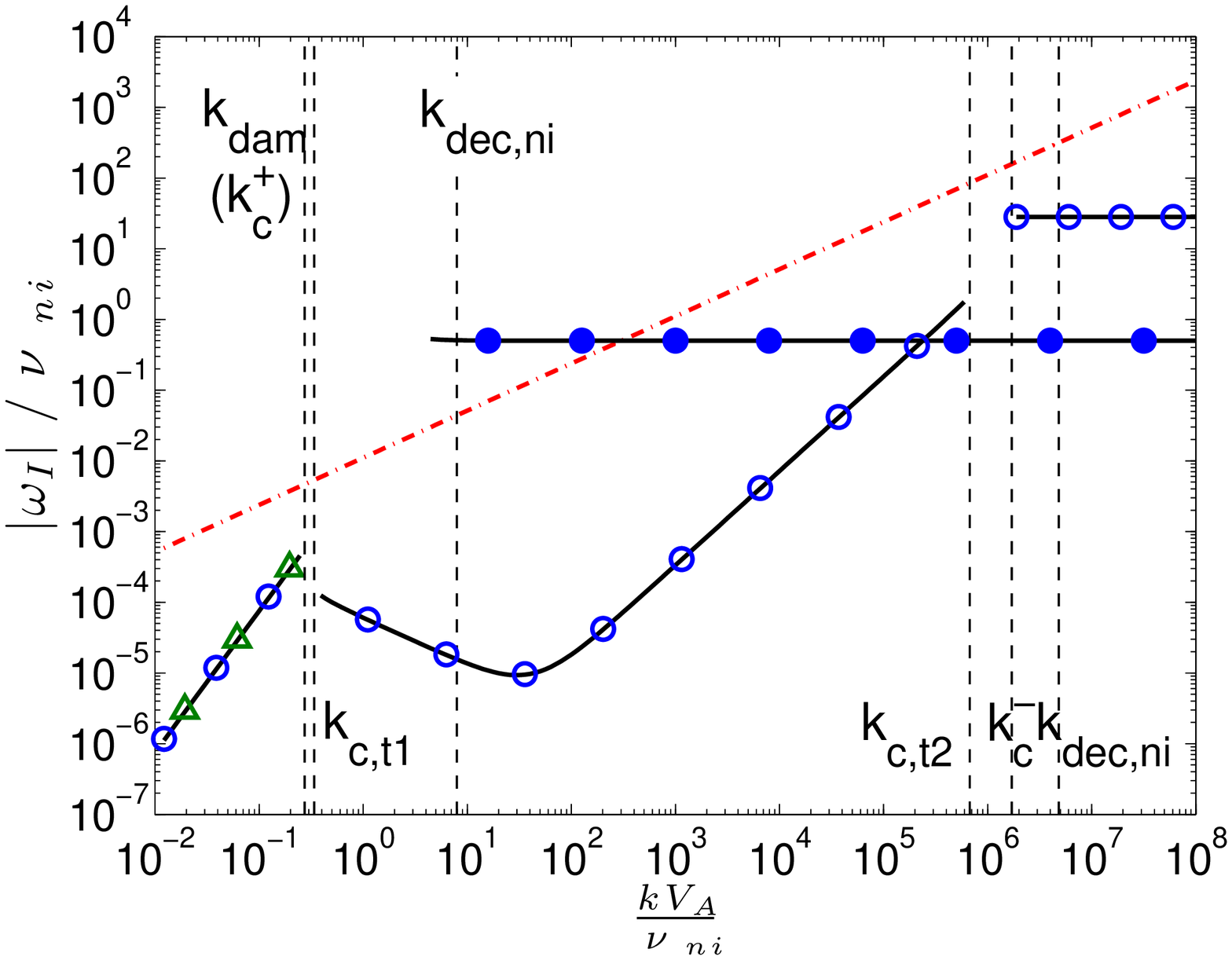}\label{figslowsc}}
\caption{Same as Fig. \ref{fig: damalf} but for slow modes. Only one branch of slow modes is present in strongly coupled regime ($k<k_\text{dec,ni}$).
On wavenumbers beyond $k_\text{dec,ni}$, there are both neutral and ion slow modes. Their analytical damping rates are represented by 
filled and open circles respectively.
Four cutoff boundaries $k_c^\pm, k_{c,t1}, k_{c,t2}$ are indicated in the figures.}
\label{fig: damslo}
\end{figure*}

\begin{figure}[htbp]
\centering
\includegraphics[width=9cm]{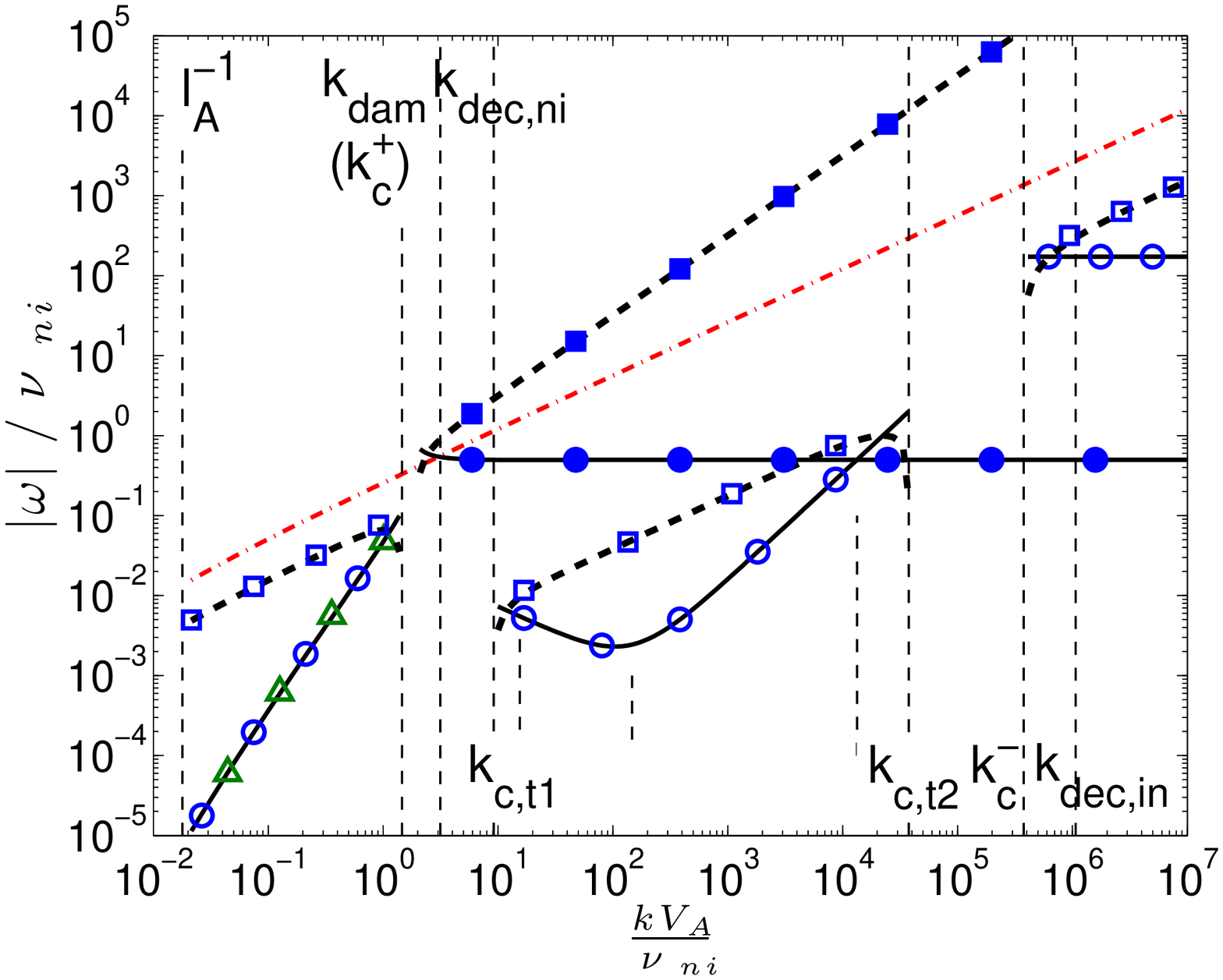}
\caption{Same as Fig. \ref{figalfmc} but for slow modes. The filled symbols correspond to the neutral slow modes. 
The short vertical dashed lines within $[k_\text{c,t1}, k_\text{c,t2}]$ 
represent $k_\text{dec,t1}$, $k_\text{tran}, k_\text{dec,t2}$ from left to right respectively. }
\label{figslowmc}
\end{figure}

Fig. \ref{figslowwnm}-\ref{figslowsc} show the results of slow modes. 
The analytical damping rate for two-fluid description is given by Eq. \eqref{eq: scsldr} for $k<k_c^+$ and Eq. \eqref{eq: comdrhw} 
for $k>k_c^-$, and the single-fluid damping rate is from Eq. \eqref{eq: sgsldr}. 
The analytical $|\omega_R|$ from Eq. \eqref{eq: scsldra} for $k<k_c^+$ and Eq. \eqref{eq: slowlbwkr} for $k>k_c^-$ is also shown in Fig. \ref{figslowmc} in the case of MC.
From Fig. \ref{figslowmc}, we find similar to other wave modes at high wave frequencies, although $\omega_R$ reemerges at the cutoff boundary $k_c^-$, 
the phase speed can only reach $c_{si}$ at the slightly smaller ion-neutral decoupling scale.

Slow modes exhibit a distinctive feature on wavenumbers above $k_\text{dec,ni}$. 
Besides the slow modes in ions, a new sort of slow modes emerges in neutrals, which is the "neutral slow mode" reported in 
\citet{Zaqa11}. 
It is also the "neutral acoustic mode" coupled with the magnetoacoustic waves discussed in 
\citet{Soler13}. 
It is generated below the neutral-ion decoupling scale
when the slow modes on larger scales impose compression and produce perturbations in neutral fluid. 
In the rest of the paper, we will term the two branches of slow modes below the scale $1/k_\text{dec,ni}$ as "neutral" and "ion" slow modes 
respectively, and term the slow modes in strongly coupled regime and "ion" slow modes together as "usual" slow modes. 
We next put our focus on the neutral and ion slow modes within the wavenumber range $[k_\text{dec,ni}, k_\text{dec,in}]$.

If we extend the wave frequencies of slow modes in strongly coupled regime (Eq. \eqref{eq: scsldrt}) to the scale 
$k_\text{dec,ni}$, we find they reach the values 
\begin{subequations}
 \begin{align}
& \omega_R^2 (k_\text{dec,ni}) \approx \nu_{ni}^2 \cos^2\theta, \label{eq: reokd}\\
& \omega_I (k_\text{dec,ni}) \approx-\frac{\nu_{ni}}{2}\sin^2\theta, \label{eq: imokd}
\end{align}
\end{subequations}
where we use the approximations $\xi_n \sim 1$ and $c_{sn} \sim c_s$. 
Starting from $k_\text{dec,ni}$, the slow modes arising in neutrals have wave frequencies as 
\begin{subequations}
 \begin{align}
& \omega_R^2=c_{sn}^2k^2, \label{eq: nsorc}\\
& \omega_I=-\frac{\nu_{ni}}{2} \sin^2\theta. \label{eq: nsoic}
\end{align}
\end{subequations}
The propagating component $\omega_R$ has the expression corresponding to pure acoustic waves in neutrals. 
The imaginary part $\omega_I$ has the same value as $\omega_I (k_\text{dec,ni})$ in Eq. \eqref{eq: imokd}. 
By setting $|\omega_R|=|\omega_I|$, we can get its cutoff wavenumber $\nu_{ni}/(2c_{sn})$, which is equal to $k_\text{dec,ni}/2$.  
Below $k_\text{dec,in}$, ions start to separate from neutrals. $\omega_I$ of the neutral slow modes becomes 
\begin{equation}\label{eq: nsoid}
    \omega_I=-\frac{\nu_{ni}}{2}.
\end{equation}
The above analytical $|\omega_I|$ (Eq. \eqref{eq: nsoic} and \eqref{eq: nsoid}) are shown by filled circles in Fig. \ref{fig: damslo} and \ref{figslowmc}.
Due to the strong anisotropy at 
relatively small scales, $\sin\theta \sim 1$ and the difference between Eq. \eqref{eq: nsoic} and \eqref{eq: nsoid}) is indistinguishable 
on the plots. 
The filled squares in Fig. \ref{figslowmc} specifically show the analytical $|\omega_R|$ (Eq. \eqref{eq: nsorc}). 
Although the slow modes in neutrals have the properties of pure acoustic waves, it is induced by the slow modes in strongly coupled regime 
and is one of the solutions to the dispersion relation of the magnetoacoustic waves in partially ionized plasma. 
We treat it as an additional branch of slow modes in our analysis. 

The relative damping efficiency of the ion and neutral slow modes at high wave frequencies depends on the ionization fraction of the 
plasma. The ratio of their damping rates at $k>k_\text{dec,in}$ is $\chi(=\rho_n/\rho_i)$.
The higher degree of ionization in WNM and SC results in a smaller ratio between the two damping rates compared with other phases. 
 
Another difference between WNM and other environments is the absence of cutoff region in WNM. 
% what is the cascade rate of "neutral slow modes"? cascade is damped means the modes is damped? 
%Since the cutoff takes place when a forward and a back- ward propagating wave merge, similar to wavelet collisions? 
%Even though the $|\omega_I|$ of the neutral slow modes intersects with $\tau_{cas}^{-1}$, since the cascade of the ion slow modes 
%remains undamped, the neutral slow modes can be regenerated at smaller scales as a result of the compression induced by the 
%persistent perturbations of magnetic field in ions. 
Also, we see from Fig. \ref{figslowwnm} that 
the cascading rate of slow modes is above the damping rate of ion slow modes over all the scales. 
Thus the usual slow modes in the case of WNM survive neutral-ion collisional damping. 
For the rest conditions, the damping scales all coincide with the cutoff boundary $k_c^+$ (see Table \ref{Tab: ctof} for the expression).

We next turn to the ion slow modes within the interval $[k_\text{dec,ni}, k_\text{dec,in}]$. In the vicinity of $k_\text{dec,ni}$, ions stay tightly 
coupled with neutrals, and their motions are overwhelmed by the acoustic perturbations in neutrals. That leads to inefficient collisional 
friction and substantially reduced damping rate of the ion slow modes. 
The damping rate is determined by the difference between $\nu_{ni}/2$ and $|\omega_I|$ of neutral slow modes (Eq. \eqref{eq: nsoic}), 
\begin{equation} \label{eq: isldrc}
   |\omega_I|=\frac{\nu_{ni}}{2}-  \frac{\nu_{ni}}{2}\sin^2\theta = \frac{\nu_{ni}}{2}\cos^2\theta. 
\end{equation}

On the other hand, ions are constrained to magnetic field lines. Driven by the magnetic perturbation, 
the slow modes with an effective sound speed 
\begin{equation}
   c_\text{s,eff}=\sqrt{\xi_i}c_{si}=\sqrt{2\gamma k_BT/m_r}
\end{equation}
as the phase speed are generated in ions, where the reduced mass is $m_r=\rho/n_i$. 
It originates from the coupling state ions remain with neutrals in this range of wavenumbers. 
Notice that $c_\text{s,eff}$ is different from $c_s=\sqrt{\gamma k_BT/m_r}$ of the slow modes in strongly coupled regime. 
The reduced mass in $c_s$ is $m_r=\rho/n$, where $n=2n_i+n_n$ is the total number density. It does not distinct the neutral and ionized 
components since neutrals and ions are frozen together and move as one fluid. But in the range $[k_\text{dec,ni}, k_\text{dec,in}]$, ions 
experience frequent collisions with neutrals and hence cannot oscillate freely. Meantime, 
the occasional collisions acting on neutrals are inadequate to transfer inertia instantly to neutrals. As a result, we see 
from the expression of $c_\text{s,eff}$ that ions alone carry the total mass of the medium to move. 

The ion slow modes can only become propagating when the real wave frequency $|\omega_R|=c_\text{s,eff}k \cos\theta $ becomes larger 
than the damping rate $|\omega_I|$ (Eq. \eqref{eq: isldrc}). By applying the cutoff condition $|\omega_R|=|\omega_I|$, we obtain the cutoff scale
\begin{equation} \label{eq: kct1}
     k_\text{c,t1}=\frac{\nu_{ni} \cos\theta}{2 c_\text{s,eff}}, 
\end{equation}
as the lower wavenumber boundary of the ion slow modes within $[k_\text{dec,ni}, k_\text{dec,in}]$. 
The ion slow modes emerge at $k_\text{c,t1}$, but the phase speed $c_\text{s,eff}$ can only be reached above the wavenumber 
\begin{equation} \label{eq: kdt1}
     k_\text{dec,t1}=\frac{\nu_{ni} \cos\theta}{c_\text{s,eff}},  
\end{equation}
satisfying $|\omega_R|=\nu_{ni}\cos^2\theta$. It actually signifies ions no longer follow the acoustic motions in neutrals, but start to develop 
their own wave modes propagating along magnetic field.

The appearance of the propagating ion slow modes induces a new component of the damping rate, associated with the wave motions. 
Since the ion slow modes are driven by magnetic perturbation and the propagation is guided by magnetic field, 
no transverse compression can be produced. Therefore different from the case for slow modes in strongly coupled regime, 
there is only damping to the parallel propagation. Correspondingly, the new component of damping rate is 
\begin{equation}
  |\omega_{I}|=\frac{\omega_{R}^2}{2\nu_{ni}}=\frac{c_\text{s,eff}^2k^2 \cos^2\theta}{2\nu_{ni}}. 
\end{equation}
We now get the complete expressions of the wave frequencies of the ion slow modes within $[k_\text{dec,ni}, k_\text{dec,in}]$
\begin{subequations}
 \begin{align}
& \omega_R^2=c_\text{s,eff}^2k^2 \cos^2\theta, \label{eq: cplisda}\\
& \omega_I=-\bigg(\frac{\nu_{ni}}{2}\cos^2\theta + \frac{c_\text{s,eff}^2k^2 \cos^2\theta}{2\nu_{ni}} \bigg). \label{eq: cplisd}
\end{align}
\end{subequations} 
The relative importance of the two terms in Eq. \eqref{eq: cplisd} changes with $k$. By equaling the two terms of the damping rate, we 
obtain the transition scale 
\begin{equation}
    k_\text{tran}=\frac{\nu_{ni}}{c_\text{s,eff}}. 
\end{equation}
When $k<k_\text{tran}$, the first term dominates the damping rate. It is independent of $\omega_R$ and 
corresponds to the decreasing $|\omega_I|$ with $k$ shown in Fig. \ref{fig: damslo} and Fig. \ref{figslowmc} under the condition of 
scale-dependent anisotropy. 
While on wavenumbers beyond $k_\text{tran}$, the second terms becomes dominant, which is proportional to $\omega_R^2$ and increases towards larger $k$. 
Notice at $k_\text{tran}$, $|\omega_R|=\nu_{ni}\cos\theta$ can be fulfilled, which coincides with the $|\omega_R|$ value 
at $k_\text{dec,ni}$ (Eq. \eqref{eq: reokd}) and is equivalent to the condition $c_\text{s,eff}k=\nu_{ni}$.   
We see $k_\text{tran}$ not only represents the transition in damping rate, but also corresponds to the critical wavelength within which 
the effective sound crossing time is equal to the neutral-ion collisional time. 
That is, below the scale $1/k_\text{tran}$, the disturbance associated with the ion slow modes cannot be effectively transmitted to neutrals 
through neutral-ion collisions. Neutrals further drift apart from ions, and ions are less effectively coupled to neutrals. 
%While most neutrals carry the sound waves, a small fraction of neutrals get involved in the wave motions of ion slow modes. 
%waves with $|\omega_R|=c_\text{s,eff}k$ is unable to propagate in neutrals. 
% this wave can propagate on larger scales? 
The differential motions between ions and neutrals become more significant, which consequently causes stronger damping to the ion slow 
modes.

The propagating component $|\omega_R|$ starts to decay when $|\omega_R|$ reaches the value of $\nu_{ni}$, corresponding to the scale 
\begin{equation} \label{eq: kdt2}
     k_\text{dec,t2}=\frac{\nu_{ni}}{c_\text{s,eff}\cos\theta}. 
\end{equation}
Neutrals become essentially unaffected by the ion slow modes. 
Subsequently $|\omega_R|$ and $|\omega_I|$ are in equality and the ion slow modes are cutoff at the scale 
\begin{equation} \label{eq: kct2}
     k_\text{c,t2}=\frac{2\nu_{ni}}{c_\text{s,eff}\cos\theta}, 
\end{equation}
which is also the upper wavenumber boundary of the ion slow modes within $[k_\text{dec,ni}, k_\text{dec,in}]$.

The analytical $|\omega_I|$ (Eq. \eqref{eq: cplisd}) is shown by open circles within $[k_\text{dec,ni}, k_\text{dec,in}]$ for all the environments 
in Fig. \ref{fig: damslo} and \ref{figslowmc}. 
And $|\omega_R|$ given by Eq. \eqref{eq: cplisda} is also displayed by open squares in Fig. \ref{figslowmc}. 
The scales $k_\text{c,t1}$ and $k_\text{c,t2}$ are indicated by vertical dashed lines in Fig. \ref{figslowwnm}-\ref{figslowsc}. The three short vertical dashed lines within $[k_\text{c,t1}, k_\text{c,t2}]$ 
in Fig. \ref{figslowmc} indicate $k_\text{dec,t1}$, $k_\text{tran}, k_\text{dec,t2}$ from large to small scales respectively. 
Their expressions using scale-dependent anisotropy are given in Appendix \ref{app:csc}.

Table \ref{Tab: sloscwf} is a summary diagram of the critical scales and wave frequencies of ion slow modes, as well as those of 
the slow modes in both strongly and weakly coupled regimes. 
All the expressions apply in low-$\beta$ environment. 
The scale expressions present here contain the propagation direction angle $\theta$. 
If we disregard the variance of $\cos\theta$ with scales, we find they are symmetrically linked by 
\begin{subequations}
\begin{align}
    k_\text{tran} &= ~~~~\frac{k_\text{dec,t1}}{\cos\theta}~~~~=~~~~\frac{2k_\text{c,t1}}{\cos\theta}~~~\approx ~\frac{\xi_n k_c^+ }{2 \sqrt{\xi_i} \cos\theta}  \label{eq: symla} \\
                        &= k_\text{dec,t2} \cos\theta = \frac{k_\text{c,t2}\cos\theta}{2}= \frac{ 2  \sqrt{\xi_i} \cos\theta k_c^-}{\xi_n},
\end{align}
\end{subequations}
where the last term in Eq. \eqref{eq: symla} is obtained by assuming $\sin^2\theta \sim 1$ and $c_s \sim c_{si}$.  
If we only focus on the relation among the critical scales smaller than $k_\text{tran}^{-1}$, we find 
\begin{equation}
     2k_\text{dec,t2}=k_\text{c,t2}= \frac{4\sqrt{\xi_i}}{\xi_n} k_c^- =  \frac{2\sqrt{\xi_i}}{\xi_n} k_\text{dec,in}. 
\end{equation}
The same relation also holds for $k_\text{dec,ni}$, $k_c^+$, $k_c^-$, and $k_\text{dec,in}$ of Alfv\'{e}n and fast modes. That is (see expressions in Table \ref{tab: decsc} and Eq. \eqref{eq: alfcfsct} and \eqref{eq: tffacfscto})
\begin{equation}
     2k_\text{dec,ni}=k_c^+= \frac{4\sqrt{\xi_i}}{\xi_n} k_c^- =  \frac{2\sqrt{\xi_i}}{\xi_n} k_\text{dec,in}
\end{equation}
when $\xi_n \sim 1$. And again we do not consider the change of $\cos\theta$ with $k$ for Alfv\'{e}n waves here. 
In fact, the wave behavior of the ion slow modes below the scale $k_\text{tran}^{-1}$ fully resembles Alfv\'{e}n waves over the whole range of scales we present, 
with $V_{Ai}$ replaced by $c_{si}$. Notice similar to $c_\text{s,eff}=\sqrt{\xi_i} c_{si}$, there is $V_A=\sqrt{\xi_i} V_{Ai}$.

Above analysis suggests that with $k_\text{tran}$ acting as a dividing line, the wave spectrum can be divided into two zones, i.e. $k<k_\text{tran}$ 
and $k>k_\text{tran}$. 
The two zones have similarities in the sense that 
the scales $k_c^+$, $k_\text{c,t1}$, $k_\text{dec,t1}$ in one zone play similar roles as $k_\text{c,t2}$, $k_c^-$, $k_\text{dec,in}$ in the other. 
But there also exist differences, resulting from the varying coupling state between neutrals and ions with scales. 
As already discussed, 
one of the main differences is for the slow modes in the strongly coupled regime, there is no damping for parallel propagation (Eq. \eqref{eq: scsldr}), 
but for the ion slow modes within $[k_\text{tran}, k_\text{c,t2}]$, damping only appears for purely parallel propagation (Eq. \eqref{eq: cplisd}). 
In addition, we observe $|\omega_I|$ within $[k_\text{c,t1}, k_\text{trans}]$ (Eq. \eqref{eq: cplisd}) can be approximately written as 
\begin{equation}
  |\omega_I| \sim \frac{\xi_i \cos^2\theta}{\xi_n}  \frac{\nu_{in}}{2}.
\end{equation}
The factor $ \xi_i \cos^2\theta/\xi_n$ reflects much weaker frictional damping of the ion slow waves 
compared with $|\omega_I|$ (Eq. \eqref{eq: comdrhw}) of the slow modes in weakly coupled regime.

To reinforce our understanding on the wave behavior, we also provide the force analysis for both fast and slow waves in Appendix \ref{app:fa}.

\begin{table*}[htbp]
\renewcommand\arraystretch{2.5}
\centering
\caption[]{Critical scales and wave frequencies of the slow modes 
}\label{Tab: sloscwf} 
\begin{tabular}{c|c|c|c|c|c|c|c|c|c}
 \hline
Regimes  & strongly coupled & \multicolumn{6}{c|}{ions coupled with neutrals } & \multicolumn{2}{c}{weakly coupled} \\
      \hline
\multirow{2}{*}{Scales} 
& $k_c^+$ & $k_\text{c,t1}$ & $k_\text{dec,t1}$ & \multicolumn{2}{c|}{$~~~k_\text{tran}~~~$}  & $k_\text{dec,t2}$ & $k_\text{c,t2}$  & $k_c^-$ & $k_\text{dec,in}$ \\
\cline{2-10}
&$\frac{2\nu_{ni}\cos\theta}{c_s\xi_n \sin^2\theta}$ &  $\frac{\nu_{ni} \cos\theta}{2c_\text{s,eff}}$ & $\frac{\nu_{ni} \cos\theta}{c_\text{s,eff}}$ &  \multicolumn{2}{c|}{$\frac{\nu_{ni}}{c_\text{s,eff}}$} & $\frac{\nu_{ni}}{c_\text{s,eff}\cos\theta}$ & $\frac{2\nu_{ni}}{c_\text{s,eff}\cos\theta}$  &  $\frac{\nu_{in}}{2c_{si} \cos\theta}$  &  $\frac{\nu_{in}}{c_{si}\cos\theta}$ \\
\hline
$|\omega_R|$ & $<c_s k \cos\theta$ & $<c_\text{s,eff} k \cos\theta$ &\multicolumn{4}{c|}{ $c_\text{s,eff} k \cos\theta$ } &$<c_\text{s,eff} k \cos\theta$ & $<c_{si}k\cos\theta$ & $c_{si}k\cos\theta$ \\
\hline
$|\omega_I|$  & $\frac{\xi_n c_s^2k^2 \sin^2\theta}{2\nu_{ni}}$  &  \multicolumn{6}{c|}{$\frac{\nu_{ni}\cos^2\theta}{2}+\frac{c_\text{s,eff}^2k^2\cos^2\theta}{2\nu_{ni}}$} & \multicolumn{2}{c}{$\frac{\nu_{in}}{2}$} \\
\hline
\end{tabular}
\end{table*}

The neutral slow modes originate from the compression produced by magnetoacoustic waves. It can only develop after neutrals decouple from 
ions and propagate with a sound speed in neutrals\footnote{In high-$\beta$ plasma, since the fast modes will behave like sound waves, the sound waves 
in neutrals will become the neutral "fast" modes.}. 
In terms of turbulence properties, below the scale $1/k_\text{dec,ni}$, for Alfv\'{e}n and fast modes in low-$\beta$ medium,
hydrodynamic turbulence starts to evolve in neutrals where no sound waves arise. 
But in the case of slow modes, neutrals begin to carry acoustic turbulence caused by interacting sound waves
\citep{Zak67, Zak70, Lvov00}
at scales smaller than $1/k_\text{dec,ni}$. % more description of this turbulence needed.
The wave motions of neutral fluid also lead to another distinctive feature of slow modes from other modes. Instead of having a purely nonpropagating 
interval, propagating ion slow waves appear within $[k_c^+, k_c^-]$. 
It suggests that in comparison with hydrodynamic motions, the wave motions of neutral slow modes 
cause weaker friction against magnetic pressure force on ions, and thus the ion slow modes can be driven by 
the persistent perturbations of magnetic field and sustained in ions even at the "cutoff" wavenumbers.

Note that in Table \ref{Tab: sloscwf} and the force analysis of slow waves in Appendix \ref{app:fa}, we adopt a fixed wave propagation direction and disregard its dependence on length scales for simplicity. It is necessary to point out, 
this treatment using fixed propagation angle can indeed provide simpler and more intuitively obvious description of wave behavior than 
that considering scale-dependent anisotropy,  but it is only applicable in analysis of linear MHD waves.

~~~~~~~~~~~~~~~~~~~~~~~~~~~~\\

The above comparisons with numerical results show that the single-fluid approach is able to depict the damping behavior of MHD modes correctly 
at large scales until reaching the cutoff boundary $k_c^+$. 
We see for Alfv\'{e}n and slow modes, the cutoff boundary $k_c^+$ of linear waves is also the $k_\text{dam}$ of turbulence in CNM, MC, DC and SC. 
However, WNM does not exhibit a cutoff region for all the three wave modes. 
The parameter space required for the existence of wave cutoffs can be confined by equaling $k_c^+$ and $k_c^-$. 
%For instance, setting $k_c^+=k_c^-$ for fast modes (Eq. \eqref{eq: tffacfsc} and \eqref{eq: tffacfscb}) approximately gives the upper limit of ionization degree $\chi \sim 16$. 
Due to the higher ionization degree than other ISM phases, MHD waves in WNM are less affected by collisions with neutrals and can avoid being cutoff. 

Another remark needs to be made is about the relation between wave cutoff and fluid decoupling. We found $k_\text{dec,ni}$ ($k_\text{dec,in}$) and $k_c^+$ 
($k_c^-$) are of the same order of magnitude, and the interval of $[k_\text{dec,ni}, k_\text{dec,in}]$ is relatively larger than $[k_c^+, k_c^-]$. 
Their physical connection is obvious. 
Only after neutrals separate themselves from the MHD wave motions of ions, can they develop their own motions and exert significant influence on ions, i.e. 
collisional friction strong enough for cutting off the MHD waves. 
On the other hand, 
although propagating waves reemerge at $k_c^-$, they can only be fully resumed after ions get free from coupling with neutrals at $k_\text{dec,in}$.

\section{Applications on CR propagation in partially ionized medium}
\label{sub: crapp}
It is necessary to introduce turbulence damping for achieving a correct understanding on the interaction of particles and MHD turbulence.
\citet{YL02,YL04,YL08} 
described CR transport in fully ionized medium and clarified fast modes are the most effective scatterer of CRs despite their damping. 
It is instructive to extend their study to cover partially ionized medium and build up a complete picture of CR propagation in the ISM. 

\subsection{Pitch-angle scattering of CRs}
We investigate pitch-angle scattering of CRs based on the formalism of diffusion coefficients obtained from Fokker-Planck theory
\citep{Jokipii1966, SchlickeiserMiller, YL02, YL04}.
The energy spectra of different modes we adopt are obtained from three-dimensional MHD simulations
\citep{CLV_incomp, YL03}.
Besides, we follow the nonlinear theory (NLT) employed in
\citet{YL08}
for our calculation on gyroresonance.
The fluctuations of $B$ in MHD turbulence result in variations of particle velocities in both perpendicular and parallel directions with regards 
to magnetic field lines. That substantially broadens the $\delta$-function resonance predicted in quasi-linear scattering theory, 
especially for large-amplitude MHD waves. 
The full set of expressions for Fokker-Planck diffusion coefficient $D_{\mu\mu}$ in low-$\beta$ medium and notations used are presented in Appendix \ref{app:c}
(see \citealt{YL04,YL08} for more details on the derivation). 

\begin{figure*}[htbp]
\centering
\subfigure[Alfv\'{e}n]{ 
   \includegraphics[width=8cm]{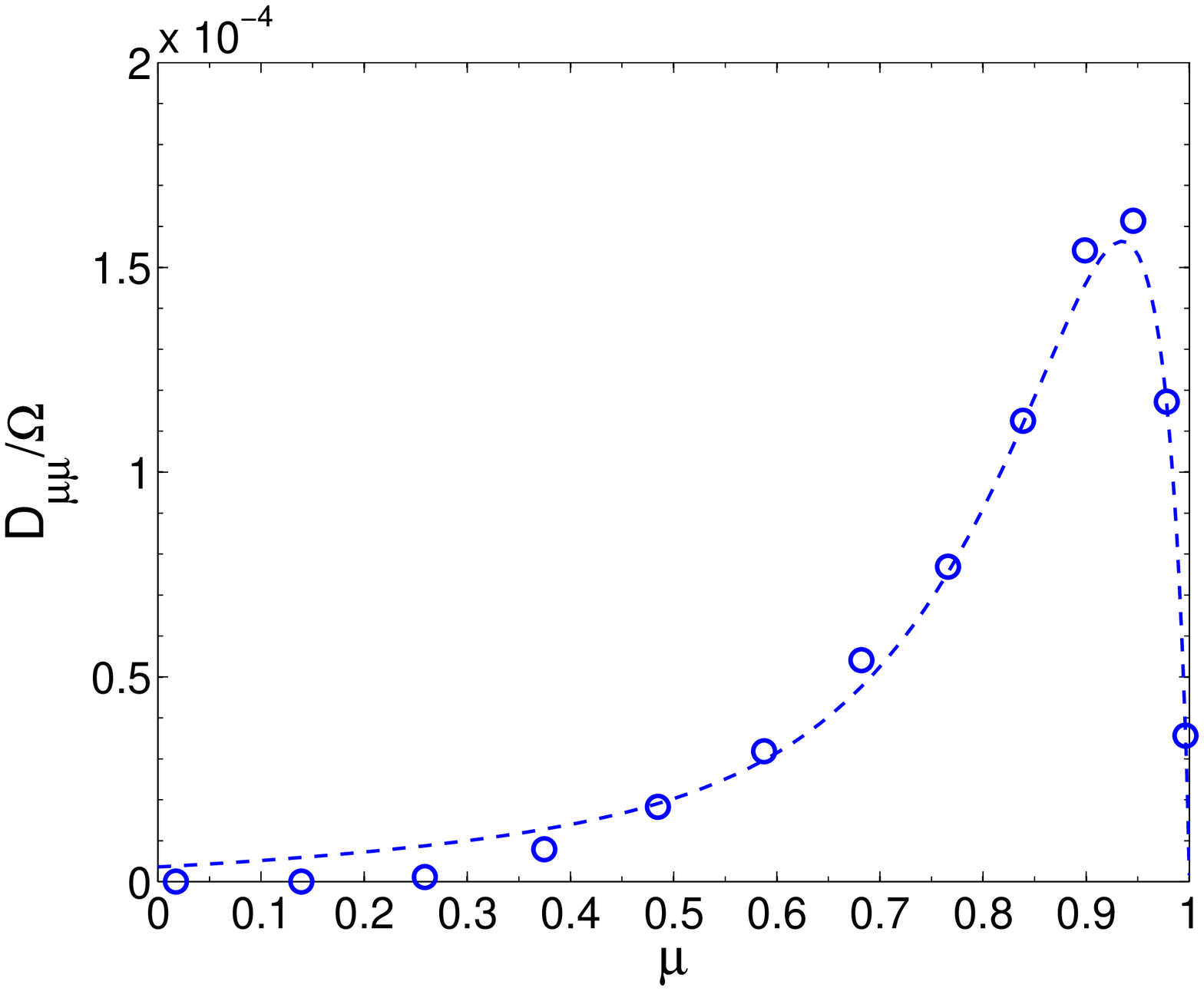}\label{figwnmduua}}
\subfigure[Fast]{
   \includegraphics[width=8cm]{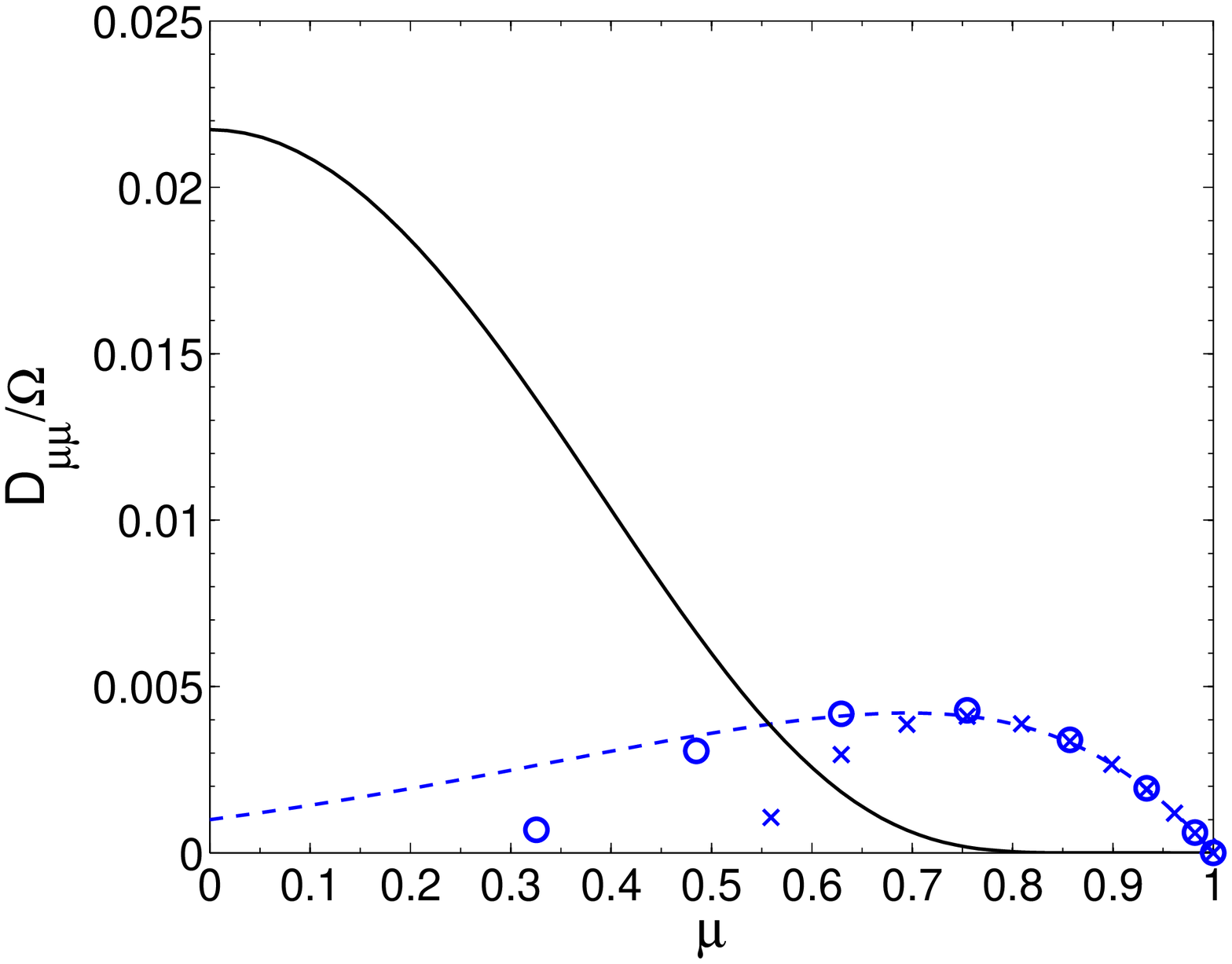}\label{figwnmduub}}
\subfigure[Slow]{  
   \includegraphics[width=8cm]{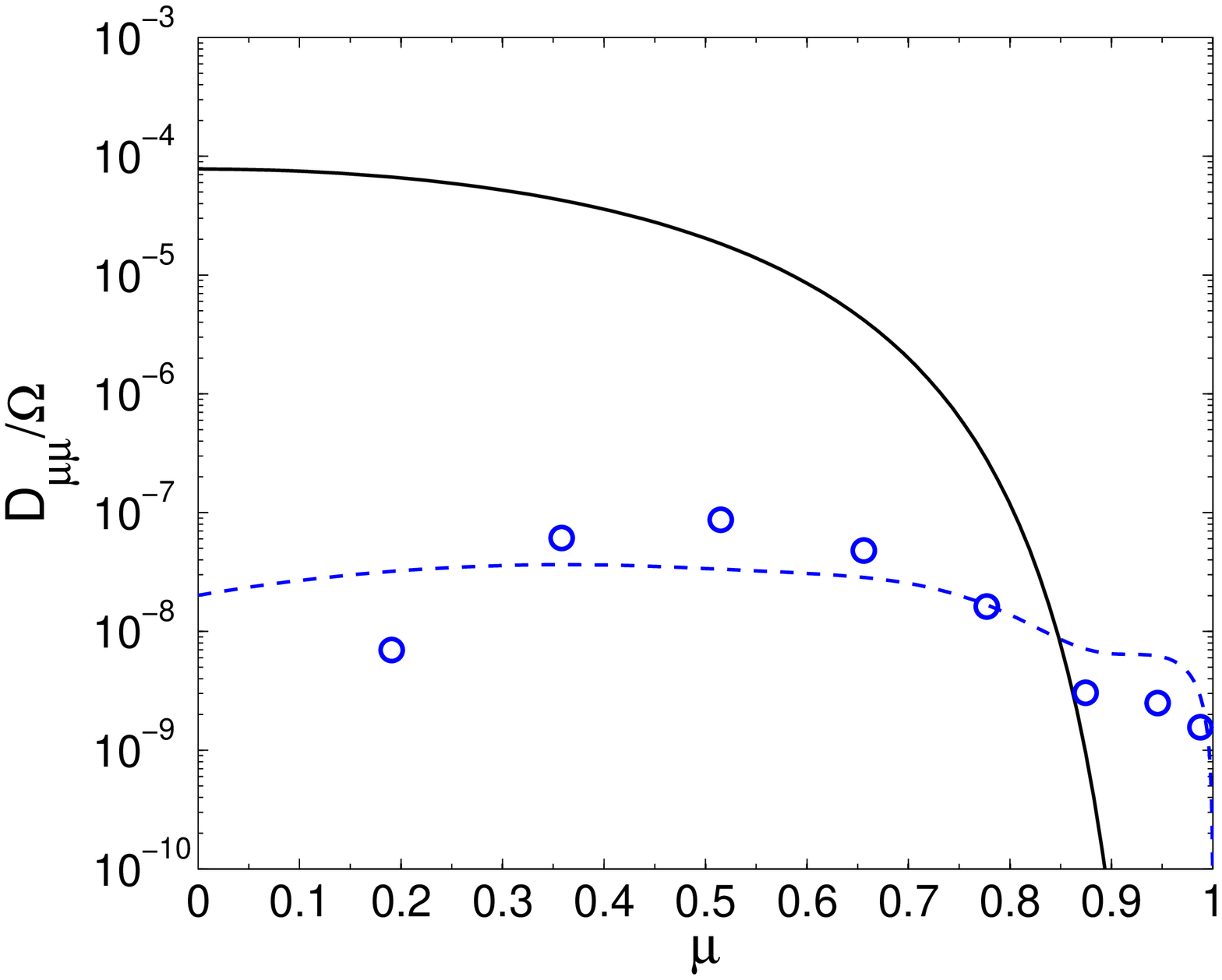}\label{figwnmduuc}}
\subfigure[Total]{ 
   \includegraphics[width=8cm]{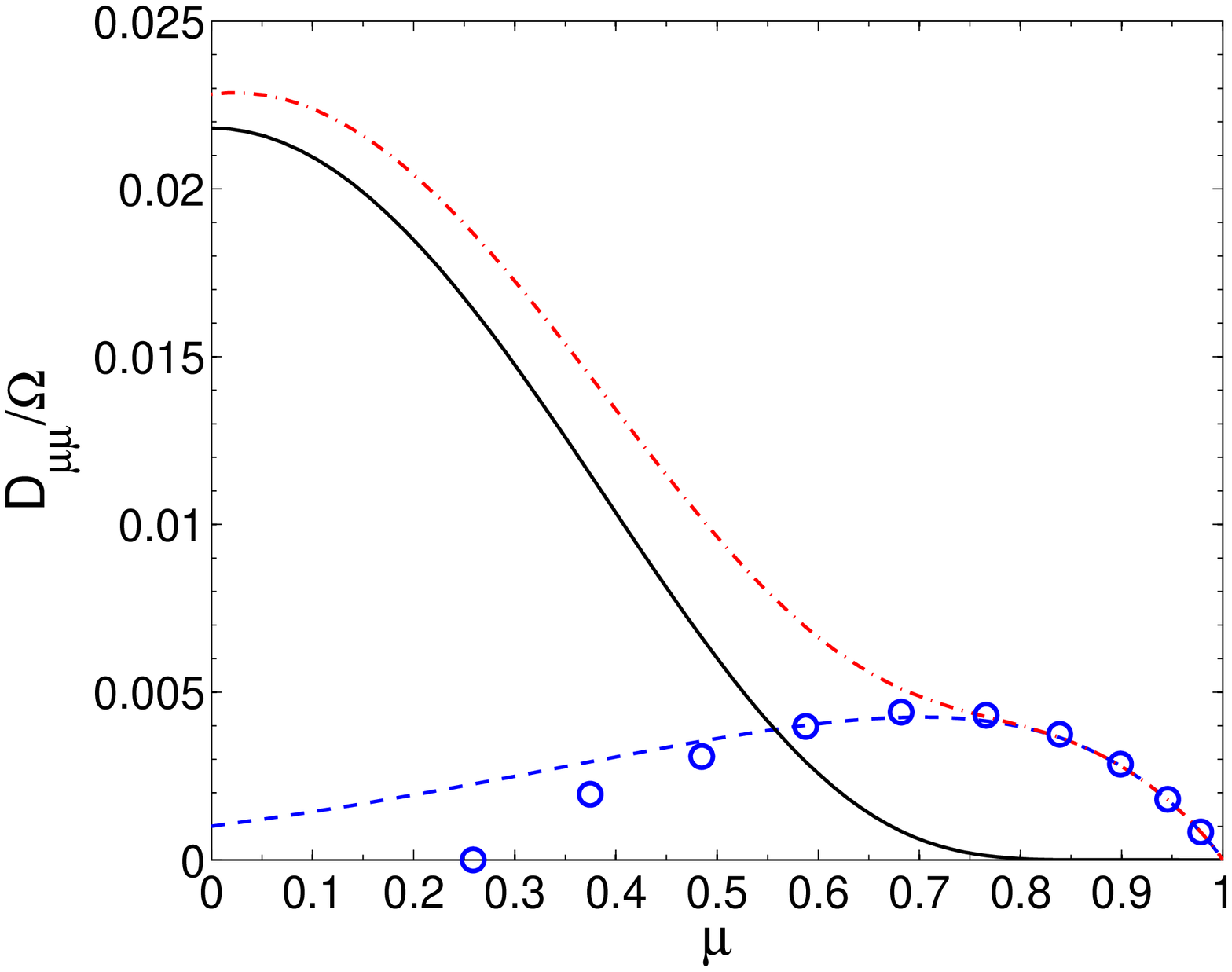}\label{figwnmduud}}
   \caption{Pitch-angle diffusion coefficients (normalized by $\Omega$) of 100 PeV CRs in (a) Alfv\'{e}n, (b) fast, and (c) slow modes in WNM. 
   Solid and dashed lines refer to TTD and gyroresonance by applying NLT. Open circles are QLT results for gyroresonance. 
   Crosses in (b) show the approximate QLT result for gyroresonance given by Eq. \eqref{eq: duufagqs}.
   (d) Total diffusion coefficients from three modes. The dash-dotted line is the sum of the solid and dashed lines. }
\label{figwnmduu}
\end{figure*}

\begin{figure*}[htbp]
\centering
\subfigure[Alfv\'{e}n]{ 
   \includegraphics[width=8cm]{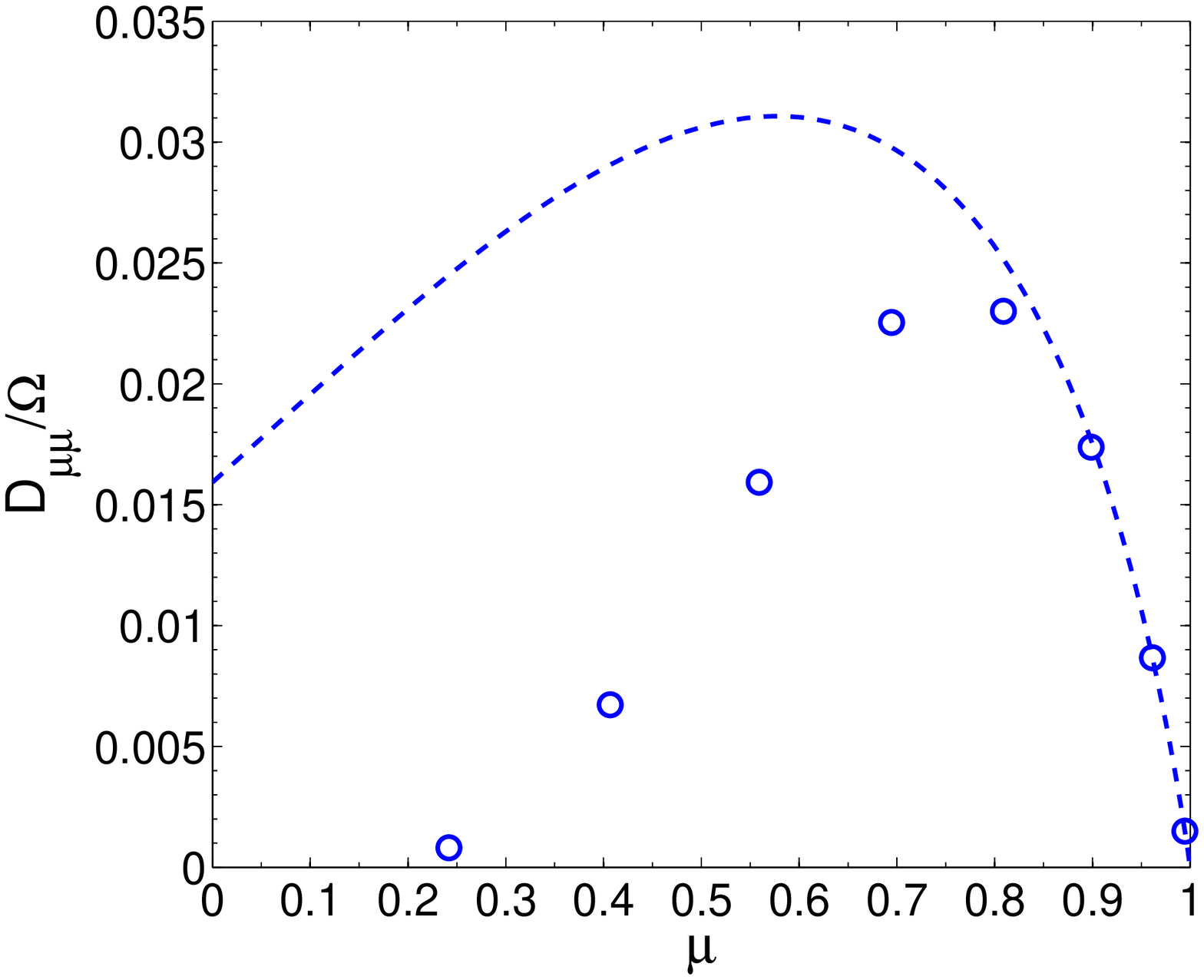}\label{figmcduua}}
\subfigure[Fast]{
   \includegraphics[width=8cm]{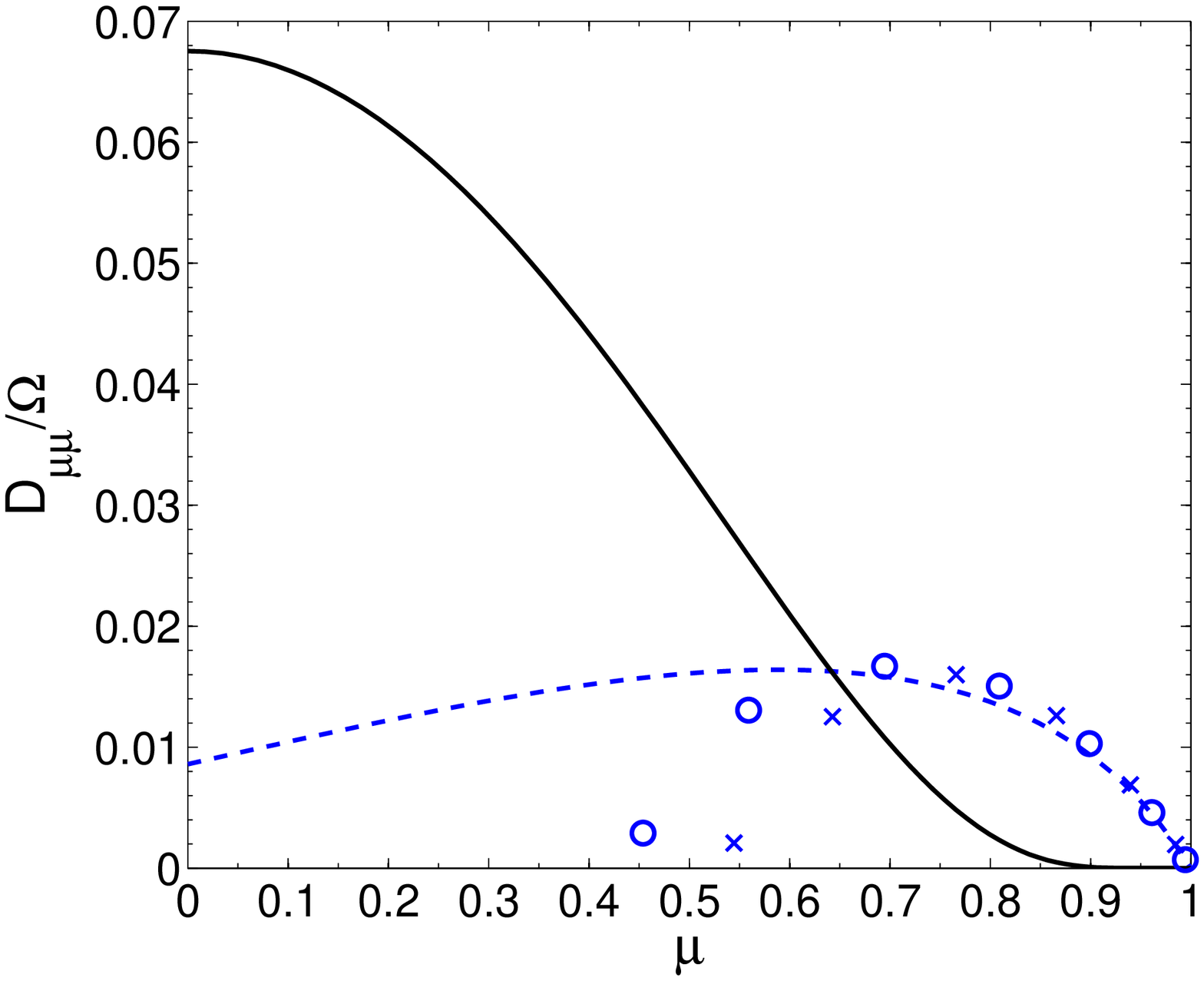}\label{figmcduub}}
\subfigure[Slow]{  
   \includegraphics[width=8cm]{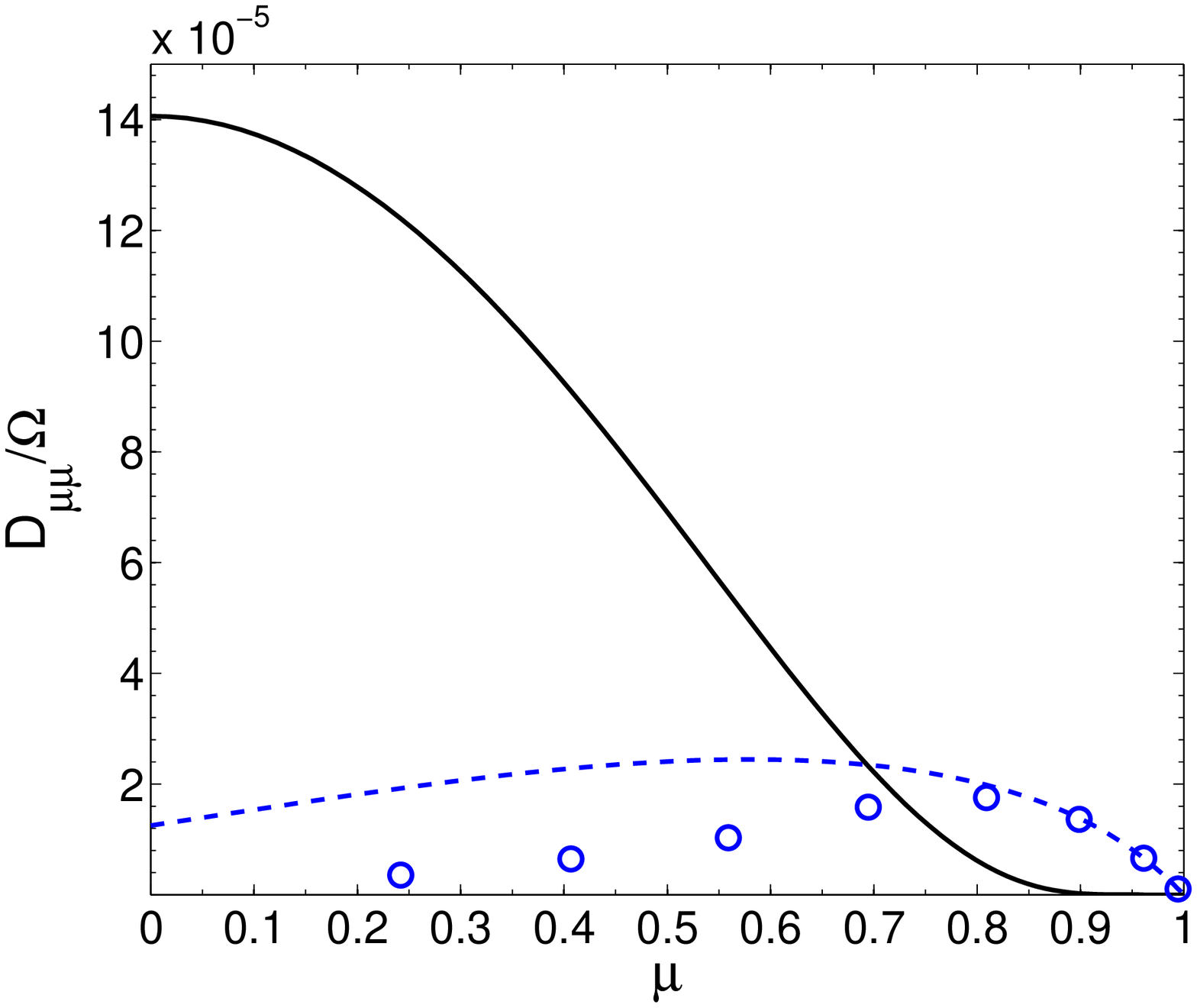}\label{figmcduuc}}
\subfigure[Total]{ 
   \includegraphics[width=8cm]{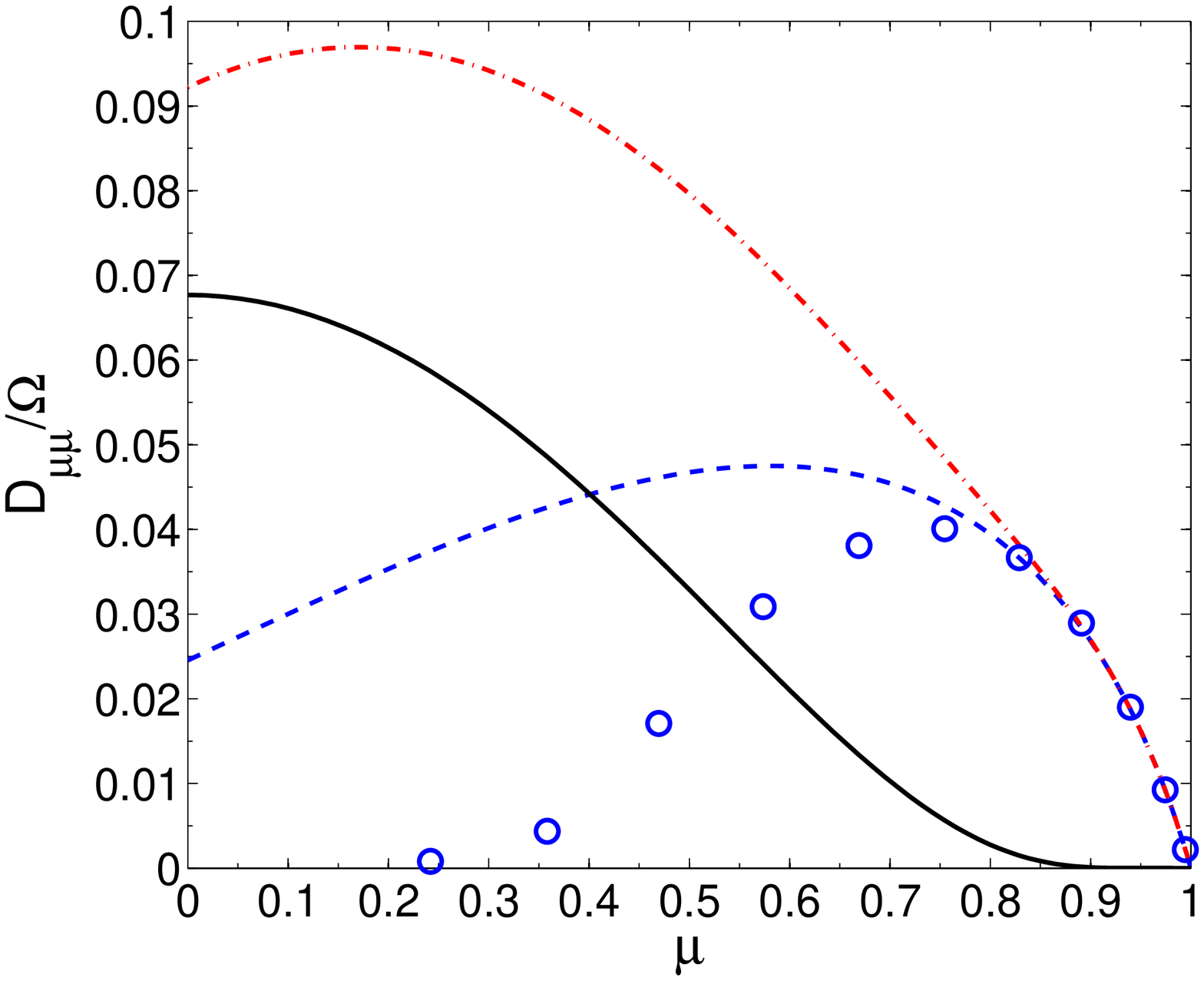}\label{figmcduud}}
   \caption{Same as Fig. \ref{figwnmduu} but for 30 TeV CRs in MC.}
\label{figmcduu}
\end{figure*}

We choose the phases WNM and MC as the representative examples of sub- and super-Alfv\'{e}nic turbulence. 
Fig. \ref{figwnmduu} and \ref{figmcduu} show the calculated  $D_{\mu\mu}$ (normalized by $\Omega$) 
as a function of $\mu$ for $100$ PeV CRs in WNM and $30$ TeV CRs in MC, 
where $\mu$ is cosine of particle pitch angle. 
CRs' energy is chosen with their Larmor radius $r_L$ larger than the maximum damping scale among the three turbulence modes. 
Here 
\begin{equation}
r_L=\frac{v_\perp}{\Omega},  ~~\Omega=\frac{eB}{\Gamma m_p c}.
\end{equation}

%Fig. (a)-(c) present the results for Alfv\'{e}n, fast and slow modes respectively.
%Fig. (d) shows the total $D_{\mu\mu} / \Omega$ from three modes. 
We separately calculate the transit-time damping (TTD, solid line) for fast and slow modes, and 
gyroresonance interactions (dashed line) for all turbulent modes.
The gyroresonance result calculated using quasi-linear theory (QLT; 
see \citealt{Jokipii1966, Schlickeiser02}) is also displayed for comparison (open circles). 
Crosses show the approximate result using QLT for gyroresonance with fast modes, given by 
\begin{equation}\label{eq: duufagqs}
D_{\mu\mu}^G(\text{QLT})=\frac{v\pi\sqrt{\mu}(1-\mu^2)}{4L\sqrt{R}} \bigg( \frac{2}{7}-\frac{2\sqrt{1-\mu^2}}{21 \mu^2}    \bigg).
\end{equation}
This simplified formula is obtained under the condition $k_{\text{dam},\perp} r_L<1$.
We see good consistency between it and the numerical integral at small pitch angles.

\begin{figure*}[htbp]
\centering
\subfigure[Fast]{ 
   \includegraphics[width=8cm]{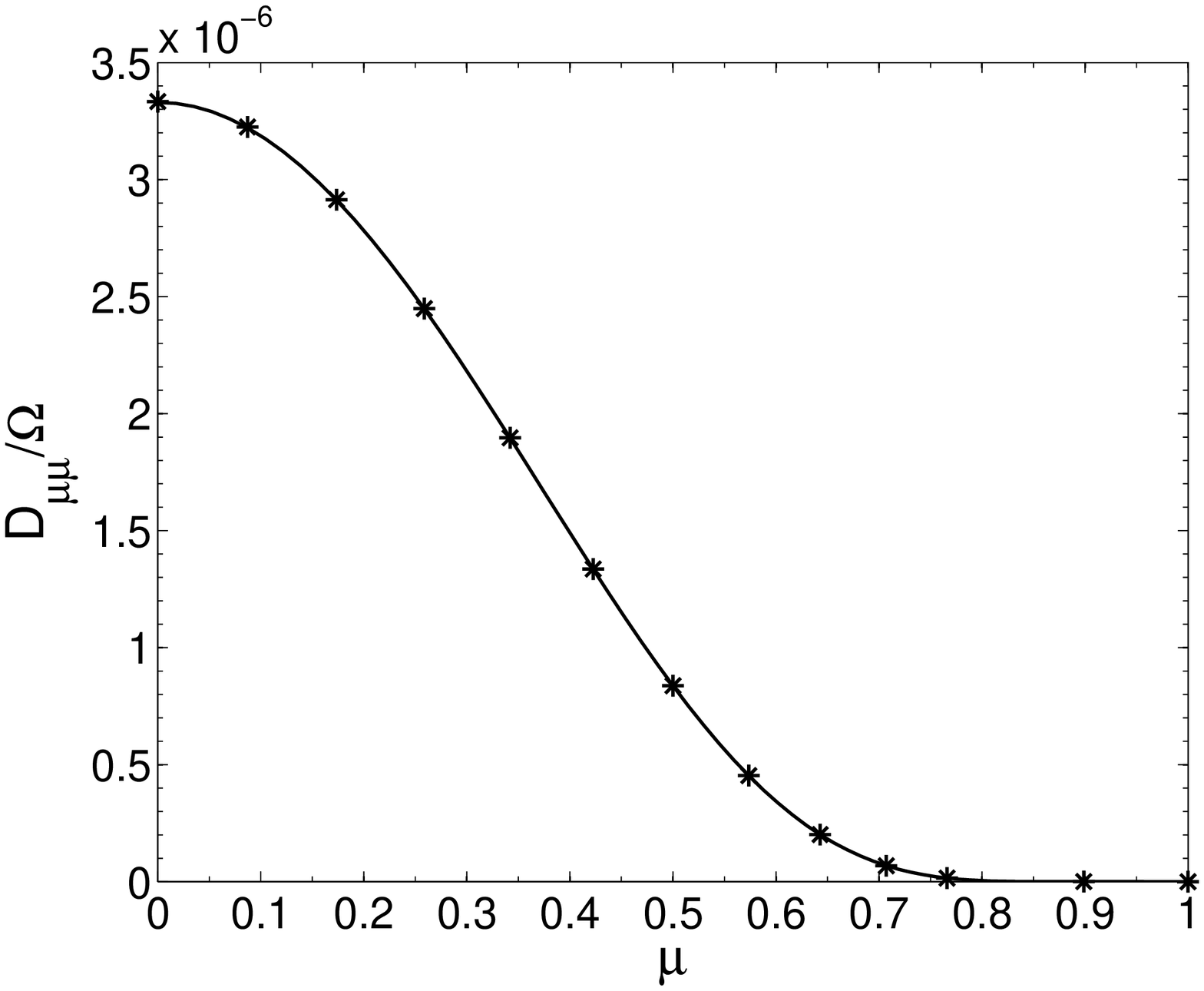}\label{figwnmduula}}
\subfigure[Slow]{
   \includegraphics[width=8cm]{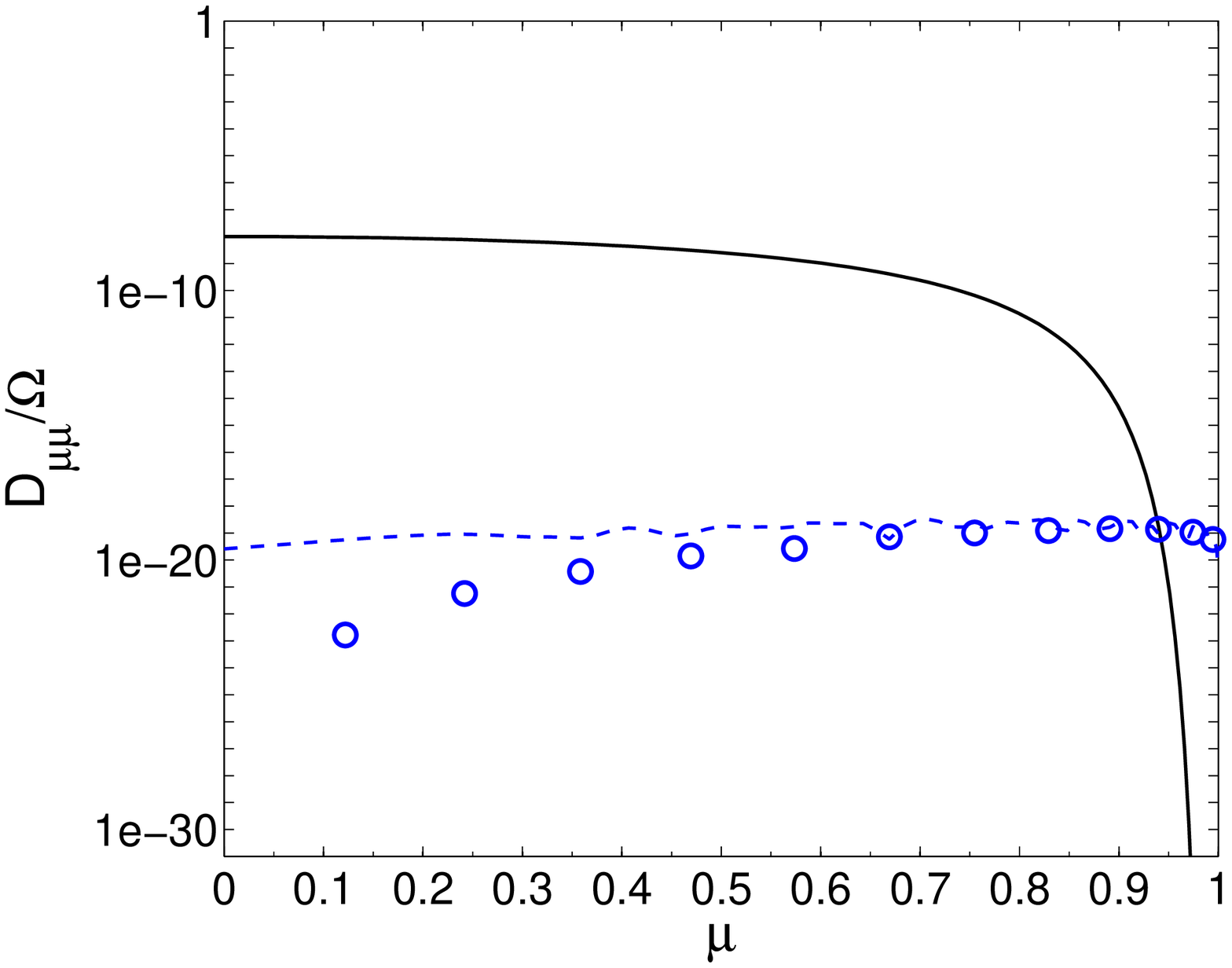}\label{figwnmduulb}}
   \caption{Same as Fig. \ref{figwnmduu} but for 10 TeV CRs in (a) fast and (b) slow modes in WNM. The asterisks in (a) are 
   analytical result given by Eq. \eqref{eq: duuftapp}. }
\label{figwnmduul}
\end{figure*}

\begin{figure*}[htbp]
\centering
\subfigure[Fast]{ 
   \includegraphics[width=8cm]{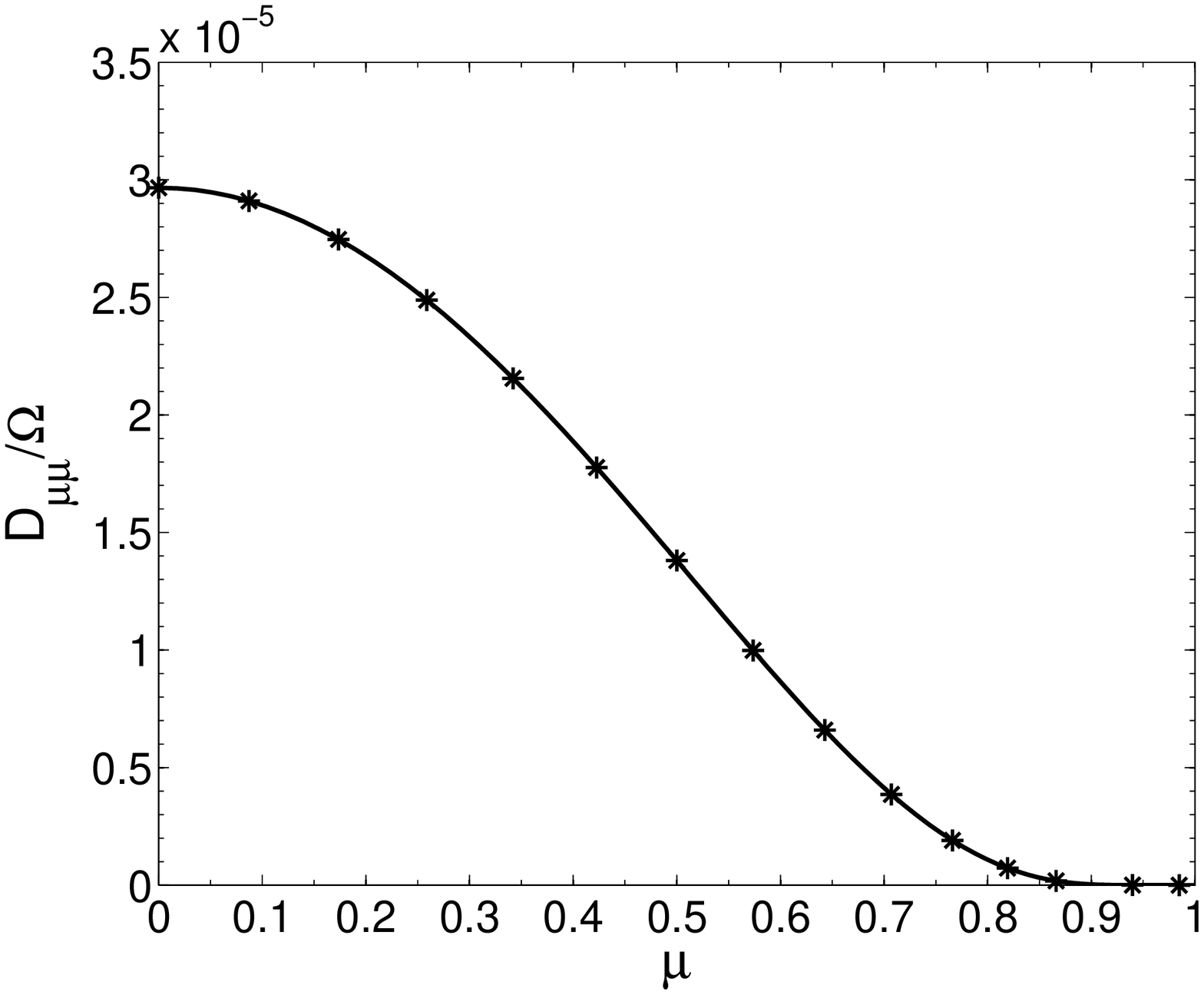}\label{figmcduula}}
\subfigure[Slow]{
   \includegraphics[width=8cm]{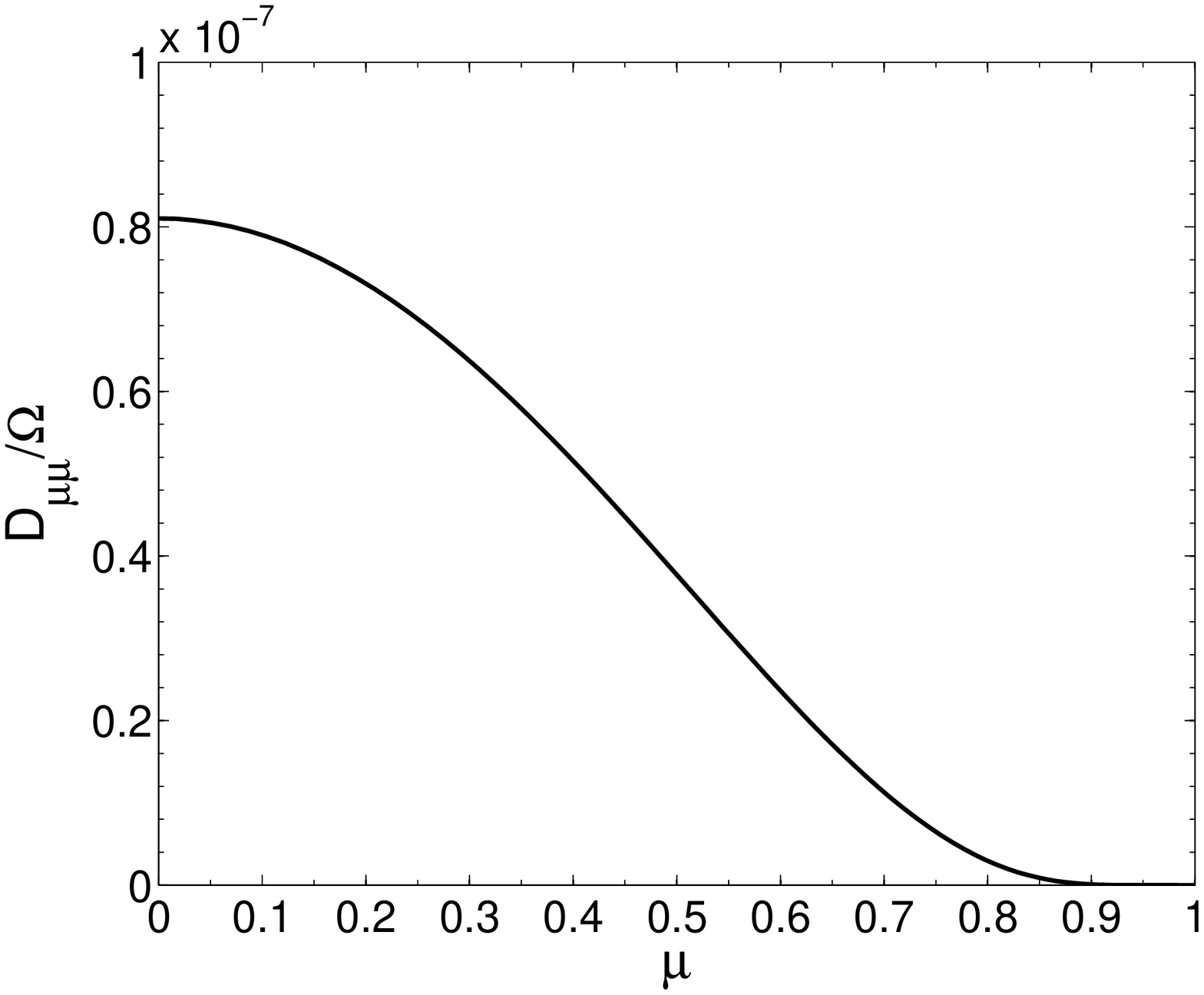}\label{figmcduulb}}
   \caption{Same as Fig. \ref{figwnmduu} but for 10 GeV CRs in (a) fast and (b) slow modes in MC. }
\label{figmcduul}
\end{figure*}

The diffusion coefficients in WNM and MC exhibit similar behavior. 
Among the three modes, fast turbulence has the largest damping scale. Despite severe damping, TTD from fast modes dominates CR scattering at large pitch angles. 
On the contrary, gyroresonance is more efficient in scattering at small pitch angles. 
The overall contribution from Alfv\'{e}n modes is smaller compared with that from fast modes. 
Especially in WNM, due to the more prominent anisotropy over all scales in sub-Alfv\'{e}n turbulence, Alfv\'{e}n modes are largely suppressed in CR scattering. 
Comparing Fig. \ref{figwnmduub} and \ref{figwnmduud}, we find fast modes contribute alone in the total diffusion coefficient. 
These results are in accordance with the early findings in 
\citet{YL02, YL04, YL08}.  
The scattering by slow modes depends on $\beta$ of the medium (see Eq. \eqref{eq: slduug}-\eqref{eq: slduugq}). 
Low $\beta$ value in ISM conditions (see Table \ref{Tab: ism}) makes the effect of slow modes negligible. 
In addition, by comparing with the results of gyroresonance from QLT, we find nonlinear effect results in a much wider range of pitch angles, while QLT gives $D_{\mu\mu}=0$ at large pitch angles due to 
its discrete resonances, $k_{\|, \text{res}} = \Omega / (v \mu)$. 
This resonance condition cannot be satisfied at small $\mu$ as the turbulence is damped before cascading down to small scales. 
Only at the large $\mu$ end, gyroresonance from QLT comes into coincidence with that from NLT. 

To examine the damping effect on CR scattering, we again calculate the diffusion coefficients but for CRs with relatively low energies. 
CRs' energy is chosen with their Larmor radius $r_L$ smaller than the minimum damping scale among the three turbulence modes. 
But since in WNM, slow modes survive neutral-ion collisional damping (see Section \ref{sec: slowsr}) and can extend to the gyroscale of ions, 
we choose 10 TeV CRs with $r_L$ smaller than the damping scale of Alfv\'{e}n modes, which is smaller than that of fast modes. 
For MC, Alfv\'{e}n modes have the smallest damping scale. Accordingly, we adopt 10 GeV as the CR energy. 

Fig. \ref{figwnmduul} and Fig. \ref{figmcduul} show the results in WNM and MC respectively. 
The asterisks represent the simplified expression for TTD with fast modes, 
\begin{equation} \label{eq: duuftapp}
   D_{\mu\mu}^{T}(\text{NLT})=\frac{v\sqrt{\pi}(1-\mu^2)^2}{8L\Delta \mu} \exp \bigg(-\frac{\mu^2}{(\Delta \mu)^2}\bigg)(\sqrt{k_\text{dam}L}-1),
\end{equation} 
which is obtained under the condition $k_{\text{dam},\perp} r_L<1$ and agrees well with integral result for low-rigidity CRs over 
all pitch angles.
Obviously, TTD with fast modes becomes the only significant scattering effect.
The remaining gyroresonance of slow modes in WNM has negligible effect.  
Compared with the results for high-energy CRs with $r_L$ larger than the damping scales of turbulence modes in Fig. \ref{figwnmduu} and \ref{figmcduu}, 
here we can clearly see the resonance gap at small pitch angles due to the absence of gyroresonance. 
TTD alone cannot fulfill the scattering of CRs with small pitch angles.

\subsection{Scattering frequency of TTD and gyroresonance}
\begin{figure*}[htbp]
\centering
\subfigure[TTD]{
   \includegraphics[width=8cm]{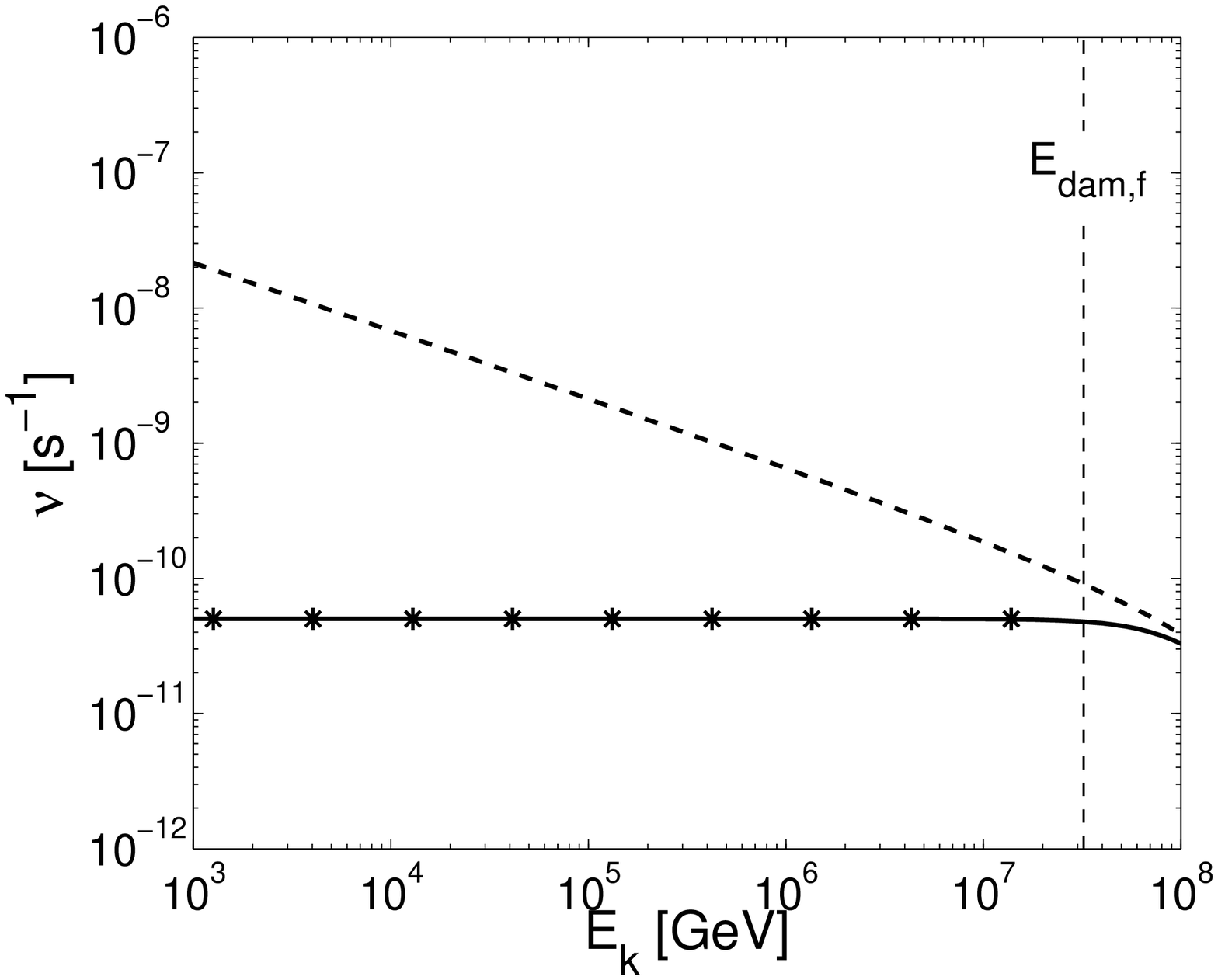}\label{figwnmsfa}}
\subfigure[Gyroresonance]{  
   \includegraphics[width=8cm]{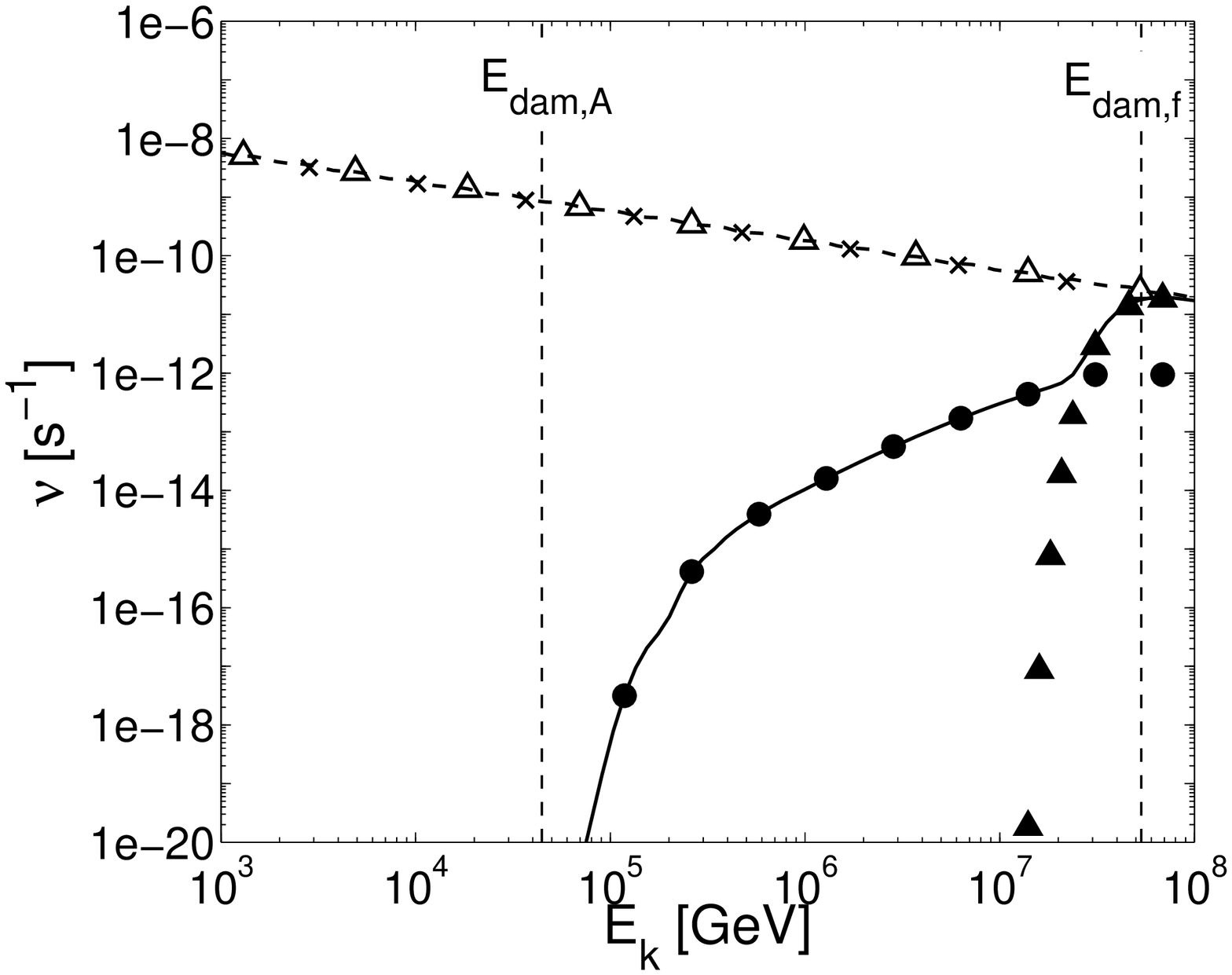}\label{figwnmsfb}}
\caption{Scattering frequency vs energy of CRs in WNM from (a) TTD at $\mu=0.1$ and (b) gyroresonance at $\mu=0.8$.
Dashed and solid lines correspond to cases in the absence of damping and in the presence of damping. 
Circles and triangles in (b) show the separate contributions from Alfv\'{e}n and fast modes. Filled and open symbols correspond to 
situations with and without damping. 
Asterisks and crosses represent the analytical results by using Eq. \eqref{eq: duuftapp} and \eqref{eq: duufagqs} respectively.}
\label{figwnmsf}
\end{figure*}

\begin{figure*}[htbp]
\centering
\subfigure[TTD]{
   \includegraphics[width=8cm]{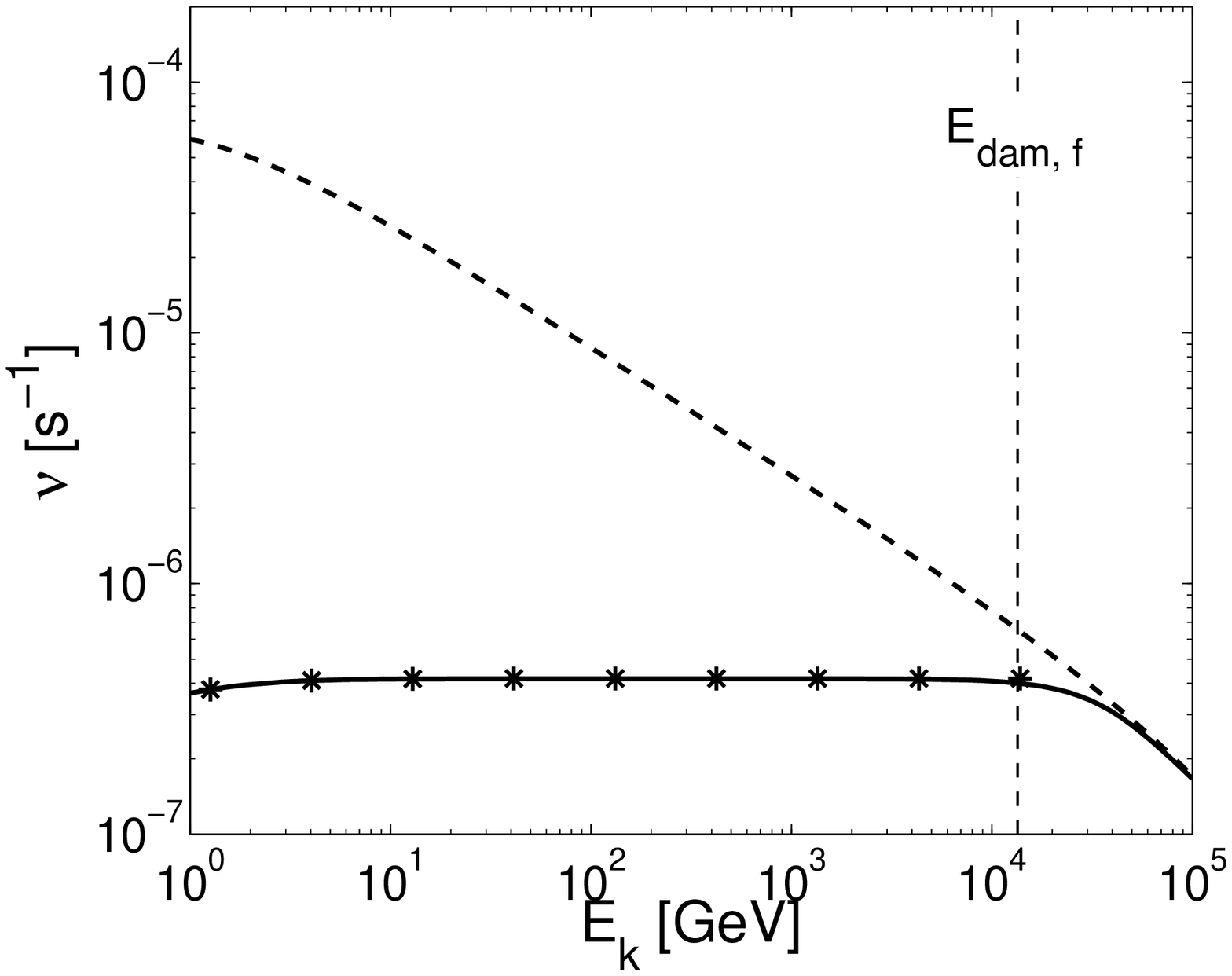}\label{figmcsfa}}
\subfigure[Gyroresonance]{  
   \includegraphics[width=8cm]{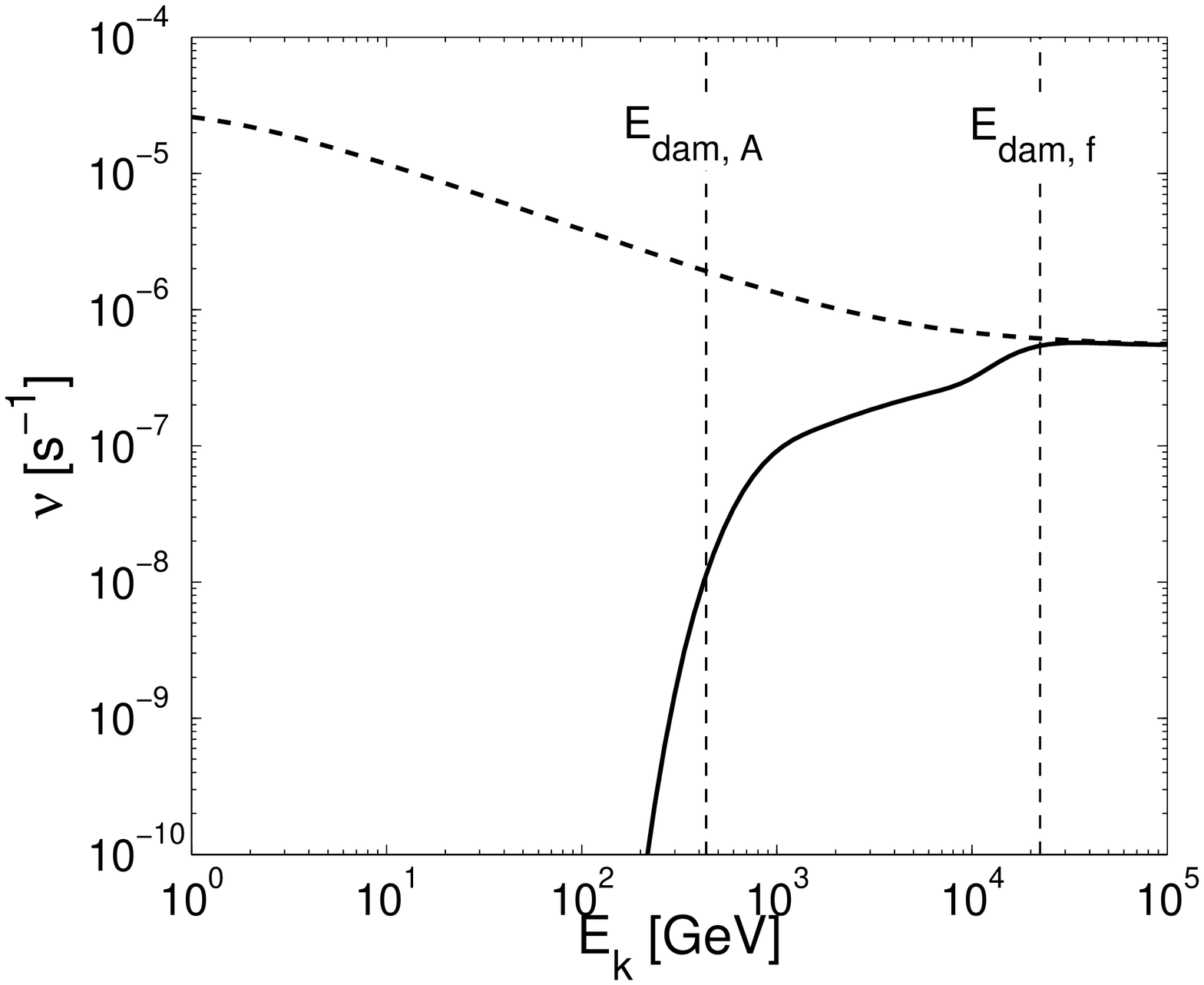}\label{figmcsfb}}
\subfigure[Gyroresonance, without damping]{  
   \includegraphics[width=8cm]{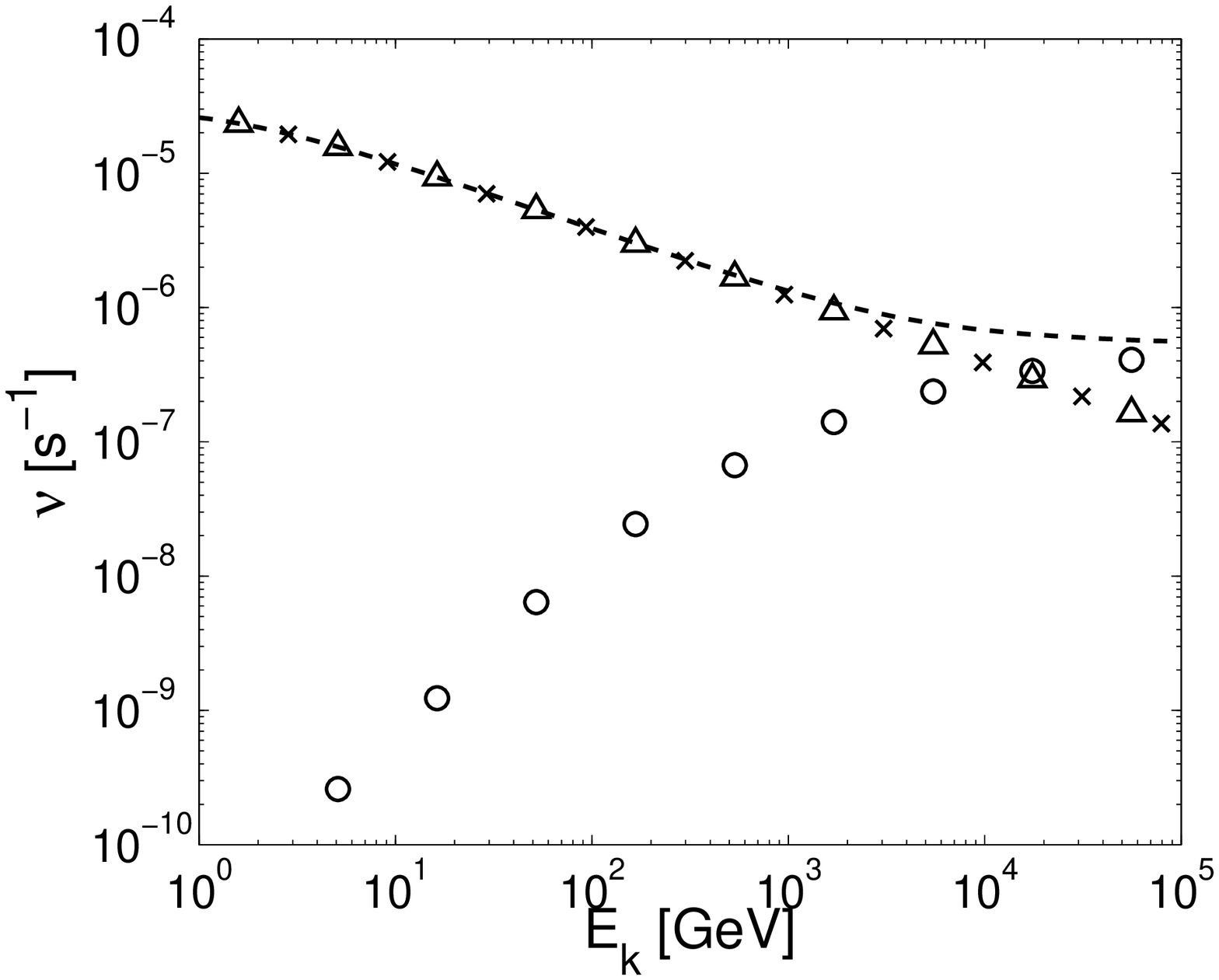}\label{figmcsfb1}}
\subfigure[Gyroresonance, with damping]{  
   \includegraphics[width=8cm]{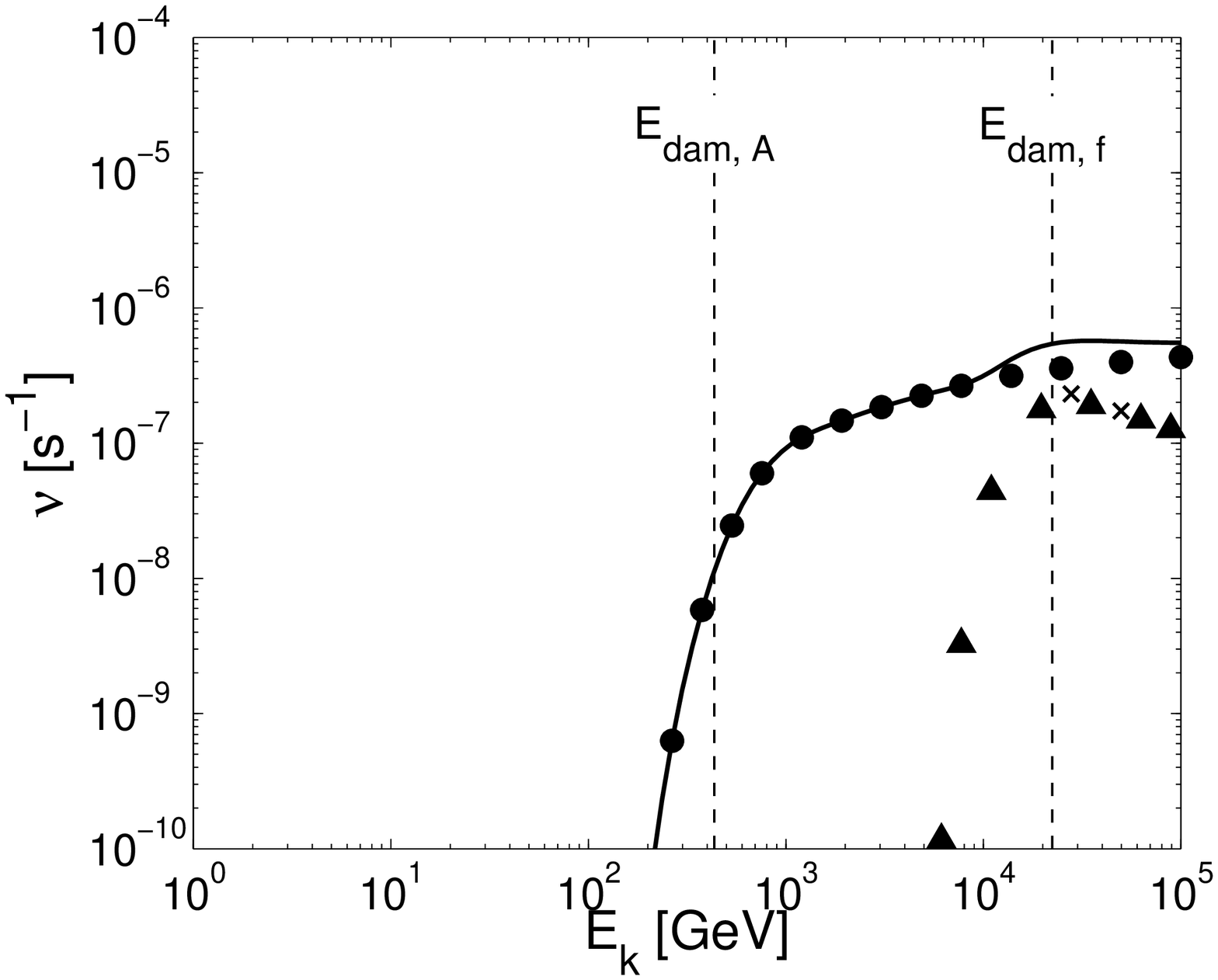}\label{figmcsfb2}}
\caption{Same as Fig. \ref{figwnmsf} but for MC. (b) is replotted in (c) and (d), showing the separate contributions from Alfv\'{e}n (circles) and fast (triangles) modes. }
\label{figmcsf}
\end{figure*}

To gain a deeper understanding on the dependence of TTD and gyroresonance on CR energy and the influence of turbulence damping,
we illustrate the scattering frequency $\nu=2 D_{\mu\mu} / (1-\mu^2)$ of TTD and gyroresonance separately in WNM (Fig.~\ref{figwnmsf}) and 
MC (Fig. \ref{figmcsf}).
Since the major scattering agent differs at small and large pitch angles, we present results of TTD at $\mu=0.1$ in Fig.~\ref{figwnmsfa} and \ref{figmcsfa},
and gyroresonance at $\mu=0.8$ in Fig.~\ref{figwnmsfb} and \ref{figmcsfb}. 
The vertical dashed lines indicate the CR energies 
with $r_L$ equal to the damping scales of Alfv\'{e}n ($E_\text{dam, A}$) and fast ($E_\text{dam, f}$) modes. 
Slow modes are not considered due to their negligible effect.

In WNM, as shown in Fig.~\ref{figwnmsfa}, when damping is absent, $\nu$ (dashed line) through TTD with fast modes decreases with CR energy. 
Otherwise $\nu$ (solid line) keeps steady until reaching $E_\text{dam, f}$ (vertical dashed line), and then coincides with the result without damping. 
The analytical scattering frequency (asterisks) is derived by using Eq. \eqref{eq: duuftapp}, and can apply to the energy range below 
$E_\text{dam, f}$. As illustrated by Eq. \eqref{eq: duuftapp}, for CRs with $r_L<k_\text{dam}^{-1}$,
$D_{\mu\mu}^T$ is determined by the damping scale. Since CR velocity is approximately equal to the light speed and its change with 
energy is marginal, Eq. \eqref{eq: duuftapp} yields a constant $D_{\mu\mu}^T$ and $\nu$ at a fixed $\mu$.
It implies only the turbulence on scales larger than $r_L$ contributes to TTD interaction. 
As a result, in the case of no turbulence damping, 
$\nu$ of lower-energy CRs has a higher value because turbulence on a larger range of scales is involved in TTD scattering. 
While in the case with damping, 
$\nu$ of TTD is independent of CR energy for CRs with $r_L<1/k_\text{dam}$, and decreases with energy when $r_L>1/k_\text{dam}$.

Fig. \ref{figwnmsfb} shows the total $\nu$ of gyroresonance by both Alfv\'{e}n and fast modes in WNM. 
The results free of damping and in the presence of damping come into coincidence when CR energy is over $E_\text{dam, f}$. 
The respective roles of Alfv\'{e}n and fast modes are explicitly displayed.  
When there is no damping, Alfv\'{e}n modes have negligible effect over the whole energy range due to strong turbulence anisotropy in WNM. So $\nu$ for gyroresonance with fast modes (open triangles) coincides with the total $\nu$. 
The crosses are analytical $\nu$ by employing Eq. \eqref{eq: duufagqs}.
Although Eq. \eqref{eq: duufagqs} is the simplified $D_{\mu\mu}^G$ using QLT for $k_{\text{dam},\perp} r_L<1$, the gyroresonance condition 
$k_\|=\Omega/(v\mu)=\sqrt{1-\mu^2}/(r_L\mu)$ can still be satisfied at large $\mu$, so that it can reach a good agreement with the 
numerical result using NLT.\footnote{In fact, Eq. \eqref{eq: duufagqs} is only valid for $\mu>0.5314$ to ensure a positive value of $D_{\mu\mu}$.}
Similar to TTD, gyroresonance with fast modes has decreasing $\nu$ with increasing CR energies, but their physical origins are different. 
For TTD, there is not a specific resonant scale. Turbulence over all the scales above $r_L$ contributes to particle scattering. 
However, gyroresonance can only become effective on specified scales which are determined by CR energy. 
We observe in Eq. \eqref{eq: duufagqs} that $D_{\mu\mu}^G \propto 1/\sqrt{R}$, where $R$ is CR rigidity. 
It demonstrates that higher-energy CRs are scattered by larger-size turbulence eddies, and consequently have larger mean free paths and 
smaller $D_{\mu\mu}^G$ and $\nu$.

When damping is taken into account, since gyroresonance has a preferential resonant scale, $\nu$ corresponding to Alfv\'{e}n (filled circles) and fast 
(filled triangles) modes only become efficient after $r_L$ exceeds their damping scales. 
The analytical $D_{\mu\mu}^G$ given by Eq. \eqref{eq: duufagqs} can only apply to CRs with energies larger than $E_\text{dam,f}$
in this situation.
Therefore, TTD is the only effective interaction and governs particle scattering at lower energies.
Since Alfv\'{e}n modes have a smaller damping scale than fast modes, their contribution can be seen within the energy range [$E_\text{dam,A}, E_\text{dam,f}$]. 
But fast modes dominate the scattering of CR with energies higher than $E_\text{dam,f}$.

In the case of super-Alfv\'{e}nic MC, we see much higher $\nu$ values for both TTD and gyroresonance. The $\nu$ of TTD in Fig. \ref{figmcsfa} shows a similar trend as that in 
Fig. \ref{figwnmsfa} for WNM. But for $\nu$ of gyroresonance (Fig. \ref{figmcsfb}), the Alfv\'{e}n modes become more important in MC compared with WNM
due to less prominent turbulence anisotropy.
Fig. \ref{figmcsfb1} and \ref{figmcsfb2} display the components of $\nu$ from Alfv\'{e}n and fast contributions for cases without and with damping.
When damping is absent, $\nu$ of gyroresonance with Alfv\'{e}n modes (open circles) increases towards higher energies. 
The opposite trend compared with gyroresonance with fast modes originates from the scale-dependent turbulence anisotropy. 
Lower-energy CRs interact with many elongated eddies with $k_\perp \gg k_\| \sim 1/r_L$
within one gyroperiod and this random walk causes inefficient scattering. 
While higher-energy CRs are scattered by larger-size eddies which are less anisotropic and hence the scattering efficiency increases
\citep{YL04}.
We see in Fig. \ref{figmcsfb1}, approaching $l_A$, 
weak turbulence anisotropy makes Alfv\'{e}n modes have comparable efficiency as fast modes in gyroresonance scattering. 
Their joint effect flattens $\nu$ at high energies.

In brief, we find both TTD and gyroresonance with fast modes decrease with CR energy, while gyroresonance with Alfv\'{e}n modes increases with energy. 
The damping effect can only make a difference in CR scattering when $r_L<1/k_\text{dam}$. 
Then TTD becomes the only scattering agent and has a scattering efficiency independent of particle energy. 
At higher energies with $r_L>1/k_\text{dam}$, the importance of gyroresonance with Alfv\'{e}n modes depends on the anisotropy degree of the turbulence. 
In super-Alfv\'{e}nic turbulence, Alfv\'{e}n modes can become more efficient than fast modes in gyroresonance scattering at large scales comparable to $l_A$.

\subsection{Results on parallel mean free path of CRs}
\begin{figure*}[htbp]
\centering
\subfigure[WNM]{ 
   \includegraphics[width=8cm]{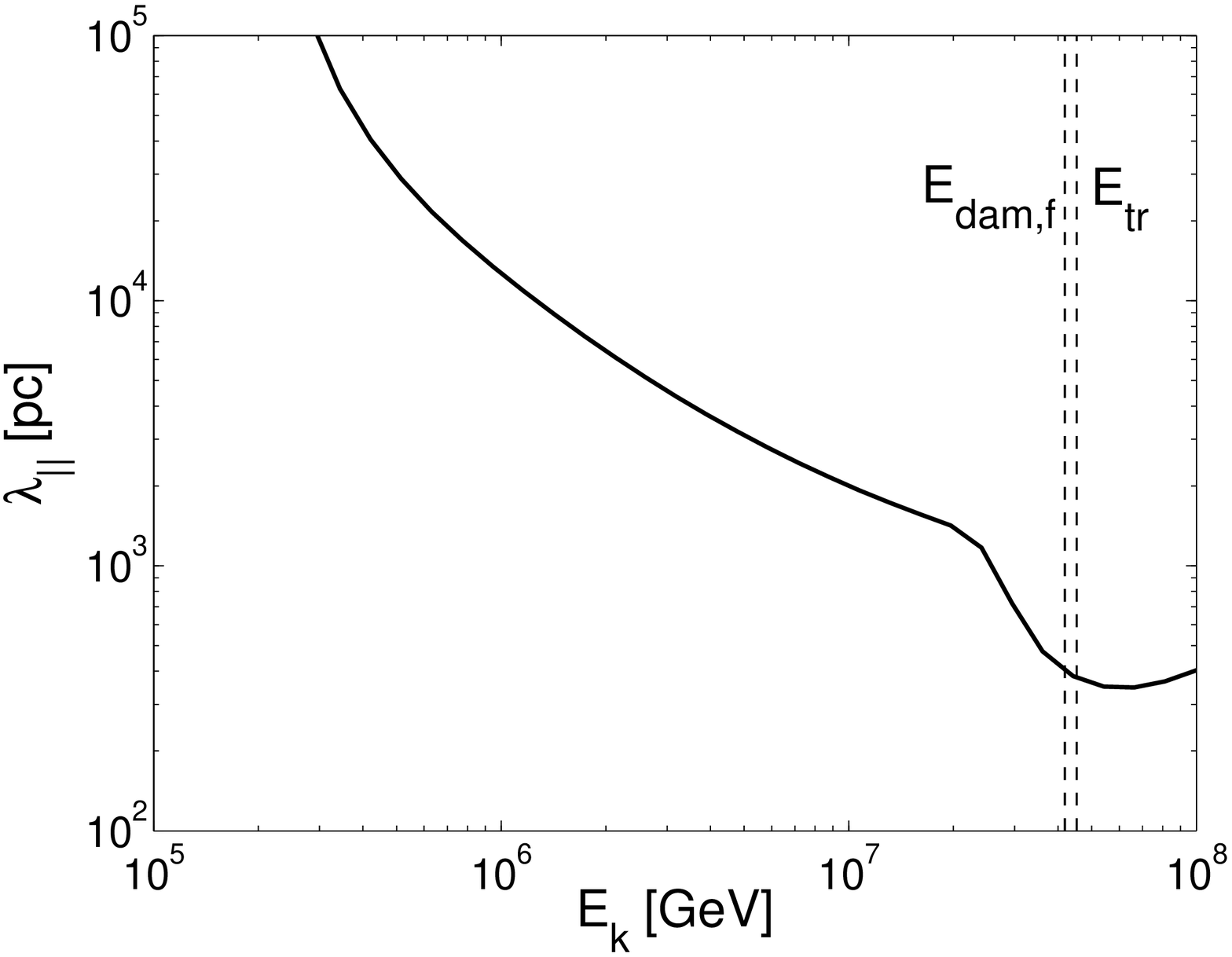}\label{figwnmpmfp}}
\subfigure[CNM]{ 
   \includegraphics[width=8cm]{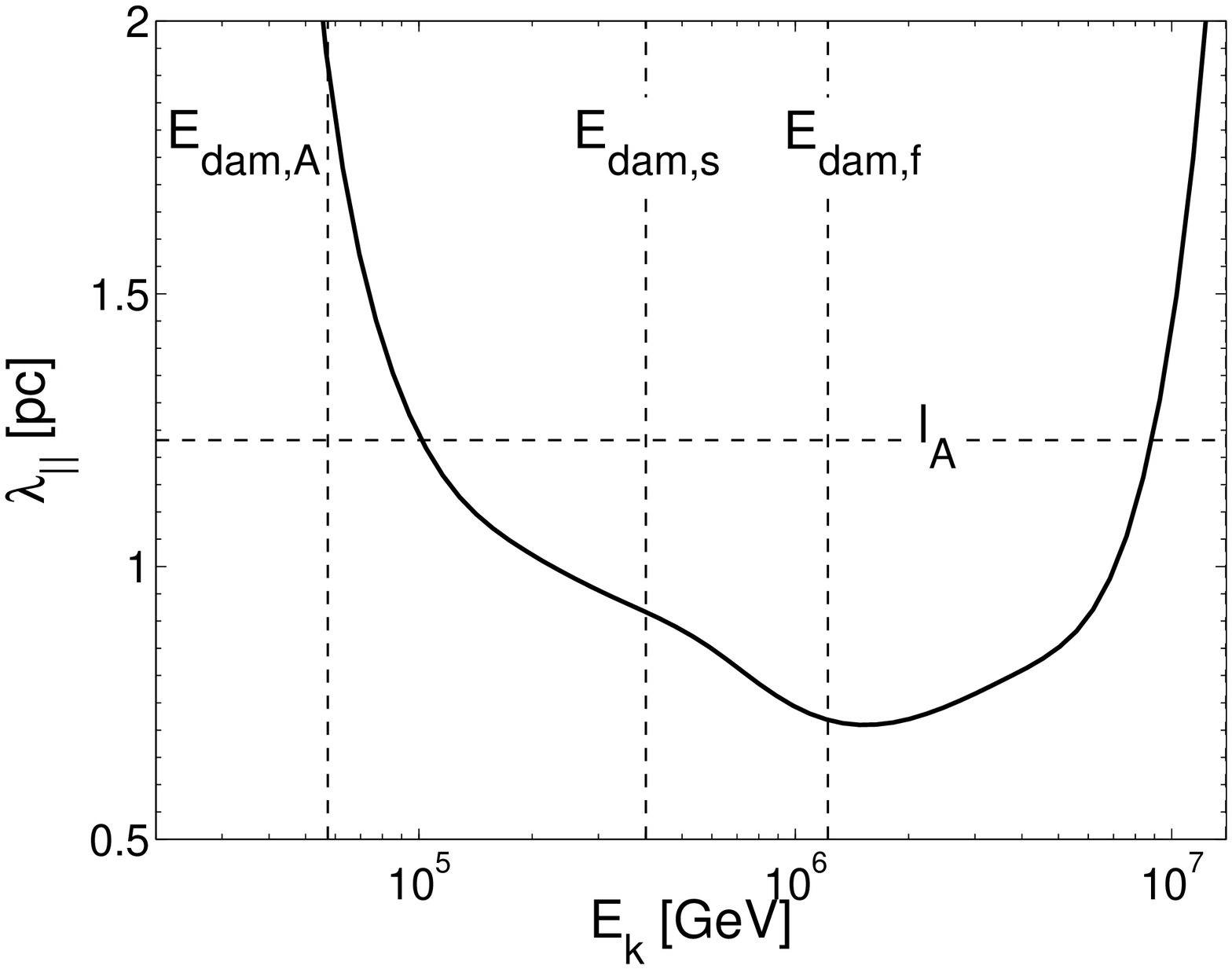}\label{figcnmpmfp}}
\subfigure[MC]{ 
   \includegraphics[width=8cm]{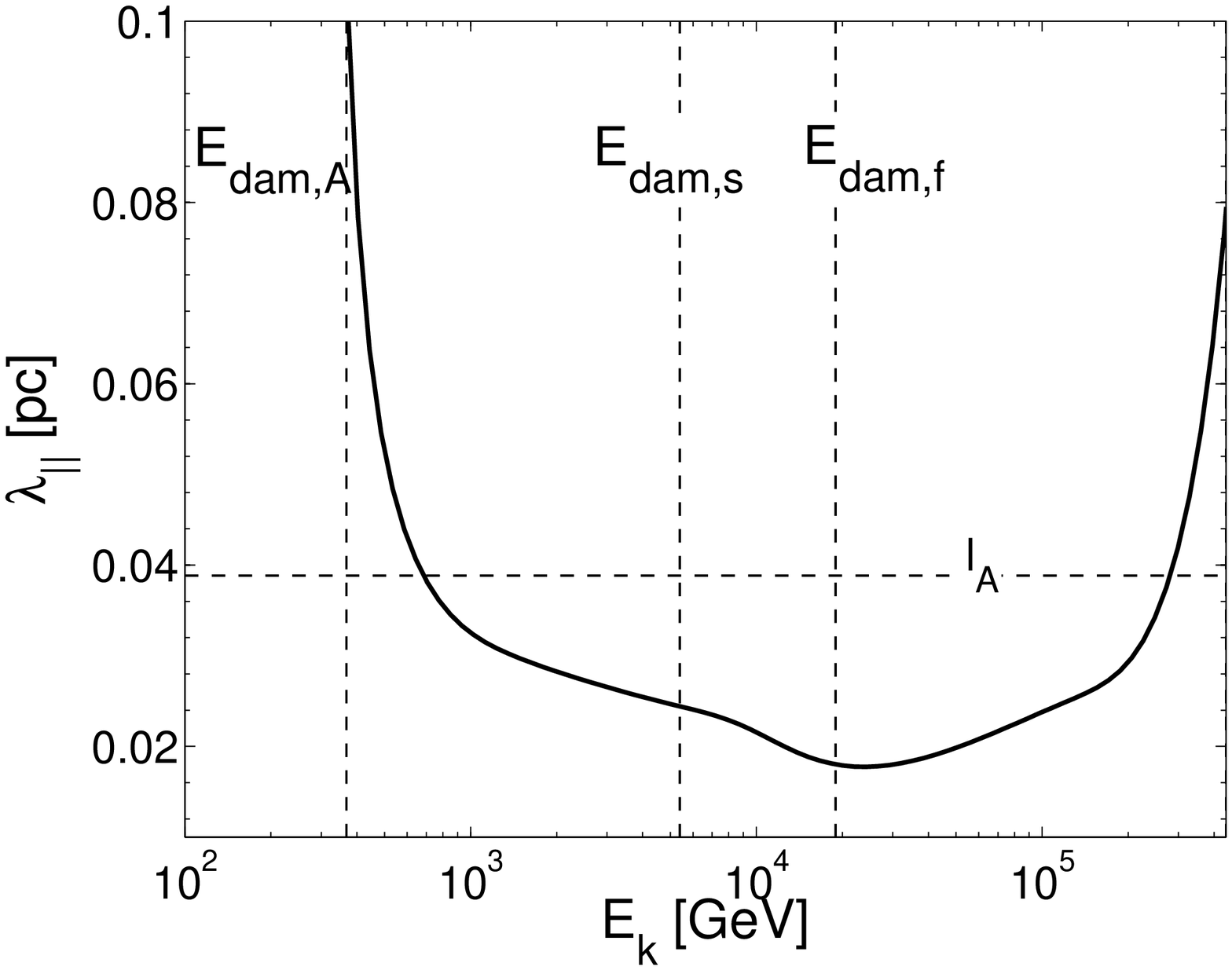}\label{figmcpmfp}}
\subfigure[DC]{ 
   \includegraphics[width=8cm]{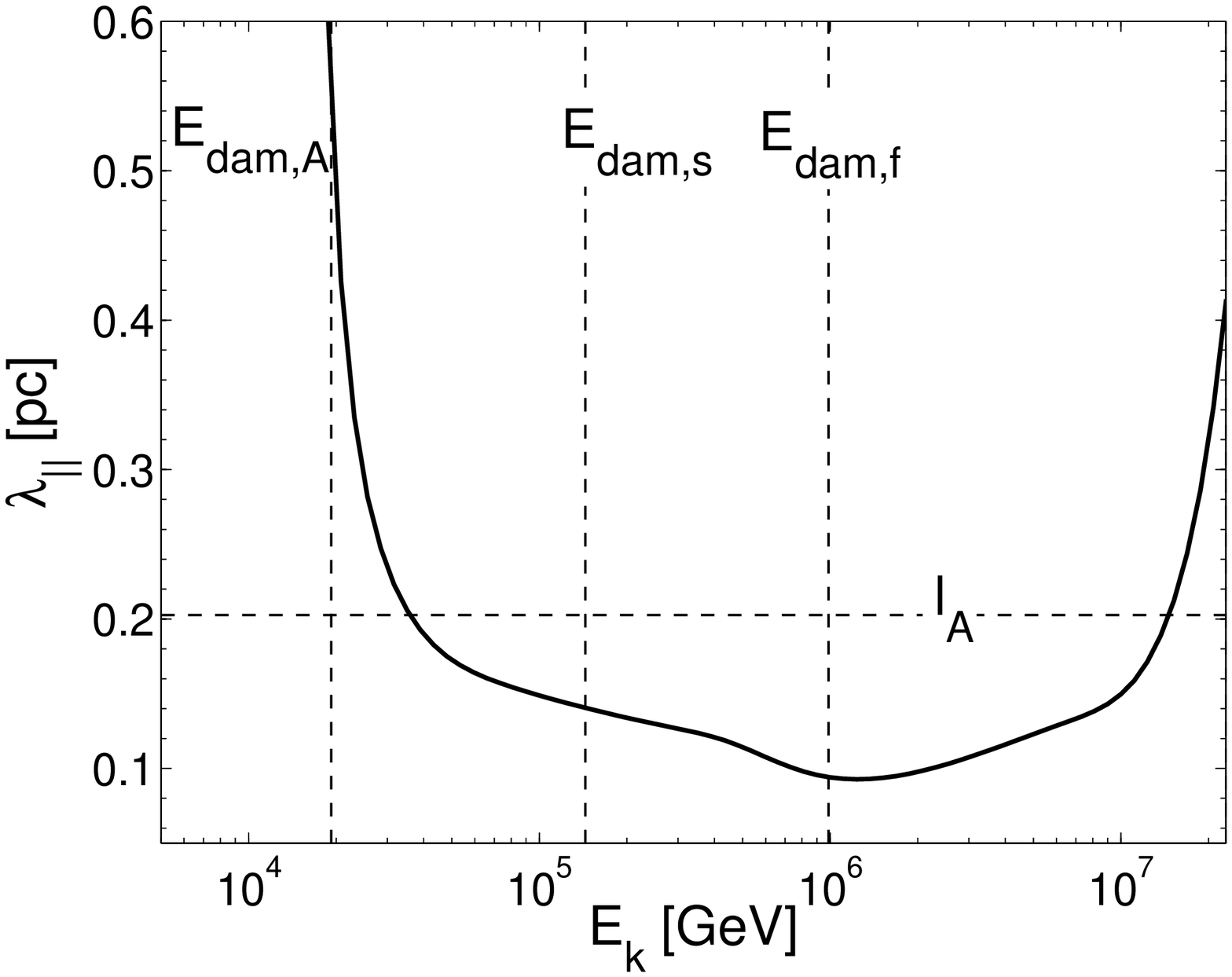}\label{figdcpmfp}}
\caption{ Parallel mean free path of CRs as a function of their energies in (a) WNM, (b) CNM, (c) MC and (d) DC.   
   Vertical dashed lines represent the CR energies with their $r_L$ equal to the damping scales of Alfv\'{e}n ($E_\text{dam, A}$), 
   slow ($E_\text{dam, s}$), and fast ($E_\text{dam, f}$) modes. 
   Notice that $E_\text{dam, A}$ is below the energy range shown in (a). $E_{tr}$ is the 
   energy corresponding to $l_{tr}$. The horizontal dashed lines in (b), (c) and (d) refer to the length scales of $l_A$. }
\label{figmfp}
\end{figure*}

Given the results of $D_{\mu\mu}$, the parallel mean free path $\lambda_\|$ of CRs can be evaluated by integrating over $\mu$, 
\begin{equation}
\lambda_\|=\frac{3v}{4}\int_0^1 d\mu \frac{(1-\mu^2)^2}{D_{\mu\mu}(\mu)}, 
\end{equation}
where $D_{\mu\mu}$ is the total contributions from both TTD and gyroresonance with the three modes calculated using NLT. 
%The $\lambda_\|$ reflects interactions between particles and turbulence and its dependence on particle energy is not necessarily monotonic. 
Fig.~\ref{figwnmpmfp}-\ref{figdcpmfp} present $\lambda_\|$ as a function of CR energy $E_k$. 
The upper limit of the energy range is restrained by $r_L<L$ in WNM and $r_L<l_A$ in other environments.

We can clearly see the impact of turbulence damping on $\lambda_\|$.
In the case of WNM (Fig. \ref{figwnmpmfp}),  
even in the undamped range of turbulence, neither Alfv\'{e}n nor slow modes can effectively scatter CRs. Only 
when CR energy approaches $E_\text{dam, f}$, can CRs be confined by gyroresonance with fast modes. 
Nevertheless, the resulting $\lambda_\|$ is larger than $l_{tr}$ over the whole energy range. 
It shows in WNM, direct interactions with turbulence modes are invalid in scattering CRs. Instead, other mechanisms, e.g. 
field line wandering, can play a more important role in determining $\lambda_\|$.

In the other ISM phases (Fig. \ref{figcnmpmfp}-\ref{figdcpmfp}), we observe a nonmonotonic U-shaped dependence of $\lambda_\|$ on CR energy. 
The dramatical decrease of $\lambda_\|$ at $E_\text{dam, A}$ and the further decrease of $\lambda_\|$ at $E_\text{dam, f}$ come 
from the contribution of gyroresonance with Alfv\'{e}n and fast modes respectively, while 
the rise of $\lambda_\|$ at $E_k>E_\text{dam, f}$ is in accordance with the decreasing $\nu$ of TTD with energy.
The marginal change of $\lambda_\|$ at $E_\text{dam, s}$ reflects the insignificant role of slow modes in CR scattering.

On the other hand, although TTD contributes over all energy range and is more efficient in CR scattering, 
it is incapable to scatter CRs with small pitch angles. 
Thus TTD alone is not enough to confine CRs and leads to infinite $\lambda_\|$ in the energy range where gyroresonance is absent. 
Therefore, the lower energy limit of effective CR scattering and finite $\lambda_\|$ is determined by the damping scale of Alfv\'{e}n mode.

%The above analysis informs us turbulence damping has a significant influence on the scattering behavior of CRs with $r_L<1/k_\text{dam}$. 
Table \ref{tab: decsccrene} summarizes the lowest energies of CRs with their scattering unaffected by turbulence damping in different partially ionized 
ISM, which are determined by the damping scales of fast modes. 
The scattering and acceleration of CRs with lower energies are subject to turbulence damping. 
Notice that different from the situation 
discussed here, \citealt{YL04} showed in fully ionized plasma, damping of fast modes strongly depends on the angle $\theta$ between $\mathbf{k}$ and 
$\mathbf{B}$, which results in anisotropic distribution of fast mode energy at small scales. 
In their case, the lowest energy for CR scattering unaffected by turbulence damping should be a function of $\theta$.

\begin{table}[h]
\renewcommand\arraystretch{2}
\centering
\begin{threeparttable}
\caption[]{The minimum energy of CRs unaffected by turbulence damping in different ISM phases. }\label{tab: decsccrene} 
  \begin{tabular}{c|c|c|c|c}
      \toprule
  \multirow{2}*{ISM phases}       &   \multicolumn{3}{c|}{$k_\text{dam}^{-1}$}       & \multirow{2}*{$E_{k, min}$} \\
                 \cline{2-4}               
                                                  &  Alfv\'{e}n    & fast        & slow                             &         \\
                 \hline         
  WNM                                       &    $0.003$ pc    & $4.0$ pc     & ---                                    & $45.3$ PeV \\        
  CNM                                        &    $0.005$ pc    & $0.1$ pc  & $0.04$ pc                     & $1.2$ PeV \\
  MC                                          &     $6.7$ AU      & $0.002$ pc & $98.2$ AU                 & $18.9$ TeV \\
  DC                                           &    $35.0$ AU    & $0.009$ pc  & $261.7$ AU             & $0.99$ PeV  \\
              \bottomrule
    \end{tabular}
 \end{threeparttable}
\end{table}

\subsection{Other mechanisms on confining CRs}
Our results show that in partially ionized medium, MHD turbulence, especially fast modes, is severely damped by neutral-ion collisions. 
Consequently, only high-energy CRs can be efficiently scattered through interactions with turbulence modes. 
Besides scattering, the diffusion of CRs can also arise from the spatial wandering of magnetic field lines
\citep{Jokipii1966}. 
In three dimensional turbulence, the wandering of field lines is induced by turbulent motions and characterized by turbulence properties.
\citet{LV99} quantitively described the field line wandering according to the scaling laws of GS95-type MHD turbulence,
and demonstrated its key role in determining the rate that magnetic reconnection proceeds. 
Based on the \citet{LV99} prescription for magnetic field wandering and the related field line diffusion,  
significant theoretical reformulations on e.g. thermal production
\citep{Narayan_Medv,Lazarian06}
and CR propagation
\citep{YL08, Yan:2011valencia,LY14},
have been achieved. 

Other mechanisms which can enhance scattering of CRs include streaming instability   % are they the same? 
\citep{Wentzel74, Cesarsky80}
and gyroresonance instability
\citep{LB06,YL11}. 
The additional MHD waves excited by CRs can in turn efficiently scatter CRs within a range of energies (e.g., $\lesssim 100$ GeV). 
\citet{YLS12} investigated the shock acceleration of CRs in the presence of streaming instability. 
They found the CR flux near supernova remnants is strongly enhanced compared with typical Galactic values, so the growth rates of streaming 
instability can overcome the background turbulence damping and boost scattering of CRs. 
This finding enables them to reproduce the observed gamma-ray emission from the supernova remnant W28.

The growth rate of the resonant waves excited by streaming CRs is 
\begin{equation}
    \Gamma_\text{g,CR} \sim \Omega_0 \frac{n_\text{CR} (>\Gamma)}{n_i} \bigg(\frac{v_\text{stream}}{V_A}-1 \bigg),
\end{equation}
where $\Omega_0=eB/m_p c$ is the non-relativistic CR gyrofrequency. 
$n_\text{CR} (>\Gamma)$ is the number density of CRs with their energies larger than $\Gamma$ GeV, 
and the $r_L$ value corresponding to $\Gamma$ gives the wavevector of the generated waves, i.e. $r_L \sim1/k_\| \sim 1/k$, which propagate closely 
parallel to magnetic field. 
And $v_\text{stream}$ is the streaming velocity of CRs, which should be larger than $V_A$ so as to amplify waves. 

The instability can only set in when $\Gamma_\text{g,CR}$ exceeds the total damping rate, which includes the neutral-ion collisional 
and neutral viscous damping 
discussed in this paper, and also the nonlinear damping by background MHD turbulence 
\citep{YL02,YL04,FG04,BL08,YL11}. 
The non-linear damping rate takes the form as 
\citep{YL11}
\begin{equation}
  \Gamma_\text{turb}=r_L^{-\frac{1}{2}} l_A^{-\frac{1}{2}} V_A
\end{equation}
for $r_L<l_A$ in super-Alfv\'{e}nic turbulence, and 
\begin{equation}
  \Gamma_\text{turb}=r_L^{-\frac{1}{2}} L^{-\frac{1}{2}} V_L 
\end{equation}
for $r_L<l_{tr}$ in sub-Alfv\'{e}nic turbulence. 

\begin{figure}[htbp]
\centering
\includegraphics[width=9cm]{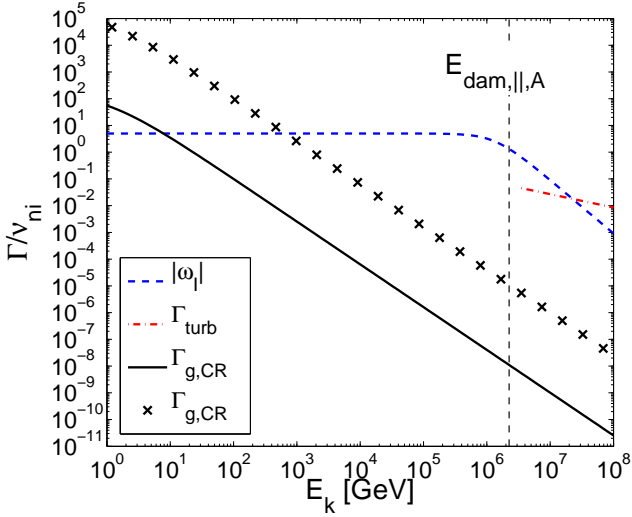}
\caption{Growth and damping rates (normalized by $\nu_{ni}$) of streaming instability as a function of CR energy in WNM. 
The dashed line is the neutral-ion collisional damping rate. The dash-dotted line is the nonlinear damping rate. 
The growth rates with different CR number densities, $n_\text{CR} (>\Gamma) = 2 \times 10^{-10} \Gamma^{-1.6} \text{cm}^{-3}$
(solid line), and $n_\text{CR} (>\Gamma) = 2 \times 10^{-7} \Gamma^{-1.6} \text{cm}^{-3}$
(crosses) are shown.
The vertical dashed line indicates the energy corresponding to the parallel damping scale of the background Alfv\'{e}n modes. }
\label{figwnmgrow}
\end{figure}

As an example, 
Fig. \ref{figwnmgrow} compares the growth and damping rates at different CR energies in the WNM environment. The neutral-ion collisional 
damping rate for parallel Alfv\'{e}n waves (Eq. \eqref{eq: anasol} with $\theta=0$) applies, which has the value of $|\omega_I|=\nu_{in}/2$ 
at large wavenumbers corresponding to low CR energies. Notice neutral viscous damping is unimportant for the generated waves in this particular case. 
The nonlinear damping only exists for CRs with their $r_L$ larger than the parallel damping 
scale of the background Alfv\'{e}nic turbulence.   % how about slow and fast 
For the growth rate represented by the solid line,
we assume $v_\text{stream}$ has the same order of magnitude as $V_A$ and adopt the CR density near the sun, 
$n_\text{CR} (>\Gamma) = 2 \times 10^{-10} \Gamma^{-1.6} \text{cm}^{-3}$
\citep{Wentzel74}. 
It shows the streaming instability can only overcome the collisional damping and effectively contribute to particle scattering 
for CRs with energies lower than $\sim 10$ GeV. 
Crosses show a higher growth rate with the CR density increased by three orders of magnitude. As a result, 
CRs with up to $\sim 1$ TeV can be significantly scattered and confined. 

From above example we see to establish a comprehensive picture of CR propagation, other scattering mechanisms should also be incorporated 
in addition to direct interactions with turbulence modes.  
For relatively low-energy CRs, streaming instability can be a promising mechanism of CR confinement by scattering. 
For high-energy CRs with their $r_L$ exceeding the damping scale of MHD turbulence, TTD and gyroresonance scattering operate. 
Regarding the CRs with intermediate energies, their confinement is mainly attributed to field line wandering as mentioned earlier. 

We again stress that in analyzing these scattering processes in a partially ionized medium, neutral-ion collisional damping is an 
essential physical ingredient, which determines the dissipation scale of turbulence cascade, and is the major constraint for the growth of instabilities. 
Taking the damping effect into account is necessary to attain realistic calculation of diffusion coefficients and proper understanding of CR 
propagation in partially ionized medium.

\section{Discussion}
In partially ionized medium, the wave behavior strongly depends on the coupling state between neutrals and ions. 
We consider low-$\beta$ medium and specifically distinguish 
the neutral-ion and ion-neutral decoupling scales of three modes and the regimes with different coupling degrees separated by them. 
The cutoff boundaries of waves are closely correlated with the decoupling scales. In fact, 
the cutoff interval approximately shares the same domain as that between the neutral-ion and ion-neutral decoupling scales.
This coincidence indicates fluid decoupling can be the physical origin of the wave cutoff. 
Unlike Alfv\'{e}n and fast waves, propagating waves can arise within the original cutoff region of slow waves, which results in two cutoff 
intervals sitting among three propagating wave branches. We for the first time provide full analytical expressions of all the cutoff boundaries and the 
wave frequencies of both "neutral" and "ion" slow waves in different coupling regimes. Slow waves on scales within $[k_\text{dec,ni},k_\text{dec,in}]$ exhibit
more complicated properties in both propagating and damping components of the wave frequencies. Accordingly, the wave spectrum on 
intermediate scales is also divided into different coupling regimes, confined by multiple critical scales which are linked together in a symmetric patten (see Section \ref{sec: slowsr}).

In order to test the validity of the single-fluid approach, we adopted NewtonÕs iteration method to solve the single-fluid dispersion relation. 
By comparing the solutions with those to the two-fluid dispersion relation under weak damping assumption, we see consistent results 
at large scales down to the lower cutoff boundary $k_c^+$. 
We further numerically confirmed the validity of single-fluid approximation in describing MHD waves in partially ionized ISM and SC environments. 
It means for practical purposes, single-fluid treatment can also be used to determine the damping scales of both incompressible and compressible 
turbulence modes in strongly coupled regime in similar conditions.

To obtain the damping scales of MHD turbulence, we follow the present-day understanding of turbulence and treat the turbulence 
cascades of Alfv\'{e}n, fast and slow modes separately. 
It turns out the scale-dependent anisotropy of turbulence plays a critical role in regulating wave behavior and deriving turbulence damping.  
We see a close relation between the cutoffs appearing in linear MHD waves and the damping of MHD turbulence. 
Particularly, because of the intrinsic critical balance of the turbulent motions (GS95), the cascade of Alfv\'{e}n modes is truncated at the lower 
cutoff boundary where the nonpropagating waves arise. 
As a result, the wave cutoff should also be taken into account when deriving turbulence damping scales. 

In addition, we apply the analytical results on turbulence damping to a variety of partially ionized ISM phases and solar chromosphere. 
\begin{itemize}
\item[-] We find {\it neutral viscosity plays a significant role in damping Alfv\'{e}n modes in WNM and SC}, while neutral-ion collisions act as the 
dominant damping effect in other environments. 

\item[-] Fast modes in all conditions are damped out in strongly coupled regime. And the damping is especially severe in WNM and SC 
because the cascading rates in both environments are substantially low 
and in addition, the damping rate in WNM is relatively high due to its low density.
Besides, 
different from the case in fully ionized medium, the damping of fast modes only has a weak dependence on wave propagation direction. 

\item[-] In the case of slow modes, since cutoff appears earlier than turbulence damping, 
the damping scales of slow modes are given by the largest cutoff scales in all conditions except for WNM. 
Due to the absence of cutoff intervals, slow modes in WNM survive neutral-ion collisional damping. 
In fact, all the three modes in WNM do not have cutoffs. 
The physical explanation is due to the relatively high ionization fraction in WNM, 
coupling of neutrals with ions on larger scales is easier since every neutral has more chances to collide with ions, 
and decoupling of ions from neutrals on smaller scales is also easier since every ion has fewer chances to collide with neutrals. 
Consequently, the separation between the neutral-ion and ion-neutral decoupling scales is much shortened and cutoff can be avoided. 
\end{itemize}

Regarding the turbulence cascade of slow modes below the neutral-ion decoupling scale in WNM, 
although the MHD turbulence in ions survives neutral-ion collisional damping, since slow modes are slaved to Alfv\'{e}n, 
it may be quenched at a smaller scale with the damping of Alfv\'{e}n modes. 
Provided the cascade of "ion" slow modes can extend to the regime where collisionless damping is dominant, 
we refer the reader to 
\citet{LG01} 
on the slow mode damping below the proton mean free path. 
For the acoustic turbulence of "neutral" slow modes, it is subject to both neutral-ion collisional and neutral viscous damping. 
The damping scale can be determined by comparing the cascading rate and both damping rates, 
which falls beyond the scope of the current work and will be discussed in future work.

As one important application of the present study, we explored the damping effect on CR propagation in partially ionized medium. 
The local reference system where GS95-type scalings stand is also very important for studying CR propagation, since the scattering 
of a CR particle is determined by its interaction with the local magnetic field perturbation instead of background field. 
That ensures realistic statistics of magnetic fluctuations which we use for calculating CR scattering.

Alfv\'{e}n modes are found to be an ineffective scatterer of CRs 
\citep{Chandran00, YL02, YL04}. 
During the energy cascade of Alfv\'{e}n modes, turbulent eddies become more and more elongated along magnetic field lines, 
with energy concentrating in the direction perpendicular to the local mean magnetic field. 
Thus a CR particle with $r_L$ comparable to the parallel scale of eddies interacts with many uncorrelated eddies in perpendicular direction 
within one gyroperiod. The random walk leads to very inefficient gyroresonance scattering of CRs by Alfv\'{e}n modes. 
But in the presence of neutral-ion collisional damping, turbulence cascade is heavily damped and truncated at a large scale, where turbulence 
anisotropy is relatively weak. 
Especially in super-Alfv\'{e}nic turbulence, turbulent eddies are nearly isotropic when $r_L$ approaches $l_A$. 
As a result, high-energy CRs can be effectively scattered by gyroresonance with Alfv\'{e}n modes, as shown above.

\section{Summary}
In this paper we continued the work in Paper \uppercase\expandafter{\romannumeral1} on damping of MHD turbulence in partially ionized medium. 
We put more emphasis on compressible modes and apply the analytical results to a variety of low-$\beta$ ISM phases and solar chromosphere.
As an important application, we studied the CR propagation in partially ionized medium considering the neutral-ion collisional damping, as well as 
neutral viscous damping, of MHD turbulence. 
Here we summarize the main results. 

1. We present explicit correlation between the cutoff boundaries of MHD waves and decoupling scales. 
Under the consideration of scale-dependent anisotropy, they are 
\begin{subequations}
 \begin{align}
 & k_c^+=(\frac{2}{\xi_n})^\frac{3}{2}k_\text{dec,ni}, \\
 & k_c^-=2^{-\frac{3}{2}} k_\text{dec,in}. 
\end{align}
\end{subequations}
for Alfv\'{e}n modes, and 
\begin{subequations} 
\begin{align}
& k_{c}^+=\frac{2}{\xi_n} k_\text{dec,ni},  \\
& k_{c}^-=\frac{1}{2} k_\text{dec,in}. 
\end{align}
\end{subequations} 
for fast modes. 
Slow modes have the same
\begin{equation}
 k_c^-=2^{-\frac{3}{2}} k_\text{dec,in}
\end{equation}
as Alfv\'{e}n modes,
but the relation between $k_c^+$ and $k_\text{dec,ni}$ depends on turbulence properties.

2. Single-fluid approach is capable of correctly describing wave behavior and damping properties in strong coupling regime. 

3. Cutoff of MHD waves should also be taken into account when deriving damping scales of turbulence cascade. 

4. We showed the importance of neutral viscosity in damping Alfv\'{e}n modes in WNM and SC. Especially for the Alfv\'{e}n modes in SC, 
neutral viscosity is the dominant damping effect instead of neutral-ion collisions.

5. We thoroughly investigated the behavior of slow waves and provided full expressions of the multiple cutoffs and wave frequencies in different 
coupling regimes.

6. The scale-dependent anisotropy of GS95-type turbulence is important for understanding wave behavior, calculating turbulence damping scales, 
and studying interactions between CRs and magnetic perturbations.

7. We evaluated the scattering efficiencies of TTD and gyroresonance with the three turbulence modes, and 
found due to the severe damping in partially ionized medium, only high-energy CRs can be effectively scattered through direct interactions 
with turbulence modes. Compared with other environments, 
CRs in WNM are poorly confined due to its prominent turbulence anisotropy, as well as the largest damping scale of 
fast modes.
As for CRs with lower energies, other effects such as field line wandering and streaming instability can set in and contribute in confining CRs.

\section*{Acknowledgement}
S. X. acknowledges the support from China Scholarship Council during her stay in University of Wisconsin-Madison.  
S. X. and H. Y. are supported by the NSFC grant AST-11473006.
S. X. also thanks Heshou Zhang for the stimulating discussions. 
A. L. acknowledges the NSF grant AST 1212096, a distinguished visitor PVE/CAPES appointment at the Physics Graduate Program of the Federal University of Rio Grande do Norte and thanks the INCT INEspao and Physics Graduate Program/UFRN for hospitality.

\appendix

\section{A summary of the notations used in this paper}
\label{app:a}
\begin{longtable}{|l|c|}
\hline
drag coefficient & $\gamma_d$ \\
ion density & $\rho_i$ \\
neutral density & $\rho_n$ \\
total density & $\rho$ \\
proton mass & $m_p$ \\
mass of hydrogen atom & $m_H$ \\
ion mass & $m_i$ \\
neutral mass & $m_n$ \\
reduced mass & $m_r$ \\
ion number density & $n_i$ \\
neutral number density & $n_n$ \\
total number density & $n$ \\
neutral-ion collisional cross-section & $\sigma_{ni}$ \\
%neutral-ion mean free path & $l_{mfp}$ \\
%slip speed & $\delta$ \\
%mean dipole polarizability of neutrals & $\alpha_n$ \\
%Bohr radius & $a_0$ \\
electronic charge & $e$ \\
neutral-ion collision frequency & $\nu_{ni}$ \\
ion-neutral collision frequency & $\nu_{in}$ \\
ion fraction & $\xi_i$ \\
neutral fraction & $\xi_n$ \\
ratio of neutral and ion densities & $\chi$ \\
wave frequency & $\omega$ \\
real part of wave frequency & $\omega_R$ \\
imaginary part of wave frequency & $\omega_I$ \\
wave number & $k$ \\
$x, y, z$ component of $k$ & $k_x, k_y, k_z$ \\
$k$ component parallel to local magnetic field & $k_\|$ \\
$k$ component perpendicular to local magnetic field & $k_\perp$ \\
wave propagation angle with regard of magnetic field & $\theta$ \\
Alfv$\acute{e}$n speed & $V_A$ \\
Alfv$\acute{e}$n speed of ion-electron gas & $V_{Ai}$ \\
magnetic field & $\mathbf{B}$ \\
sound speed & $c_s$ \\
sound speed of neutral gas & $c_{sn}$ \\
sound speed of ion-electron gas & $c_{si}$ \\
ion gas pressure & $P_i$ \\
neutral gas pressure & $P_n$ \\
%electron gas pressure & $P_e$ \\
%ion gas temperature & $T_i$ \\
%electron gas temperature & $T_e$ \\
temperature & $T$ \\
Boltzmann constant & $k_B$ \\
adiabatic constant & $\gamma$ \\
ratio of gas pressure to magnetic pressure & $\beta$ \\
injection scale of turbulence & $L$ \\
injection scale of MHD turbulence in super-Alfv$\acute{e}$nic turbulence & $l_A$ \\
transition scale from weak to strong MHD turbulence in sub-Alfv$\acute{e}$nic turbulence & $l_{tr}$ \\
turbulent velocity at $L$ & $V_L$ \\
turbulent velocity at $l$  & $v_l$ \\
turbulent velocity at $l_{tr}$  & $v_{tr}$ \\
Alfv$\acute{e}$nic Mach number & $M_A$ \\
damping scale & $l_d$ \\
ion velocity & $\mathbf{v_i}$ \\
neutral velocity & $\mathbf{v_n}$ \\
mean free path for a neutral particle & $l_n$ \\
cross section for a neutral-neutral collision & $\sigma_{nn}$ \\
%rate of damping by neutral-ion collisions & $\Gamma_{ni}$ \\
%rate of damping by neutral viscosity & $\Gamma_{nn}$ \\
speed of light & $c$ \\
Lorentz factor & $\Gamma$ \\
CR particle's velocity & $v$ \\
CR particle's gyrofrequency & $\Omega$ \\
CR particle's non-relativistic gyrofrequency & $\Omega_0$ \\
CR particle's pitch angle cosine & $\mu$ \\
pitch angle diffusion coefficient & $D_{\mu\mu}$ \\
scattering frequency & $\nu$ \\  
growth rate of streaming instability & $\Gamma_\text{g,CR}$ \\
streaming speed of CRs & $v_\text{stream}$ \\
number density of CRs & $n_\text{CR}$ \\ 
nonlinear damping rate & $\Gamma_\text{turb}$ \\
\hline
\end{longtable}

% the table should be added 

\section{Approximate solutions to single-fluid dispersion relation}
\label{app:b}
Here we list the analytical solutions to the single-fluid dispersion relation at some tractable limits following the method described in Section \ref{sec: sgfapp}.

1. $\theta\rightarrow0^{\circ}$  
\begin{equation}
        \lambda_{1,2}=\pm c_n, \lambda_{3,4}=\frac{V_A}{2}(-i\tilde{k}\pm\sqrt{4-\tilde{k}^2}).
\end{equation}

2. $\theta\rightarrow90^{\circ}$
\begin{equation}
        \lambda_{1,2}=\pm c_n.
\end{equation}

3. $\beta\rightarrow0$ 
\begin{equation}
\begin{array}{lcl}
        \lambda_{1,2}=\pm c_n\cos{\theta} \\
        \lambda_{3,4}=\frac{V_A}{2}(-i\tilde{k}\sec{\theta}\pm\sqrt{4-\tilde{k}^2\sec^2{\theta}}).\\  
\end{array}       
\end{equation}

4. $\beta\rightarrow\infty$ 
\begin{equation}
        \lambda_{1,2}=\pm c_n,
        \lambda_{3,4}=-\frac{iV_A\cos{\theta}(\tilde{k}\sec^2{\theta}\pm\sqrt{\tilde{k}^2\sec^4{\theta}-4})}{2}.
\end{equation}

5. $\tilde{k}\rightarrow0$ (ideal MHD case)
\begin{equation}
        \lambda_{1,2}=\pm\frac{\sqrt{c_n^2+V_A^2+\Delta}}{\sqrt{2}},
        \lambda_{3,4}=\pm\frac{\sqrt{c_n^2+V_A^2-\Delta}}{\sqrt{2}}.
\end{equation}
where
\begin{equation}
\Delta=\sqrt{(c_n^2+V_A^2)^2-4c_n^2V_A^2\cos^2{\theta}}.
\end{equation}

6. Small $\theta$ and $\beta\rightarrow0$
\begin{equation}
\begin{array}{lcl}
\lambda_{1,2}=\pm\frac{1+\cos^2{\theta}}{2}c_n\\
\lambda_3=-\frac{V_A\zeta}{2}i-\frac{4\zeta-2\tilde{k}\sec{\theta}\zeta^2+\zeta^3}{4(2\zeta^2+3\tilde{k}\sec{\theta}\zeta-4)}i\\
\lambda_4=-\frac{V_A\kappa}{2}i-\frac{4\kappa-2\tilde{k}\sec{\theta}\kappa^2+\kappa^3}{4(2\kappa^2-3\tilde{k}\sec{\theta}\kappa-4)}i
\end{array}   
\end{equation}
where 
\begin{equation}
  \zeta=\tilde{k}-i\sqrt{4-\tilde{k}^2}, \kappa=\tilde{k}+i\sqrt{4-\tilde{k}^2}.
\end{equation} 

7. Small $\theta$ and $\beta\rightarrow\infty$
\begin{equation}
\begin{array}{lcl}
\lambda_{1,2}=\pm c_n\\
\lambda_3=-\frac{V_A\zeta}{2}i-\frac{4\cos^2{\theta}-2\tilde{k}\sec{\theta}\zeta+\zeta^2}{4(\tilde{k}\sec{\theta}-\zeta)}i\\
\lambda_4=-\frac{V_A\kappa}{2}i-\frac{4\cos^2{\theta}-2\tilde{k}\sec{\theta}\kappa+\kappa^2}{4(\tilde{k}\sec{\theta}-\kappa)}i
\end{array}   
\end{equation}

8. $\theta\leq90^{\circ}$
\begin{equation}
\begin{array}{lcl}
        \lambda_1=c_n+\frac{c_n\sin^2{\theta}V_A^2}{2(c_n^2+ic_n\tilde{k}\sec{\theta}V_A-V_A^2)},\\
        \lambda_2=-c_n-\frac{c_n\sin^2{\theta}V_A^2}{2(c_n^2-ic_n\tilde{k}\sec{\theta}V_A-V_A^2)},\\
        \lambda_3=-\frac{iV_A}{\tilde{k}\sec^3{\theta}}.
\end{array}
\end{equation} 

9. Low $\beta$ and $\theta\rightarrow0$
\begin{equation}
\lambda_{1,2}=\pm c_n,
\lambda_3=-\frac{V_A\zeta}{2}i,
\lambda_4=-\frac{V_A\kappa}{2}i.
\end{equation} 

10. Low $\beta$ and $\theta\rightarrow90^{\circ}$
\begin{equation}
\lambda_{1,2}=\frac{V_A}{2}(-i\tilde{k}\tan{\theta}\pm\sqrt{4-\tilde{k}^2\tan{\theta}^2}).
\end{equation} 

11. High $\beta$ and $\theta\rightarrow0$
\begin{equation}
\lambda_{1,2}=\pm c_n,
\lambda_{3,4}=-\frac{V_A(\tilde{k}\pm\sqrt{\tilde{k}^2-4})}{2}i.
\end{equation} 

12. High $\beta$ and $\theta\rightarrow90^{\circ}$
\begin{equation}
\lambda_{1,2}=\pm c_n,
\lambda_3=-iV_A\tilde{k}\sec{\theta}.
\end{equation} 

13. Small $\tilde{k}$ and $\beta\rightarrow0$
\begin{equation}
\begin{array}{cl}
\lambda_{1,2}=\pm \frac{c_n}{\sqrt{2}},\\
\lambda_3=V_A-\frac{i\tilde{k}V_A\sec{\theta}}{3i\tilde{k}\sec{\theta}+2},
\lambda_4=-V_A+\frac{\tilde{k}V_A\sec{\theta}}{3\tilde{k}\sec{\theta}+2i}.
\end{array}   
\end{equation}

14. Small $\tilde{k}$ and $\beta\rightarrow\infty$
\begin{equation}
\begin{array}{cl}
\lambda_{1,2}=\pm c_n,\\
\lambda_3=\frac{V_A}{\sqrt{2}}-\frac{i\tilde{k}V_A\sec{\theta}}{\sqrt{2}(i\tilde{k}\sec{\theta}+\sqrt{2})},
\lambda_4=-\frac{V_A}{\sqrt{2}}-\frac{i\tilde{k}V_A\sec{\theta}}{\sqrt{2}(-i\tilde{k}\sec{\theta}+\sqrt{2})}.
\end{array}   
\end{equation}

15. Small $\tilde{k}$ and $\theta\rightarrow0$
\begin{equation}
\begin{array}{cl}
\lambda_{1,2}=\pm c_n,\\
\lambda_3=V_A+\frac{i\tilde{k}V_A^2(V_A^2-c_n^2)}{i\tilde{k}V_A(c_n^2-3V_A^2)+2V_A(c_n^2-V_A^2)}, 
\lambda_4=-V_A+\frac{i\tilde{k}V_A^2(V_A^2-c_n^2)}{-i\tilde{k}V_A(c_n^2-3V_A^2)+2V_A(c_n^2-V_A^2)}.
\end{array}   
\end{equation}

16. Small $\tilde{k}$ and $\theta\rightarrow90^{\circ}$
\begin{equation}
\lambda_{1,2}=\pm \frac{2(V_A^2+c_n^2)^{\frac{3}{2}}}{3V_A^2+2c_n^2}. 
\end{equation}

\section{Critical scales of ion slow modes considering scale-dependent anisotropy}
\label{app:csc}

By substituting the scaling relations described in Eq. \eqref{eq: supscal} and \eqref{eq: subscal} into Eq. \eqref{eq: kct1}, \eqref{eq: kdt1}, \eqref{eq: kdt2}, 
and \eqref{eq: kct2}, we get expressions 
\begin{equation}
     k_\text{c,t1}=\bigg(\frac{\nu_{ni}}{2\sqrt{\xi_i} c_{si}}\bigg)^\frac{3}{4}L^{-\frac{1}{4}}M_A^\frac{3}{4},  ~~ k_\text{c,t2}=\bigg(\frac{2\nu_{ni}}{\sqrt{\xi_i}c_{si}}\bigg)^\frac{3}{2}L^\frac{1}{2}M_A^{-\frac{3}{2}}, 
\end{equation}
\begin{equation}
     k_\text{dec,t1}=\bigg(\frac{\nu_{ni}}{\sqrt{\xi_i} c_{si}}\bigg)^\frac{3}{4}L^{-\frac{1}{4}}M_A^\frac{3}{4},  ~~ k_\text{dec,t2}=\bigg(\frac{\nu_{ni}}{\sqrt{\xi_i}c_{si}}\bigg)^\frac{3}{2}L^\frac{1}{2}M_A^{-\frac{3}{2}}, 
\end{equation}
for $1/k<l_A$ in super-Alfv\'{e}nic turbulence, and 
\begin{equation}
     k_\text{c,t1}=\bigg(\frac{\nu_{ni}}{2\sqrt{\xi_i} c_{si}}\bigg)^\frac{3}{4}L^{-\frac{1}{4}}M_A,  ~~ k_\text{c,t2}=\bigg(\frac{2\nu_{ni}}{\sqrt{\xi_i}c_{si}}\bigg)^\frac{3}{2}L^\frac{1}{2}M_A^{-2}, 
\end{equation}
\begin{equation}
     k_\text{dec,t1}=\bigg(\frac{\nu_{ni}}{\sqrt{\xi_i} c_{si}}\bigg)^\frac{3}{4}L^{-\frac{1}{4}}M_A,  ~~ k_\text{dec,t2}=\bigg(\frac{\nu_{ni}}{\sqrt{\xi_i}c_{si}}\bigg)^\frac{3}{2}L^\frac{1}{2}M_A^{-2}, 
\end{equation}
for $1/k<l_{tr}$ in sub-Alfv\'{e}nic turbulence, under the consideration of strong turbulence anisotropy.

\section{Force analysis of magnetoacoustic waves}
\label{app:fa}
The wave behavior can be better understood with the aid of force analysis.
Here we follow the method in 
\citet{Sol13} for Alfv\'{e}n waves to 
perform an analysis on the forces for magnetoacoustic waves. 
We start from the momentum equations of ions and neutrals
(see e.g. \citealt{Zaqa11}), 
\begin{subequations}
\begin{align}
 -i \omega v_{ix}&=-i \frac{c_{si}^2 k_x^2}{\omega} (v_{ix}+v_{iz})-i\frac{2V_{Ai}^2k_x^2}{\omega} v_{ix} \nonumber \\
 & ~~~~~  -\nu_{in} (v_{ix}-v_{nx}), \\
 -i \omega v_{iz}&=-i \frac{c_{si}^2 k_x^2}{\omega} (v_{ix}+v_{iz})-\nu_{in}(v_{iz}-v_{nz}), \\
 -i \omega v_{nx}&=-i \frac{c_{sn}^2 k_x^2}{\omega} (v_{nx}+v_{nz})+\nu_{ni}(v_{ix}-v_{nx}), \\
  -i \omega v_{nz}&=-i \frac{c_{sn}^2 k_x^2}{\omega} (v_{nx}+v_{nz})+\nu_{ni}(v_{iz}-v_{nz}). 
\end{align}
\end{subequations}
Here $v_i$ and $v_n$ are velocity perturbations in ions and neutrals. We consider the constant magnetic field is along z-direction, and waves 
propagate in x-z plane. 
For a qualitative study on wave behavior, we fix the wave propagation direction as $45^\circ$ with regards to magnetic field, i.e. $k_x=k_z$, 
to simplify the algebraic procedures. 
To conduct the force analysis for ion and neutral particles in different directions, from above equations we get 
\begin{subequations} \label{eq: moeq}
\begin{align}
 f_{i,x}&=|P_i|+|M|-|F_x|, \\
 f_{i,z}&=|P_i|-|F_z|, \\
 f_{n,x}&=|P_n|+|F_x|, \\
 f_{n,z}&=|P_n|+|F_z|. 
\end{align}
\end{subequations}
Here $|P_i|$ and $|P_n|$ are gas pressure forces acting on ions and neutrals. $|M|$ is magnetic pressure force and only acts on ions in perpendicular 
direction to magnetic field. $|F_x|$ and $|F_z|$ are friction forces between neutrals and ions in x and z directions. 
The expressions of forces are not displayed for simplicity. 
We next take the parameters of MC phase and use the numerically solved $\omega$ of fast and slow waves to calculate the corresponding forces. 
To further simply the problem, we also set $m_i=2m_n$ to have equal sound speeds in neutrals and ions. 

\begin{figure*}[htbp]
\centering
\subfigure[Wave frequencies of fast modes]{
   \includegraphics[width=8cm]{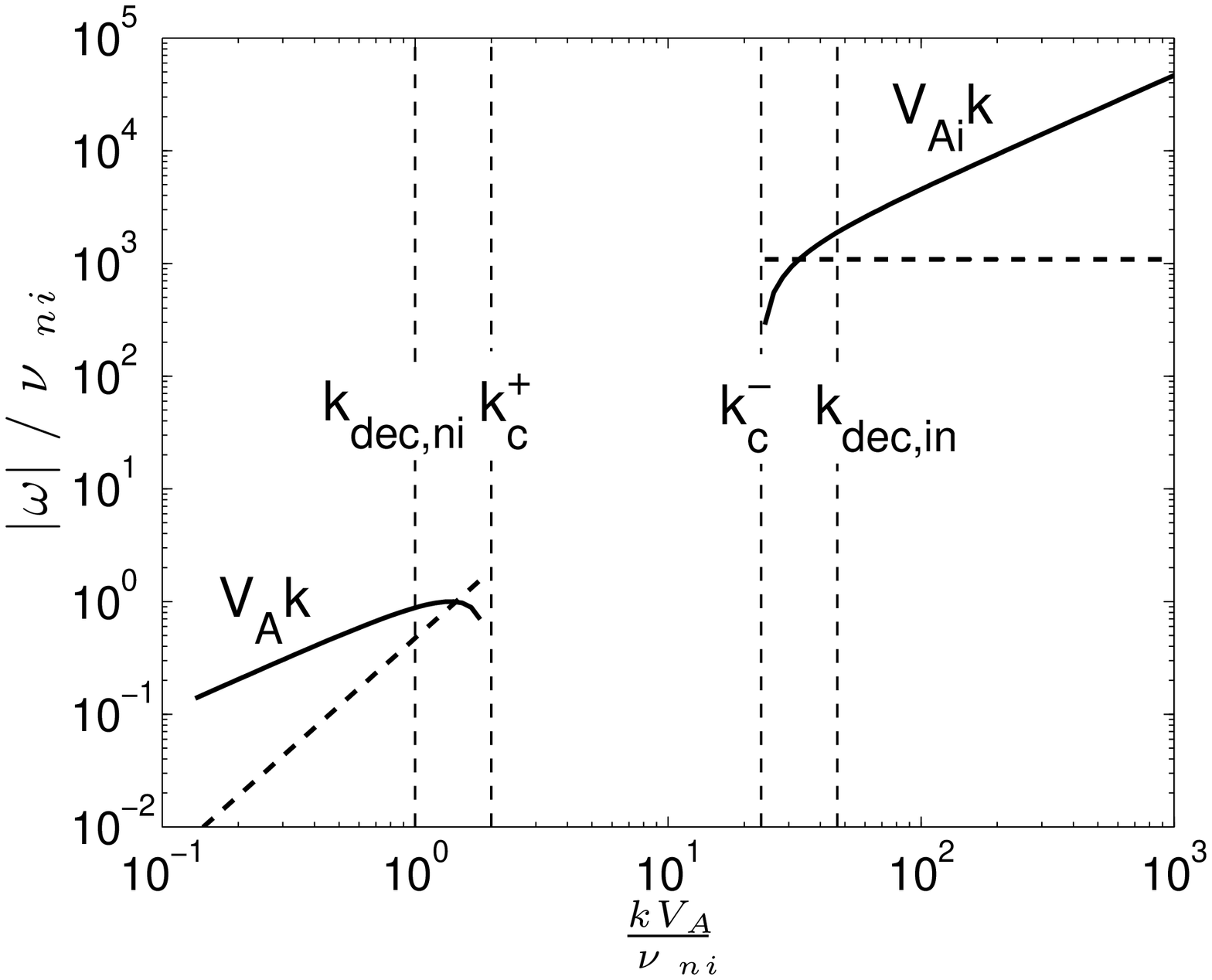}\label{figsolfast}}
\subfigure[Forces of fast modes]{
   \includegraphics[width=8cm]{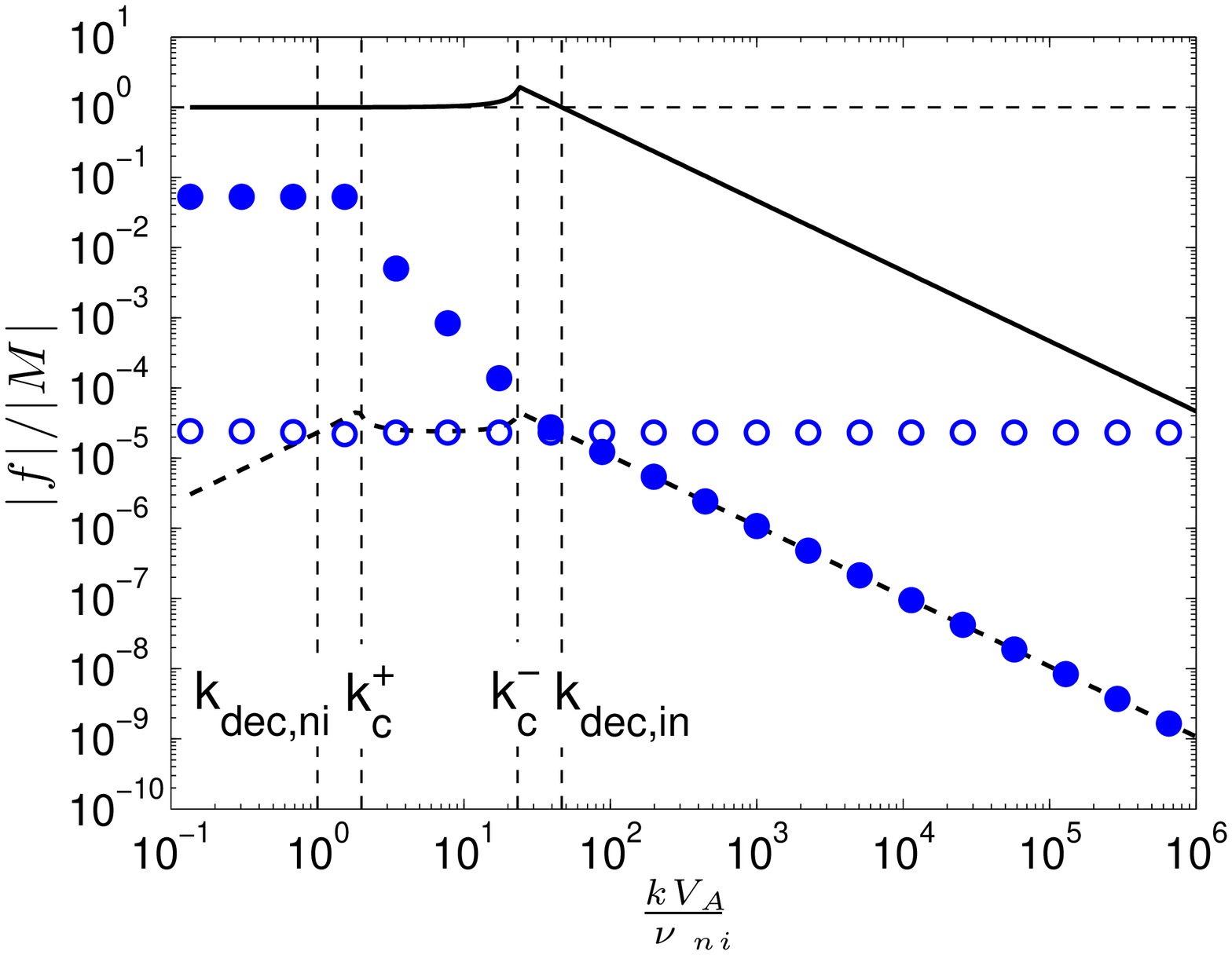}\label{figfastfc}}   
\subfigure[Wave frequencies of usual slow modes]{
   \includegraphics[width=8cm]{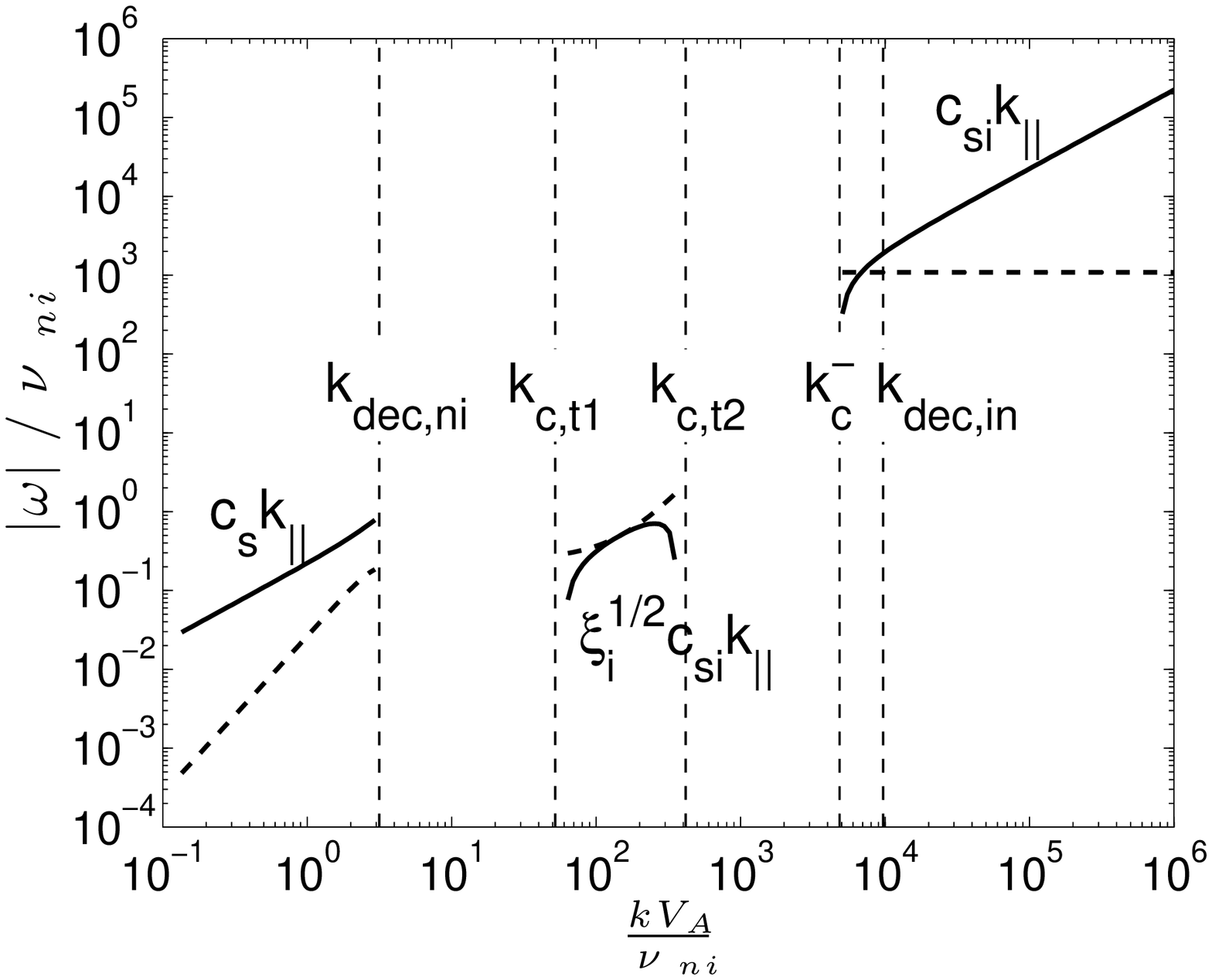}\label{figsolsli}}
\subfigure[Forces of usual slow modes]{
   \includegraphics[width=8cm]{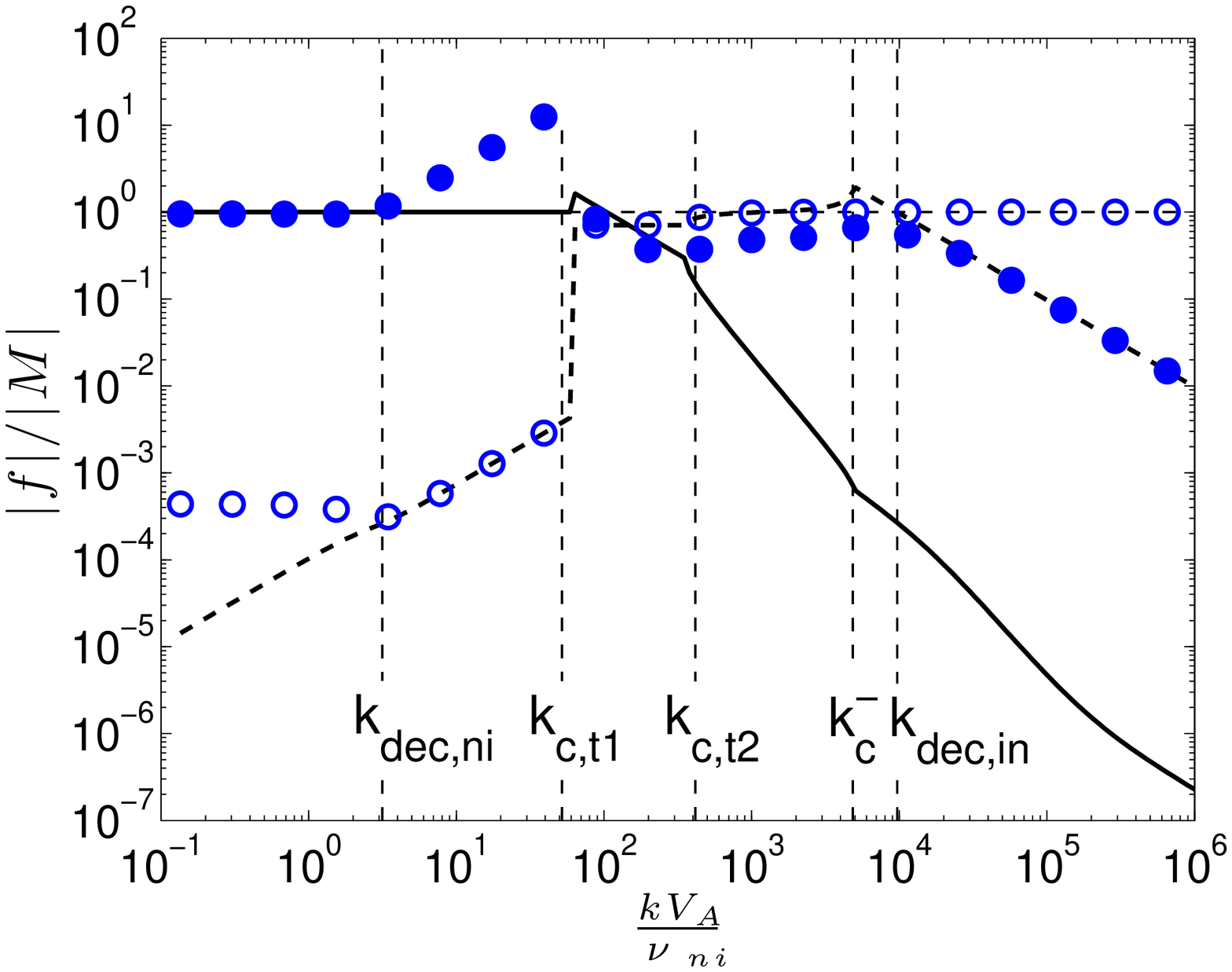}\label{figslowifc}}
\subfigure[Wave frequencies of neutral slow modes]{
   \includegraphics[width=8cm]{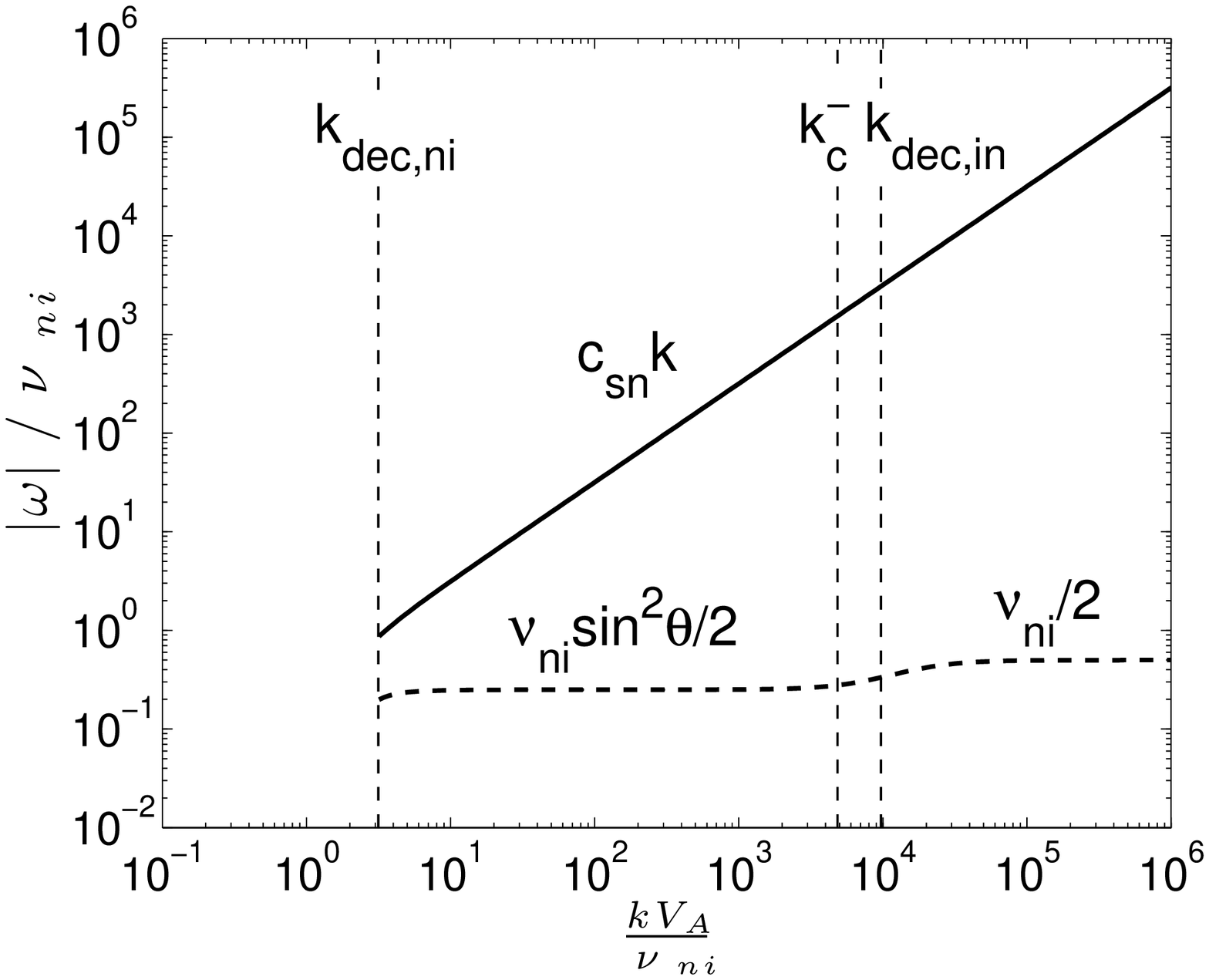}\label{figsolsln}}
\subfigure[Forces of neutral slow modes]{
   \includegraphics[width=8cm]{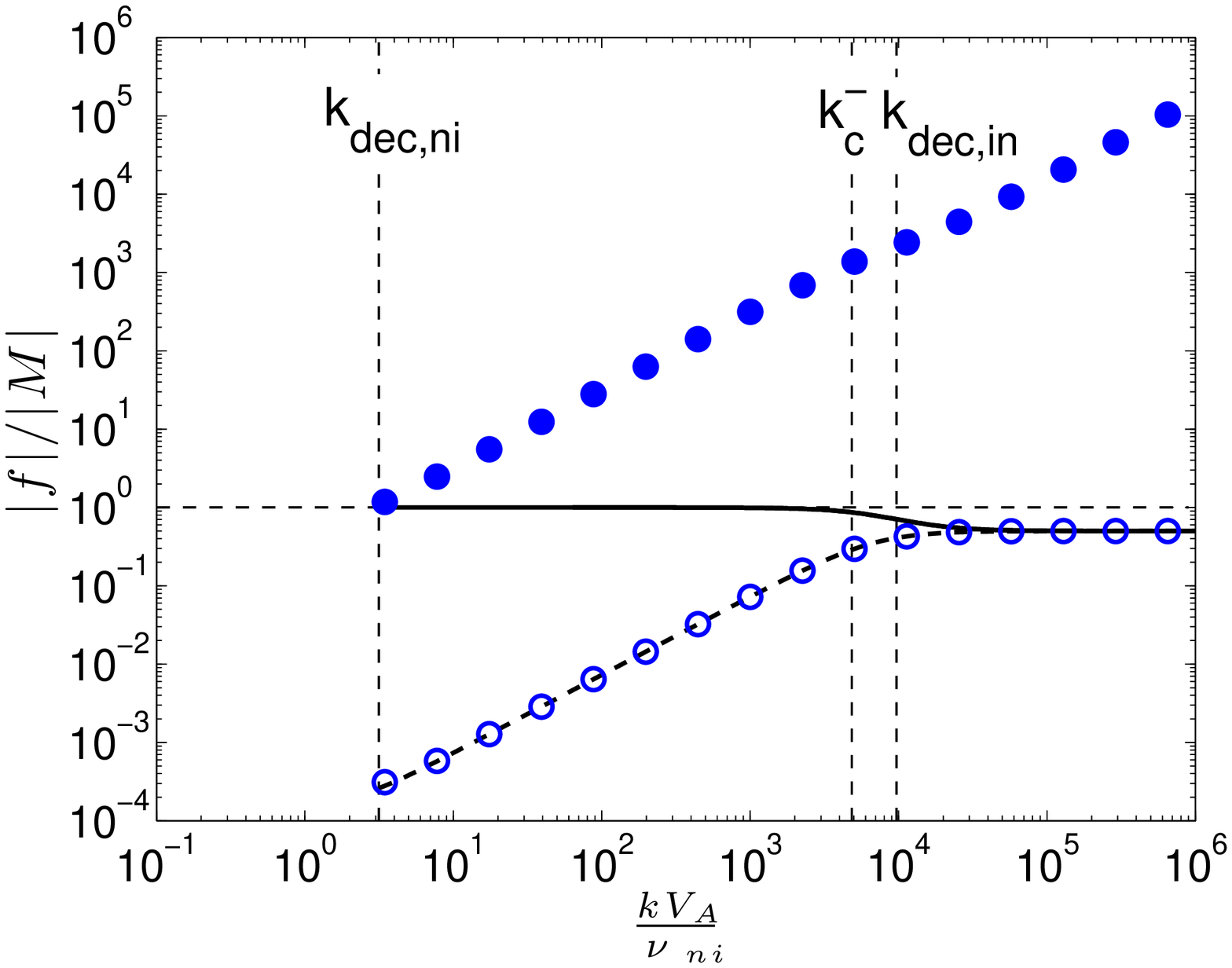}\label{figslownfc}}
   
\caption{ Wave frequencies and force ratios vs. normalized $k$ of magnetoacoustic waves using MC parameters. 
In (a), (c), and (e), the propagating ($|\omega_R|$) and damping ($|\omega_I|$) components of wave frequencies are represented by solid and dashed lines respectively. 
In (b), (d), and (f),  
the displayed force ratios are $|F_x|/|M|$ (solid line), $|F_z|/|M|$ (dashed line), $|P_i|/|M|$ (open circles), and $|P_n|/|M|$ (filled circles). 
The horizontal dashed lines indicate the position $|f|/|M|=1$. }
\label{fig: wffr}
\end{figure*}

In figure \ref{fig: wffr}, we display the wave frequencies and different forces $|f|$ normalized by $|M|$ for magnetoacoustic waves using the parameters 
of MC. The analytical expressions of $|\omega_R|$ and $|\omega_I|$ of some branches are present specifically.

(a) {\it Fast waves~~} 
We see from Fig. \ref{figfastfc}, $|F_x|\sim|M|$ (solid line) at large scales makes two fluids strongly coupled and oscillate together. 
The forces in Eq. \eqref{eq: moeq} satisfy 
\begin{equation}\label{eq: scfrela}
    \frac{f_{i,x}}{\rho_i} \approx \frac{f_{n,x}}{\rho_n} > \frac{f_{i,z}}{\rho_i} \approx \frac{f_{n,z}}{\rho_n} 
\end{equation}
to make neutrals move together with ions. The perpendicular component of force is larger due to the involvement of magnetic pressure.
$|F_z|/|M|$ (dashed line) increases towards $k_\text{dec,ni}$ and exceeds $|P_i|/|M|$ (open circles) at $k_\text{dec,ni}$. 
Within the cutoff interval, $|F_x|/|M|$ also increases to be larger than $1$. 
The strong friction force acting on ions efficiently dissipates the oscillatory motions of ions.
In the mean time, 
due to the weak coupling with ions, $|P_n|/|M|$ (filled circles) in neutrals substantially drops to the same level as $|P_i|/|M|$ after the cutoff interval, 
and further decreases together with $|F_z|/|M|$ towards smaller scales. Accordingly, no wave motions can be developed in neutrals after they 
decouple from ions at $k_\text{dec,ni}$.

At scales smaller than $1/k_\text{dec, in}$, the forces have relations 
\begin{subequations} 
\begin{align}
   &  \frac{f_{i,x}}{\rho_i} \gg \frac{f_{n,x}}{\rho_n}, ~\frac{f_{i,z}}{\rho_i} \gg \frac{f_{n,z}}{\rho_n}, \\
   &  f_{i,x} \gg f_{i,z}, ~f_{n,x} \gg f_{n,z}. 
\end{align}
\end{subequations}
With decreasing frictions in both directions, 
ions oscillate under the pressure forces, and marginally affected by neutrals. The damped fast waves reemerge, but only in ions, with a phase 
speed $V_{Ai}$ instead of $V_A$ (shown in Fig. \ref{figsolfast}).

(b) {\it Usual slow waves~~}
Owning to the constant propagation direction ($\theta=45^\circ$) adopted, we observe different wave behavior of the 
slow waves here (Fig. \ref{figsolsli}) from that in Fig. \ref{figslowmc} despite the same environment parameters used. 
First, since the relation $k_\| \ll k_\perp \sim k$ breaks down, $k_\text{dec,ni}$ is smaller than $k_c^+$ (see Eq. \eqref{eq: comsdc} and \eqref{eq: comscf}). 
Thus $k_\text{dec,ni}$ becomes the cutoff scale of the usual slow waves in this case. 
Second, according to the expression in Table \ref{Tab: sloscwf}, 
the ratio $k_\text{c,t1}/k_\text{c,t2}$ substantially increases due to the relatively large value of $\cos\theta$. Meanwhile, the ratio $|\omega_I|/|\omega_R|$ 
of the ion slow waves within $[k_\text{c,t1}, k_\text{c,t2}]$ also increases, which leads to more efficient damping and a shorter wavenumber interval of this wave branch. 
In addition, with a constant $\cos\theta$, we cannot observe a decreasing $|\omega_I|$ with $k$. 
From above comparison we clearly see that the wave propagation direction can significantly influence the slow wave behavior. 

Similar analysis to fast modes applies to the usual slow modes on scales larger than $1/k_\text{dec,ni}$. 
Within the interval $[k_\text{dec,ni}, k_\text{c,t1}]$, ions are still strongly coupled with neutrals and undergo the same forces as neutrals, 
which we will discuss later for the neutral slow waves. 
The propagating ion slow waves arise at $k_\text{c,t1}$ after $|P_i|/|M|$ substantially increases and $|M|$ dominates over $|F_x|$.  
But due to the rise of $|F_z|/|M|$, the wave is strongly damped with $|\omega_I|>|\omega_R|$ (see Fig. \ref{figsolsli}) and cut off shortly at $k_\text{c,t2}$.
The ion slow modes can only be fully resumed beyond $k_\text{dec,in}$ when $|F_z|/|M|$ decreases to $1$. 
Then friction forces further decrease and become negligible compared with $|M|$ towards smaller scales. The forces have relations as 
\begin{equation}
    \frac{f_{i,x}}{\rho_i}>\frac{f_{i,z}}{\rho_i} \gg \frac{f_{n,z}}{\rho_n} > \frac{f_{n,x}}{\rho_n}.
\end{equation}

(c) {\it Neutral slow waves~~} 
As shown in Fig. \ref{figsolsln}, with $\theta=45^\circ$ used, we can clearly see the change in the 
damping rate $|\omega_I|$ of the neutral slow modes (see Eq. \eqref{eq: nsoic} and \eqref{eq: nsoid}).
Fig. \ref{figslownfc} illustrates the forces corresponding to the neutral slow modes. 
$|P_n|/|M|$ significantly increases and results in a nearly isotropic force on neutrals. The compression produced by the usual slow modes at larger scales 
drives oscillations in neutrals. Without much interference from ions, isotropically propagating sound waves are able to arise and maintain in the neutral fluid. 
Before reaching $k_\text{dec,in}$, ions are still coupled with neutrals. As a results, the forces are 
\begin{equation}
      \frac{f_{n,x}}{\rho_n} \approx \frac{f_{n,z}}{\rho_n} > \frac{f_{i,x}}{\rho_i} \gg \frac{f_{i,z}}{\rho_i}. 
\end{equation}
At scales smaller than $k_\text{dec,in}^{-1}$, $|F_x|/|M|$ drops and converges with $|F_z|/|M|$. Accordingly $f_{i,x}/\rho_i$ further decreases and 
becomes significantly smaller compared with the forces on neutrals. 

\section{The Fokker-Planck diffusion coefficients}
\label{app:c}
(a) {\it Super-Alfv\'{e}nic turbulence}~~
Since the hydrodynamic turbulence over $[L, l_A]$ does not contribute to particle scattering, 
the integral interval of $D_{\mu\mu}$ is from the injection scale of GS95 turbulence $l_A$ to the damping scale. 
In the following expressions of $D_{\mu\mu}$ for the case of super-Alfv\'{e}nic turbulence, we use $l_A$ as the injection scale $L$ and 
$M_A=1$ applies correspondingly. 

In Alfv$\acute{e}$nic turbulence, by applying NLT, we can get
%alfven mode gyro
\begin{equation}\label{eq: duusaalf}
    D_{\mu\mu}^G(\text{NLT})=\frac{4v\sqrt{\pi}(1-\mu^2) }{3LR^2} \int_1^{k_\text{dam}L} x x_\perp^{-10/3} dx \int_{0}^{1} (1+\frac{\mu V_A}{v })^2
\frac{d\eta}{\eta \Delta \mu} \frac{J_1^2(w)}{w^2} \exp\left[-\frac{x_\|}{x_\perp^{2/3}}-\frac{(\mu-\frac{1}{\eta x R})^2}{\Delta \mu ^2}\right], 
\end{equation}
where 'G' refers to gyroresonance. 
Here the dimensionless quantities are: $R=v/(\Omega L)$ (CR rigidity), $x=kL$, $\eta=\cos\theta$. 
And $\Delta \mu=\Delta v_\|/v$ represents the dispersion in particle pitch angles due to the nonlinear effect. $J_n(w)$ is Bessel function with $w=k_\perp v_\perp/\Omega$.
In QLT, with the $\delta$-function resonance condition used, $D_{\mu\mu}^G$ becomes
\begin{equation}
    D_{\mu\mu}^G(\text{QLT})=\frac{4v\pi(1-\mu^2) }{3LR^2} \int_1^{k_\text{dam}L} x^2 x_\perp^{-10/3} dx \int_{0}^{1} (1+\frac{\mu V_A}{v })^2
\frac{d\eta}{\eta \mu} \frac{J_1^2(w)}{w^2} \exp\left(-\frac{x_\|}{x_\perp^{2/3}}\right)   \delta\left(x-\frac{1}{\eta \mu R}\right).
\end{equation}

For fast modes, we obtain 
%fast mode gyro
\begin{equation}\label{eq: duufag}
D_{\mu\mu}^G(\text{NLT})=\frac{v\sqrt{\pi}(1-\mu^2)}{LR^2}\int_{1}^{k_\text{dam}L} x^{-5/2} dx  \int_{0}^{1} (1+\frac{\mu V_A}{v \eta})^2
\frac{\eta d\eta}{\Delta \mu} (J_1^{\prime}(w))^2\exp\left[-\frac{(\mu-\frac{1}{\eta x R})^2}{\Delta \mu ^2}\right],
\end{equation}
%fast mode TTD
\begin{equation}
D_{\mu\mu}^{T}(\text{NLT})=\frac{v\sqrt{\pi}(1-\mu^2)}{LR^2}\int_{1}^{k_\text{dam}L} x^{-5/2} dx  \int_{0}^{1} (1+\frac{\mu V_A}{v \eta})^2
\frac{\eta d\eta}{\Delta \mu} J_1^2(w)\exp\left[-\frac{(\mu-\frac{V_A}{\eta v})^2}{\Delta \mu ^2}\right],
\end{equation}
corresponding to gyroresonance and TTD interaction. 
For low-rigidity CRs with their $r_L$ shorter than the damping scale of turbulence, 
$J_n(w)$ can take the asymptotic form $(w/2)^n/n!$ at small argument. Then $D_{\mu\mu}^{T}(\text{NLT})$ approximately becomes
\begin{equation} 
   D_{\mu\mu}^{T}(\text{NLT})=\frac{v\sqrt{\pi}(1-\mu^2)^2}{8L\Delta \mu} \exp \bigg(-\frac{\mu^2}{(\Delta \mu)^2}\bigg)(\sqrt{k_\text{dam}L}-1).
\end{equation} 

The $D_{\mu\mu}^G$ in QLT takes the form 
\begin{equation}
D_{\mu\mu}^G(\text{QLT})=\frac{v\pi(1-\mu^2)}{LR^2}\int_{1}^{k_\text{dam}L} x^{-3/2} dx  \int_{0}^{1} (1+\frac{\mu V_A}{v \eta})^2
\frac{\eta d\eta}{\mu} (J_1^{\prime}(w))^2 \delta\left(x-\frac{1}{\eta \mu R}\right).
\end{equation}
Its simplified form for low-rigidity CRs is 
\begin{equation}
D_{\mu\mu}^G(\text{QLT})=\frac{v\pi\sqrt{\mu}(1-\mu^2)}{4L\sqrt{R}} \bigg( \frac{2}{7}-\frac{2\sqrt{1-\mu^2}}{21 \mu^2}    \bigg).
\end{equation}

Diffusion coefficients in the case of slow modes are 
%slow mode gyro
\begin{equation} \label{eq: slduug}
    D_{\mu\mu}^G(\text{NLT})=\frac{v\sqrt{\pi}(1-\mu^2) \beta^2}{3LR^2} \int_1^{k_\text{dam}L} x x_\perp^{-10/3} dx \int_{0}^{1} (1+\frac{\mu c_s}{v })^2 (1-\eta^2)
\frac{\eta^3d\eta}{ \Delta \mu} (J_1^{\prime}(w))^2 \exp\left[-\frac{x_\|}{x_\perp^{2/3}}-\frac{(\mu-\frac{1}{\eta x R})^2}{\Delta \mu ^2}\right],
\end{equation}
%slow mode TTD
\begin{equation} \label{eq: slduut}
    D_{\mu\mu}^T(\text{NLT})=\frac{v\sqrt{\pi}(1-\mu^2) \beta^2}{3LR^2} \int_1^{k_\text{dam}L}x x_\perp^{-10/3} dx \int_{0}^{1} (1+\frac{\mu c_s}{v })^2 (1-\eta^2)
    \frac{\eta^3 d\eta}{\Delta \mu}J_1^2(w)\exp\left[-\frac{x_\|}{x_\perp^{2/3}}-\frac{(\mu-\frac{c_s}{v})^2}{\Delta \mu ^2}\right],
\end{equation}
\begin{equation} \label{eq: slduugq}
    D_{\mu\mu}^G(\text{QLT})=\frac{v\pi(1-\mu^2) \beta^2}{3LR^2} \int_1^{k_\text{dam}L} x^2 x_\perp^{-10/3} dx \int_{0}^{1} (1+\frac{\mu c_s}{v })^2 (1-\eta^2)
\frac{\eta^3d\eta}{ \mu} (J_1^{\prime}(w))^2  \exp\left(-\frac{x_\|}{x_\perp^{2/3}}\right)     \delta\left(x-\frac{1}{\eta \mu R}\right).
\end{equation}

(b) {\it  Sub-Alfv\'{e}nic turbulence}~~
Sub-Alfv\'{e}nic turbulence have different energy spectra for weak and strong turbulence regimes. Accordingly, the diffusion coefficient is a sum 
of the components in both weak and strong turbulence. 

For the gyroresonance with Alfv\'{e}n modes, we have $D_{\mu\mu}^G=D_{\mu\mu,w}^G+D_{\mu\mu,s}^G$. Here 
\begin{equation}
    D_{\mu\mu,w}^G(\text{NLT})=\frac{2v\sqrt{\pi}(1-\mu^2) M_A^{2}}{LR^2} \int_1^{M_A^{-2}} x_\perp^{-2} dx_\perp (1+\frac{\mu V_A}{v })^2
\frac{1}{\Delta \mu} \frac{J_1^2(w)}{w^2} \exp\left[-\frac{(\mu-\frac{1}{ R})^2}{\Delta \mu ^2}\right],
\end{equation}
is $D_{\mu\mu}^G$ in weak turbulence, and 
\begin{equation}
    D_{\mu\mu,s}^G(\text{NLT})=\frac{4v\sqrt{\pi}(1-\mu^2) M_A^{4/3}}{3LR^2} \int_{M_A^{-2}}^{k_\text{dam}L} x x_\perp^{-10/3} dx \int_{0}^{1} (1+\frac{\mu V_A}{v })^2
\frac{d\eta}{\eta \Delta \mu} \frac{J_1^2(w)}{w^2} \exp\left[-\frac{x_\|}{x_\perp^{2/3}M_A^{4/3}}-\frac{(\mu-\frac{1}{\eta x R})^2}{\Delta \mu ^2}\right],
\end{equation}
is $D_{\mu\mu}^G$ in strong turbulence. We can see $D_{\mu\mu,s}^G(\text{NLT})$ is similar to the $D_{\mu\mu}^G(\text{NLT})$ in super-Alfv\'{e}nic 
case (Eq. \eqref{eq: duusaalf}), but with a factor $M_A^{4/3}$ added ($M_A<1$). 

In QLT, since weak turbulence only has smaller structures developed in perpendicular direction, i.e. $k_\|=L^{-1}$ over $[L, l_{tr}]$, 
weak turbulence does not contribute to gyroresonance scattering unless CRs have $r_L$ comparable to $L$. Thus we only consider the contribution of strong turbulence and get 
\begin{equation}
    D_{\mu\mu,s}^G(\text{QLT})=\frac{4v\pi(1-\mu^2) M_A^{4/3}}{3LR^2} \int_{M_A^{-2}}^{k_\text{dam}L} x^2 x_\perp^{-10/3} dx \int_{0}^{1} (1+\frac{\mu V_A}{v })^2
\frac{d\eta}{\eta \mu} \frac{J_1^2(w)}{w^2} \exp\left(-\frac{x_\|}{x_\perp^{2/3}M_A^{4/3}}\right)   \delta\left(x-\frac{1}{\eta \mu R}\right).
\end{equation}

The energy spectrum of fast modes are not dependent on turbulence regimes. Fast modes have similar diffusion coefficients in sub-Alfv\'{e}nic turbulence 
as those in super-Alfv\'{e}nic case. The expressions Eq. \eqref{eq: duufag}-\eqref{eq: duufagqs} still apply, but here $L$ is the driving scale of turbulence 
and $M_A^2$ should be multiplied accounting for the ratio between kinetic and magnetic energies of turbulence.

Slow modes have 
\begin{equation} \label{eq: slduug}
    D_{\mu\mu,w}^G(\text{NLT})=\frac{v\sqrt{\pi}(1-\mu^2) \beta^2 M_A^{2}}{2LR^2} \int_1^{M_A^{-2}} x_\perp^{-2} dx_\perp (1+\frac{\mu c_s}{v })^2 (1-\eta^2)
\frac{\eta^4}{ \Delta \mu} (J_1^{\prime}(w))^2 \exp\left[-\frac{(\mu-\frac{1}{ R})^2}{\Delta \mu ^2}\right],
\end{equation}
\begin{equation} \label{eq: slduut}
    D_{\mu\mu,w}^T(\text{NLT})=\frac{v\sqrt{\pi}(1-\mu^2) \beta^2 M_A^{2}}{2LR^2} \int_1^{M_A^{-2}}x_\perp^{-2} dx_\perp (1+\frac{\mu c_s}{v })^2 (1-\eta^2)
    \frac{\eta^4 }{\Delta \mu}J_1^2(w)\exp\left[-\frac{(\mu-\frac{c_s}{v})^2}{\Delta \mu ^2}\right].
\end{equation}
in weak turbulence, and 
\begin{equation} \label{eq: slduug}
    D_{\mu\mu,s}^G(\text{NLT})=\frac{v\sqrt{\pi}(1-\mu^2) \beta^2 M_A^{4/3}}{3LR^2} \int_{M_A^{-2}}^{k_\text{dam}L} x x_\perp^{-10/3} dx \int_{0}^{1} (1+\frac{\mu c_s}{v })^2 (1-\eta^2)
\frac{\eta^3d\eta}{ \Delta \mu} (J_1^{\prime}(w))^2 \exp\left[-\frac{x_\|}{x_\perp^{2/3}M_A^{4/3}}-\frac{(\mu-\frac{1}{\eta x R})^2}{\Delta \mu ^2}\right],
\end{equation}
\begin{equation} \label{eq: slduut}
    D_{\mu\mu,s}^T(\text{NLT})=\frac{v\sqrt{\pi}(1-\mu^2) \beta^2 M_A^{4/3}}{3LR^2} \int_{M_A^{-2}}^{k_\text{dam}L}x x_\perp^{-10/3} dx \int_{0}^{1} (1+\frac{\mu c_s}{v })^2 (1-\eta^2)
    \frac{\eta^3 d\eta}{\Delta \mu}J_1^2(w)\exp\left[-\frac{x_\|}{x_\perp^{2/3}M_A^{4/3}}-\frac{(\mu-\frac{c_s}{v})^2}{\Delta \mu ^2}\right],
\end{equation}
in strong turbulence. 

QLT in strong turbulence gives 
\begin{equation} \label{eq: slduugq}
    D_{\mu\mu}^G(\text{QLT})=\frac{v\pi(1-\mu^2) \beta^2 M_A^{4/3}}{3LR^2} \int_{M_A^{-2}}^{k_\text{dam}L} x^2 x_\perp^{-10/3} dx \int_{0}^{1} (1+\frac{\mu c_s}{v })^2 (1-\eta^2)
\frac{\eta^3d\eta}{ \mu} (J_1^{\prime}(w))^2  \exp\left(-\frac{x_\|}{x_\perp^{2/3} M_A^{4/3}}\right)     \delta\left(x-\frac{1}{\eta \mu R}\right).
\end{equation}

\bibliographystyle{apj.bst}
\bibliography{yan}

\end{document}